%% file: paper_twcccreport.tex
\documentclass[nobibstyle,review]{twcccreport}
\reportnumber{2024-04}
\usepackage{amsmath,amsfonts,amssymb}
\usepackage{hyperref}
\usepackage{graphics,graphicx}
\newcommand{\email}[1]{\protect\href{mailto:#1}{#1}}

\usepackage[authoryear,round]{natbib}

\usepackage{titlesec}
\titleformat{\paragraph}[runin]{\normalfont\bfseries}{\theparagraph.}{}{}[:]   %%<--- this one

\usepackage{ifthen}
\newboolean{LongVersion}
\setboolean{LongVersion}{true} % true or false
\newboolean{OneColumn}
\setboolean{OneColumn}{true} % true or false

\usepackage[caption=false, font=footnotesize]{subfig}
\usepackage{comment}
\usepackage{braket}
\usepackage{mathtools}
\usepackage[shortlabels]{enumitem}
\usepackage{orcidlink}

\RequirePackage[amsthm,amsmath,thmmarks,hyperref]{ntheorem}
\usepackage[capitalize,nameinlink]{cleveref}

\newtheorem{theorem}{Theorem}

\newtheorem{proposition}{Proposition}
\newtheorem{lemma}{Lemma}

\theoremstyle{definition}
\newtheorem{definition}{Definition}
\newtheorem{assumption}{Assumption}
\crefname{assumption}{Assumption}{Assumptions}

\newtheorem{remark}{Remark}
\newtheorem{example}{Example}

\newcommand{\QED}{$\square$}

\DeclareMathOperator*{\argmin}{argmin}

\newcommand{\lev}{\textnormal{lev}}
\newcommand{\id}{\textsc{id}}

\newcommand{\defas}{:=}
\newcommand{\asdef}{=:}

\newcommand{\real}{\mathbb{R}}
\newcommand{\posreal}{\mathbb{R}_{>0}}
\newcommand{\nnegreal}{\mathbb{R}_{\geq 0}}

\newcommand{\realm}[2]{\mathbb{R}^{#1\times #2}}

\newcommand{\allint}{\mathbb{I}}
\newcommand{\nnegint}{\mathbb{I}_{\geq 0}}
\newcommand{\posint}{\mathbb{I}_{>0}}
\newcommand{\intinterval}[2]{\mathbb{I}_{#1:#2}}

\newcommand{\norm}{\mathrm{N}}

\newcommand{\calK}{\mathcal{K}}
\newcommand{\calKinf}{{\mathcal{K}}_{\infty}}

\newcommand{\calKL}{\mathcal{KL}}

\newcommand{\ssp}{s_{\textnormal{sp}}}
\newcommand{\rsp}{r_{\textnormal{sp}}}

\newcommand{\usp}{u_{\textnormal{sp}}}
\newcommand{\ysp}{y_{\textnormal{sp}}}
\newcommand{\zsp}{z_{\textnormal{sp}}}

\newcommand{\fP}{f_{\textnormal{P}}}
\newcommand{\FP}{F_{\textnormal{P}}}
\newcommand{\hP}{h_{\textnormal{P}}}
\newcommand{\wP}{w_{\textnormal{P}}}
\newcommand{\xP}{x_{\textnormal{P}}}
\newcommand{\xPs}{x_{\textnormal{P,s}}}

\newcommand{\zP}{z_{\textnormal{P}}}
\newcommand{\zPs}{z_{\textnormal{P,s}}}

\newcommand{\bfd}{\mathbf{d}}
\newcommand{\bfe}{\mathbf{e}}

\newcommand{\bfr}{\mathbf{r}}
\newcommand{\bfu}{\mathbf{u}}

\newcommand{\bfw}{\mathbf{w}}
\newcommand{\bfx}{\mathbf{x}}
\newcommand{\bfy}{\mathbf{y}}
\newcommand{\bfz}{\mathbf{z}}
\newcommand{\bfbeta}{\boldsymbol{\beta}}
\newcommand{\bfxi}{\boldsymbol{\xi}}
\newcommand{\bfomega}{\boldsymbol{\omega}}
\newcommand{\bfepsilon}{\boldsymbol{\varepsilon}}
\newcommand{\bfupsilon}{\boldsymbol{\upsilon}}
\newcommand{\bfzeta}{\boldsymbol{\zeta}}

\newcommand{\bbA}{\mathbb{A}}
\newcommand{\bbB}{\mathbb{B}}
\newcommand{\bbD}{\mathbb{D}}
\newcommand{\bbU}{\mathbb{U}}
\newcommand{\bbW}{\mathbb{W}}
\newcommand{\bbV}{\mathbb{V}}
\newcommand{\bbX}{\mathbb{X}}
\newcommand{\bbY}{\mathbb{Y}}
\newcommand{\bbZ}{\mathbb{Z}}

\newcommand{\calA}{\mathcal{A}}
\newcommand{\calB}{\mathcal{B}}

\newcommand{\calS}{\mathcal{S}}
\newcommand{\calU}{\mathcal{U}}
\newcommand{\calX}{\mathcal{X}}
\newcommand{\calZ}{\mathcal{Z}}

\begin{document}

\title{Offset-free model predictive control: stability under plant-model mismatch%
  %% TODO Cite published version
  \thanks{This report is an extended version of a submitted paper. This work was
    supported by the National Science Foundation (NSF) under Grant 2138985.
    (e-mail: \email{skuntz@ucsb.edu}; \email{jbraw@ucsb.edu})} %
  \thanks{Version 2 includes additional technical discussions
    (\Cref{rem:quad,rem:ras,rem:ras:ctrl:est,rem:mpc:nominal,rem:mpc:robust,%
      rem:mpc:robust:tradeoffs,rem:mpc:mismatch,rem:mpc:mismatch:tradeoffs,%
      rem:sstp,rem:sstp:mismatch}) and a new section (\Cref{sec:linear}) where
    \Cref{lem:sstp,lem:sstp:mismatch} have been moved, and with additional
    commentary on connections to linear systems
    (\Cref{rem:sstp,rem:sstp:mismatch}). The main technical results remain
    unchanged.}%
}
\author{Steven J.~Kuntz and James B.~Rawlings \\
  Department of Chemical Engineering \\
  University of California, Santa Barbara}

\maketitle

\input{paper_abstract.tex}

\input{paper_body.tex}

\appendix
\input{paper_appendix.tex}

\bibliographystyle{plainnat}
\bibliography{paper_twcccreport}

\end{document}

%% file: paper_abstract.tex
\begin{abstract}
  We present the first general stability results for nonlinear offset-free model
  predictive control (MPC). Despite over twenty years of active research, the
  offset-free MPC literature has not shaken the assumption of closed-loop
  stability for establishing offset-free performance. In this paper, we present
  a nonlinear offset-free MPC design that is robustly stable with respect to the
  tracking errors, and thus achieves offset-free performance, despite
  plant-model mismatch and persistent disturbances. Key features and assumptions
  of this design include quadratic costs, differentiability of the plant and
  model functions, constraint backoffs at steady state, and a robustly stable
  state and disturbance estimator. We first establish nominal stability and
  offset-free performance. Then, robustness to state and disturbance estimate
  errors and setpoint and disturbance changes is demonstrated. Finally, the
  results are extended to sufficiently small plant-model mismatch. The results
  are illustrated %
  \ifthenelse{\boolean{LongVersion}}{%
    by numerical examples. %
  }{%
    in a nonlinear reactor example. %
  }%
\end{abstract}

%% file: paper_body.tex
\section{Introduction}\label{sec:intro}
%% Offset-free MPC
\ifthenelse{\boolean{OneColumn}}{%
  Offset-free %
}{%
  \IEEEPARstart{O}{ffset-free} %
}%
model predictive control (MPC) is a popular advanced control method for
offset-free tracking of setpoints despite plant-model mismatch and persistent
disturbances. This is accomplished by combining regulation, estimation, and
steady-state target problems, each designed with a state-space model that is
augmented with uncontrollable integrating modes, called \emph{integrating
  disturbances}, that provide integral action through the estimator. Despite
over twenty years of applied use and active research, there are no results on
the stability of nonlinear offset-free MPC.

%% Offset-free conditions: linear models
Sufficient conditions for which linear offset-free MPC stability implies
offset-free performance were first established
by~\cite{muske:badgwell:2002,pannocchia:rawlings:2003}. %
\ifthenelse{\boolean{LongVersion}}{%
  While \cite{muske:badgwell:2002,pannocchia:rawlings:2003} do not explicitly
  mention control of nonlinear plants, the results are widely applicable to both
  linear and nonlinear plants with asymptotically constant disturbances, as
  controller stability is assumed rather than explicitly demonstrated. In
  fact,~\cite{pannocchia:rawlings:2003} demonstrate offset-free control on a
  highly nonlinear, non-isothermal reactor model. %
}{%
  The results are widely applicable to both linear and nonlinear plants, as
  linearity of the plant is not specifically required. In
  fact,~\cite{pannocchia:rawlings:2003} demonstrate offset-free control on a
  highly nonlinear, non-isothermal reactor model. %
}

%% Offset-free conditions: nonlinear models and beyond
Offset-free MPC designs with \emph{nonlinear models}
\ifthenelse{\boolean{LongVersion}}{%
  and tracking costs %
}{}%
were first considered by~\cite{morari:maeder:2012}. For the special case of
state feedback, \cite{pannocchia:gabiccini:artoni:2015} give a disturbance model
and estimator design for which the offset-free MPC is provably asymptotically
stable and offset-free. However, no general stability results are given. %
\ifthenelse{\boolean{LongVersion}}{%
  In~\cite{pannocchia:gabiccini:artoni:2015}, the state-feedback observer design
  is generalized to economic cost functions, and convergence to the optimal
  steady state is demonstrated. A general, output-feedback %
}{%
  Output-feedback %
}%
offset-free \emph{economic} MPC was first proposed
by~\cite{vaccari:pannocchia:2017}, and later extended
by~\cite{pannocchia:2018,% modifier-adaptation OF-EMPC
  faulwasser:pannocchia:2019% modifier-adaptation offset-free economic MPC
  % without terminal constraints
  % vaccari:bonvin:pelagagge:pannocchia:2021% comparing gradient estimation
  % methods for modifier-adaptation OF-EMPC
}, where gradient corrections ensure closed-loop stability implies optimal
steady-state performance.

%% Output- and reference-tracking MPC
There are no stability results for offset-free MPC.\@ The results discussed thus far
assume, rather than establish, closed-loop stability. While some authors have
proposed stable nonlinear MPC designs for output tracking~\citep{%
  falugi:2015,% nonlinear reference tracking MPC
  limon:ferramosca:alvarado:alamo:2018,% nonlinear tracking MPC
  % kohler:muller:allgower:2020,% nonlinear tracking MPC
  % berberich:kohler:muller:allgower:2022a,% linear tracking MPC for nonlinear systems
  galuppini:magni:ferramosca:2023% semidefinite stage cost case
  % soloperto:kohler:allgower:2023% output-feedback nonlinear tracking MPC without
                                % terminal ingredients
}, they do not consider plant-model mismatch and disturbance estimation.

%% Statement of results
In this paper, we propose a nonlinear offset-free MPC design that has
offset-free performance and asymptotic stability subject to plant-model
mismatch, persistent disturbances, and changing references. As in
\cite{kuntz:rawlings:2024d}, we use quadratic costs and assume differentiability
of the plant and model equations. We also consider we softened regulator output
constraints and tightened steady-state target problem constraints.

%% Outline
The remainder of this section outlines the paper and establishes notation. In
\Cref{sec:prob}, the offset-free MPC design is presented. In
\Cref{sec:stability}, we present the relevant stability theory. In
\Cref{sec:nominal}, we establish asymptotic stability of the nominal system. In
\Cref{sec:robust}, we establish robust stability with respect to estimate
errors, setpoint changes, and disturbance changes. In \Cref{sec:mismatch}, we
extend these results to the mismatched. In \Cref{sec:linear}, we make
connections to linear systems and linearization results in the literature. In
\Cref{sec:example}, the results are illustrated via numerical simulations. In
\Cref{sec:conclusion}, we conclude with a discussion of future work.

\ifthenelse{\boolean{LongVersion}}{}{%
  For brevity, additional examples and remarks are deferred to an extended
  technical report~\citep{kuntz:rawlings:2024f}. %
}

\paragraph*{Notation}
%% Numbers, vectors, matrices
\ifthenelse{\boolean{LongVersion}}{%
  Let \(\real\), \(\nnegreal\), and \(\posreal\) denote the real, nonnegative
  real, and positive real numbers, respectively. Let \(\allint\), \(\nnegint\),
  \(\posint\), and \(\intinterval{m}{n}\) denote the integers, nonnegative
  integers, positive integers, and integers from \(m\) to \(n\) (inclusive),
  respectively. Let \(\real^n\) and \(\realm{n}{m}\) denote real \(n\)-vectors
  and \(n\times m\) matrices, respectively. %^
}{}%
Let \(\underline{\sigma}(A)\) and \(\overline{\sigma}(A)\) denote the smallest
and largest singular values of \(A\in\realm{n}{m}\). %
\ifthenelse{\boolean{LongVersion}}{%
  We say a symmetric matrix \(P=P^\top\in\realm{n}{n}\) is positive definite
  (semidefinite) if \(x^\top Px>0\) (\(x^\top Px\geq 0\)) for all nonzero
  \(x\in\real^n\). %
  % TODO vector inequality \(a<b\Leftrightarrow a_i<b_i\forall i\)?
  For convenience, we write, for each \(a,b\in\real^n\), \(a>b\) (\(a\geq b\))
  if \(a_i>b_i\) (\(a_i\geq b_i\)) for all \(i\in\intinterval{1}{n}\). %
  %% TODO matrix inequality \(A\succeq B\leftrightarrow A-B\) is psd?
}{} %
%% Norms
For each positive semidefinite \(Q\), we define %
\ifthenelse{\boolean{LongVersion}}{%
  the Euclidean and \(Q\)-weighted norms by \(|x|\defas\sqrt{x^\top x}\) and %
}{}%
\(|x|_Q\defas\sqrt{x^\top Qx}\) for all \(x\in\real^n\). Let \(\delta\bbB^n
\defas \set{x\in\real^n | |x|\leq\delta}\) for \(\delta>0\). %
\ifthenelse{\boolean{LongVersion}}{%
  %% Norm rules
  For any positive definite \(Q\in\realm{n}{n}\), we have
  \(\underline{\sigma}(Q)|x|^2 \leq |x|_Q^2 \leq \overline{\sigma}(Q)|x|^2\) for
  all \(x\in\real^n\). %
}{}
%% Level sets
Given \(V:X\rightarrow\real\) and \(\rho>0\), define \(\lev_\rho
V\defas\set{x\in X | V(x)\leq\rho}\).
%% Signals and signal norms
For any signal \(a(k)\), denote%
\ifthenelse{\boolean{LongVersion}}{%
  , with slight abuse of notation, both finite and infinite sequences in bold
  font %
}{ %
  finite and infinite sequences %
}%
by \(\mathbf{a} \defas (a(0),\ldots,a(k))\) and \(\mathbf{a} \defas
(a(0),a(1),\ldots)\)%
\ifthenelse{\boolean{LongVersion}}{%
  , where length is specified or implied from context, %
}{, }%
and a subsequence by \(\mathbf{a}_{i:j} \defas (a(i),\ldots,a(j))\), where
\(i\leq j\). Define the infinite and length-\(k\) signal norms as
\(\|\mathbf{a}\| \defas \sup_{k\geq 0} |a(k)|\) and \(\|\mathbf{a}\|_{0:k}
\defas \max_{0\leq i\leq k} |a(i)|\).
%% K and KL functions
Let \(\calK\) be the class of strictly increasing
\(\alpha:\nnegreal\rightarrow\nnegreal\) such that \(\alpha(0)=0\). Let
\(\calKinf\) be the class of unbounded class-\(\calK\) functions. Let \(\calKL\)
be the class of \(\beta:\nnegreal\times\nnegint\rightarrow\nnegreal\) such that
\(\beta(\cdot,k)\in\calK\), \(\beta(r,\cdot)\) is nonincreasing, and
\(\lim_{i\rightarrow\infty}\beta(r,i) = 0\), for all \(r\geq 0\) and
\(k\in\nnegint\). Denote the identity map by \(\id(\cdot) \defas (\cdot) \in
\calKinf\).

\section{Problem statement}\label{sec:prob}
% \subsection{System of interest}\label{ssec:sys}
Consider the following discrete-time plant:
\begin{subequations}\label{eq:plant}
  \begin{align}
    \xP^+ &= \fP(\xP,u,\wP) \\
    y &= \hP(\xP,u,\wP)
  \end{align}
\end{subequations}
where \(\xP\in\bbX\subseteq\real^n\) is the \emph{plant} state,
\(u\in\bbU\subseteq\real^{n_u}\) is the input, \(y\in\bbY\subseteq\real^{n_y}\)
is the output, and \(\wP\in\bbW\subseteq\real^{n_w}\) is the \emph{plant}
disturbance. The functions \(\fP\) and \(\hP\) are not known. Instead, we assume
access to a model of the plant,
\begin{subequations}\label{eq:model}
  \begin{align}
    x^+ &= f(x,u,d) \label{eq:model:process} \\
    % d^+ &= d \\
    y &= h(x,u,d)
  \end{align}
\end{subequations}
where \(x\in\bbX\subseteq\real^n\) is the \emph{model} state and
\(d\in\bbD\subseteq\real^{n_d}\) is the \emph{model} disturbance. Without loss
of generality, we assume the nominal plant and model functions are consistent,
i.e.,
\begin{align}\label{eq:nominal}
  f(x,u,0) &= \fP(x,u,0),
  & h(x,u,0) &= \hP(x,u,0)
\end{align}
for all \((x,u)\in\bbX\times\bbU\). The plant disturbance \(\wP\) may include
exogenous disturbances, parameter errors, discretization errors, and even
unmodeled dynamics. The model disturbance \(d\) is intended to correct for
steady-state output errors, and may include individual plant disturbances
\((\wP)_i\) as well as fictitious signals specially designed to correct for
steady-state errors.

\ifthenelse{\boolean{LongVersion}}{%
  \begin{example}
    Consider a single-state linear plant with parameter
    errors,
    \begin{align*}
      \fP(\xP,u,\wP) &= (\hat{a} + (\wP)_1)\xP + (\hat{b}+(\wP)_2)u \\
      \hP(\xP,u,\wP) &= \xP + (\wP)_3
    \end{align*}
    and a single-state linear model with an input disturbance:
    \begin{align*}
      f(x,u,d) &= \hat{a}x + \hat{b}(u + d),
      & h(x,u,d) &= x.
    \end{align*}
    For this example, the plant disturbance \(\wP\) includes both parameter
    errors and measurement noise, whereas the model disturbance only provides
    the means to shift the model steady states in response to plant
    disturbances.
  \end{example}
}{}

The control objective is to drive the reference signal,
\begin{equation}\label{eq:ref}
  r = g(u,y)
\end{equation}
to the setpoint \(\rsp\) using only knowledge of the model~\cref{eq:model}, past
\((u,y)\) data, and auxiliary setpoints \((\usp,\ysp)\) (to be defined). The
setpoints \(\ssp \defas (\rsp,\usp,\ysp)\) are possibly time-varying, but only
the current value is available at a given time. The controller should be
\emph{offset-free} when the setpoint and plant disturbances are asymptotically
constant, i.e.,
\begin{align*}
  (\Delta\ssp(k),\Delta\wP(k)) &\rightarrow 0
  & &\Rightarrow & r(k) - \rsp(k) &\rightarrow 0
\end{align*}
where \(\Delta\ssp(k) \defas \ssp(k) - \ssp(k-1)\) and \(\Delta\wP(k) \defas
\wP(k) - \wP(k-1)\). Otherwise, the amount of offset should be robust to
setpoint and disturbance \emph{increments} \((\Delta\ssp,\Delta\wP)\).

\ifthenelse{\boolean{LongVersion}}{%
  \begin{remark}
    To achieve the nominal consistency assumption~\cref{eq:nominal} and track
    the reference~\cref{eq:ref}, we typically need the dimensional constraints
    \(n_y\leq n_d\) and \(n_r\leq n_u\), respectively. Otherwise their are
    insufficient degrees of freedom to manipulate the output and reference at
    steady state with the disturbance and input, respectively.
  \end{remark}

  %% NOTE If \(\rsp(k) = \sqrt(1/k)\), then the increments go to zero but there
  %% is no limit. However, the trajectory should be increasingly closer to
  %% constant, so it should be increasingly easier to track, and the MPC is
  %% again ``offset-free.''
  \begin{remark}
    We do not strictly require an asymptotically constant disturbance. For
    example, if \(\rsp(k) = 1/\sqrt{k}\) and \(\wP\equiv 0\), then the setpoint
    has no limit but increments go to zero \(\Delta\rsp(k) = 1/\sqrt{k} -
    1/\sqrt{k-1} = O(1/\sqrt{k})\). However, the setpoint signal becomes
    approximately constant as \(k\rightarrow\infty\), so we should expect the
    offset-free MPC to be approximately offset-free.
  \end{remark}
}{}

Throughout, we make the following assumptions on plant, model, and reference
functions.
\begin{assumption}[Continuity]\label{assum:cont}
  The functions \(g:\bbU\times\bbY\rightarrow\real^{n_r}\),
  \((\fP,\hP):\bbX\times\bbU\times\bbW\rightarrow\bbX\times\bbY\), and
  \((f,h):\bbX\times\bbU\times\bbD\rightarrow\bbX\times\bbY\) are continuous,
  and \(f(0,0,0)=0\), \(h(0,0,0)=0\), \(g(0,0)=0\), and \cref{eq:nominal} holds
  for all \((x,u)\in\bbX\times\bbU\).
\end{assumption}

\subsection{Constraints}\label{ssec:cons}
The sets \((\bbX,\bbY,\bbD,\bbW)\) are physical constraints (e.g., actuation
limits, non-negativity of pressures and chemical concentrations) that the
systems \cref{eq:plant,eq:model} automatically satisfy. These constraints only
need to be enforced during state estimation. Hard input constraints \(u\in\bbU\)
are enforced during both regulation and target selection. We additionally
consider joint input-output constraints of the form %
\ifthenelse{\boolean{LongVersion}}{%
  \[
    \bbZ_y \defas \set{ (u,y)\in\bbU\times\bbY | \overline{c}(u,y) \leq 0 }
  \]
}{%
  \(\overline{c}(u,y)\leq 0\) %
}%
where \(\overline{c}:\bbU\times\bbY\rightarrow\real^{n_c}\) is the constraint
function. In regulation, \(\overline{c}\) serves as a soft constraint function.
Having active constraints at steady state may cause regulator instability
(cf.~\Cref{rem:cons}), so the constraints are tightened during target selection
as follows:
\[
  \overline{\bbZ}_y \defas \set{ (u,y) \in\bbU\times\bbY | \overline{c}(u,y) +
    \overline{b} \leq 0 }
\]
where \(\overline{b}\in\posreal^{n_c}\) contains back-off constants. %
\ifthenelse{\boolean{LongVersion}}{%
  No such constraint tightening is required for the input constraints. %
}{}%
We assume the constraints satisfy the following properties.
%% NOTE: The assumption that \(\bbU\) is convex complicates the picture for
%% discrete actuators. However, it suffices to differentiate between an
%% extended, continuous definition of the actuator \(u\in\bbU_c\) and a
%% constrained, discrete definition \(u\in\bbU_d\subseteq\bbU_c\). The
%% continuous definition must be convex and compact, but the discrete definition
%% need only be compact. Convexity of the state space cannot be helped unless I
%% provide an analytic continuation of \((f,h,\fP,\hP)\) to all of
%% \(\real^{n+n_u+n_d}\) or \(\real^{n+n_u+n_w}\). Maybe I can use the
%% assumptions sans convexity to guarantee the existence of a continuously
%% differentiable continuation?
%%
%% Whitney extension theorem:
%% https://en.wikipedia.org/wiki/Whitney_extension_theorem
%% https://math.stackexchange.com/questions/2401340/on-a-differentiable-extension-of-a-function
%% https://math.wvu.edu/~kciesiel/prepF/129.DifferentiableExtensionThm/129.DifferentiableExtensionThm.pdf
%% https://encyclopediaofmath.org/wiki/Whitney_extension_theorem
%%
%% Partition of unity:
%% https://en.wikipedia.org/wiki/Partition_of_unity
%% https://math.stackexchange.com/questions/3380252/can-i-apply-whitneys-extension-theorem-to-arbitrary-smooth-functions
%% Lemma 2.26: Lee (2012) ``Introduction to Smooth Manifolds''
%% https://link-springer-com.proxy.library.ucsb.edu/book/10.1007/978-1-4419-9982-5
\begin{assumption}[Constraints]\label{assum:cons}%
  The sets \((\bbX,\bbY)\) are closed, \((\bbU,\bbW,\bbD)\) are compact, and all
  contain the origin. The function \(\overline{c} : \bbU\times\bbY \rightarrow
  \real^{n_c}\) is continuous and \(\overline{c}(0,0) + \overline{b} < 0\).
\end{assumption}

\subsection{Offset-free model predictive control}\label{sec:ofmpc}
Offset-free MPC consists of three parts or subroutines: target selection,
regulation, and state and disturbance estimation. Given a disturbance \(d\) and
setpoint \(\rsp\), the steady-state target problem (SSTP) identifies the
\emph{steady-state targets} \((x_s,u_s)\) that reach the setpoint \(\rsp\) and
satisfy constraints. The regulator is a finite horizon optimal control problem
(FHOCP) that steers the system from the current state \(x\) to the steady-state
targets \((x_s,u_s)\). Finally, the SSTP and FHOCP are implemented with
estimates of \(x\) and \(d\), rather than the true values.

\subsubsection{Steady-state target problem}\label{ssec:sstp}
Given \(d\in\bbD\) and \(\rsp\in\real^{n_r}\), we define the set of offset-free
steady-state pairs by %
\ifthenelse{\boolean{OneColumn}}{%
  \begin{equation}\label{eq:ss:ofpairs}
    \calZ_O(\rsp,d) \defas \set{ (x,u) \in \bbX\times\bbU | x=f(x,u,d), \;
      y=h(x,u,d),\; (u,y)\in\overline{\bbZ}_y,\; \rsp = g(u,y) }.
  \end{equation}
}{%
  %% NOTE For some reason multline is producing a lot of space above the
  %% equation, so I simulated the behavior with an align environment. If the
  %% equation is changed, just be careful to adjust the alignment.
  % \begin{multline}\label{eq:ss:ofpairs}
  %   \calZ_O(\rsp,d) \defas \{\; (x,u) \in \bbX\times\bbU \;|\; x=f(x,u,d), \\
  %   y=h(x,u,d),\; (u,y)\in\overline{\bbZ}_y,\; \rsp = g(u,y) \;\}.
  % \end{multline}
  \begin{align}\label{eq:ss:ofpairs}
    \calZ_O(\rsp,d) &\defas \{\; (x,u) \in \bbX\times\bbU \;|\; x = f(x,u,d), \nonumber \\
    y &= h(x,u,d),\; (u,y)\in\overline{\bbZ}_y,\; \rsp = g(u,y) \;\}.
  \end{align}
}%
To optimally select a steady-state pair from \(\calZ_O(\rsp,d)\), we minimize
the distance from some auxiliary setpoint pair \(\zsp \defas (\usp,\ysp) \in
\overline{\bbZ}_y\) (typically such that \(\rsp = g(\usp,\ysp)\)). We define the
set of feasible SSTP parameters as
\begin{equation}\label{eq:sstp:params}
  \calB \defas \set{ (\rsp,\zsp,d) \in \real^{n_r} \times \overline{\bbZ}_y
    \times \bbD | \calZ_O(\rsp,d) \neq \varnothing }.
\end{equation}
For each \(\beta = (\rsp,\usp,\ysp,d) \in \calB\), we define the SSTP by
\begin{equation}\label{eq:sstp}
  V_s^0(\beta) \defas \min_{(x,u)\in\calZ_O(\rsp,d)} \ell_s(u-\usp,h(x,u,d)-\ysp)
\end{equation}
where %
\ifthenelse{\boolean{LongVersion}}{%
  \(\beta \defas (\rsp,\usp,\ysp,d)\) are the SSTP parameters and
  \(\ell_s:\real^{n_u}\times\real^{n_y}\rightarrow\nnegreal\) is a steady-state
  cost cost function, typically a positive definite quadratic. For infeasible
  problems (\(\beta\not\in\calB\)), we let \(V_s^0(\beta) \defas \infty\). To
  guarantee the existence of solutions to the SSTP~\cref{eq:sstp} for all
  feasible \(\beta\in\calB\), the following assumption is required. %
}{%
  \(\ell_s:\real^{n_u}\times\real^{n_y}\rightarrow\nnegreal\) is the
  steady-state cost function. We also impose the following assumption on the
  SSTP~\cref{eq:sstp}. %
}%
\begin{assumption}[SSTP existence]\label{assum:sstp:exist}
  The function \(\ell_s:\real^{n_u}\times\real^{n_y}\rightarrow\nnegreal\) is
  continuous. For each \(\beta = (\rsp,\usp,\ysp,d)\in\calB\), at least one of
  the following properties holds:
  \begin{enumerate}[(a)]
  \item \(\calZ_O(\rsp,d)\) is compact;
  \item with \(V_s(x,u,\beta) \defas \ell_s(u-\usp,h(x,u,d)-\ysp)\), the
    function \(V_s(\cdot,\beta)\) is coercive in \(\calZ_O(\rsp,d)\),
    i.e., for any sequence \(\bfz\in(\calZ_O(\rsp,d))^\infty\) such that
    \(|z(k)|\rightarrow\infty\), we have \(V_s(z(k),\beta)\rightarrow\infty\).
  \end{enumerate}
\end{assumption}

Under \Cref{assum:cont,assum:cons,assum:sstp:exist}, \(\calB\) is nonempty and
the SSTP~\cref{eq:sstp} has solutions for all \(\beta\in\calB\). To ensure
uniqueness, we assume some selection rule has been applied and denote the
functions returning solutions to \cref{eq:sstp} by \(z_s \defas (x_s,u_s) :
\calB \rightarrow \bbX\times\bbU\).

\subsubsection{Regulator}\label{ssec:reg}
We consider a horizon length \(N\in\posint\), stage cost
\(\ell:\bbX\times\bbU\times\calB\rightarrow\nnegreal\), and terminal cost
\(V_f:\bbX\times\calB\rightarrow\nnegreal\). For each
\(\beta=(\ssp,d)\in\calB\), we define the terminal constraint
\cref{eq:mpc:terminal}, feasible initial state and input sequence pairs
\cref{eq:mpc:admit}, feasible input sequences at \(x\in\bbX\)
\cref{eq:mpc:admit:inputs}, feasible initial states \cref{eq:mpc:admit:states},
and feasible state-parameter pairs \cref{eq:mpc:admit:state:params} by the sets
\begin{align}
  \bbX_f(\beta)
  &\defas \lev_{c_f} V_f(\cdot,\beta) \label{eq:mpc:terminal} \\
  \calZ_N(\beta)
  &\defas \set{ (x,\bfu) \in \bbX\times\bbU^N |
    \phi(N;x,\bfu,d) \in \bbX_f(\beta) } \label{eq:mpc:admit} \\
  \calU_N(x,\beta)
  &\defas \set{ \bfu\in\bbU^N | (x,\bfu)\in\calZ_N(\beta) }
    \label{eq:mpc:admit:inputs} \\
  \calX_N(\beta)
  &\defas \set{ x\in\bbX | \calU_N(x,\beta) \neq \varnothing }
    \label{eq:mpc:admit:states} \\
  \calS_N
  &\defas \set{ (x,\beta) \in \bbX\times\calB | \calU_N(x,\beta) \neq \varnothing }
    \label{eq:mpc:admit:state:params}
\end{align}
where \(c_f>0\) and \(\phi(k;x,\bfu,d)\) denotes the solution to
\cref{eq:model:process} at time \(k\) given an initial state \(x\), constant
disturbance \(d\), and sufficiently long input sequence \(\bfu\).
%% TODO I know \(\calX_f(\beta),\calZ_N(\beta),\calU_N(x,\beta),\calX_N(\beta)\)
%% are closed~\cite[Props.~2.4,~2.10]{rawlings:mayne:diehl:2020}, but I also
%% need to show \(\calS_N\) is closed. But maybe not; \(\calS_N^\rho\) compact
%% follows the same way as the beginning of the proof of
%% \Cref{prop:mpc:robust:feas} (lift the set to include inputs; compactness of
%% the set constraints and continuity of the constraints implies compactness of
%% the lifted set; the projection back down to \((x,\beta)\) space is therefore
%% compact).
For each \((x,\bfu,\beta)\in\bbX\times\bbU^N\times\calB\), we
define the FHOCP objective by %
\ifthenelse{\boolean{OneColumn}}{%
  \begin{equation}\label{eq:mpc:obj}
    V_N(x,\bfu,\beta) \defas V_f(\phi(N;x,\bfu,d), \beta)
    + \sum_{k=0}^{N-1} \ell(\phi(k;x,\bfu,d), u(k), \beta).
  \end{equation}
}{%
  %% NOTE For some reason multline is producing a lot of space above the
  %% equation, so I simulated the behavior with an align environment. If the
  %% equation is changed, just be careful to adjust the alignment.
  % \begin{multline}\label{eq:mpc:obj}
  %   V_N(x,\bfu,\beta) \defas V_f(\phi(N;x,\bfu,d), \beta) \\
  %   + \sum_{k=0}^{N-1} \ell(\phi(k;x,\bfu,d), u(k), \beta).
  % \end{multline}
  \begin{align}\label{eq:mpc:obj}
    V_N(x,\bfu,\beta) \defas V_f(\phi&(N;x,\bfu,d), \beta) \nonumber \\
    &+ \sum_{k=0}^{N-1} \ell(\phi(k;x,\bfu,d), u(k), \beta).
  \end{align}
}%
For each \((x,\beta)\in\calS_N\), we define the FHOCP by
\begin{equation}\label{eq:mpc}
  V_N^0(x,\beta) \defas \min_{\bfu\in\calU_N(x,\beta)}
  V_N(x,\bfu,\beta).
\end{equation}
For infeasible problems (\((x,\beta)\not\in\calS_N\)), we let
\(V_N^0(x,\beta)\defas\infty\).

To guarantee closed-loop stability and robustness, we consider the following
assumptions.
\begin{assumption}[Terminal control law]\label{assum:stabilizability}%
  There exists a function \(\kappa_f:\bbX\times\calB\rightarrow\bbU\) such that
  \[
    V_f(f(x,\kappa_f(x,\beta),d),\beta) - V_f(x,\beta) \leq
    -\ell(x,\kappa_f(x,\beta),\beta)
  \]
  for each \(x\in\bbX_f(\beta)\) and \(\beta = (\ssp,d) \in \calB\).
\end{assumption}
\begin{assumption}[Quadratic costs]\label{assum:quad}
  There exist positive definite matrices \(Q\) and \(R\), a function \(P_f :
  \calB \rightarrow \realm{n}{n}\), and constants \(w_i>0,
  i\in\intinterval{1}{n_c}\) such that \(P_f(\beta)\) is positive definite and
  the stage and terminal costs can be written as %
  \ifthenelse{\boolean{OneColumn}}{%
    \begin{subequations}\label{eq:quad}
      \begin{align}
        \ell(x,u,\beta)
        &= |x-x_s(\beta)|_Q^2 + |u-u_s(\beta)|_R^2 +
          \sum_{i=1}^{n_c} w_i\max\set{0,\overline{c}_i(u,h(x,u,d))} \\
        V_f(x,\beta) &= |x-x_s(\beta)|_{P_f(\beta)}^2
      \end{align}
    \end{subequations}
  }{%
    \begin{subequations}\label{eq:quad}
      \begin{align}
        \ell(x,u,\beta)
        &= |x-x_s(\beta)|_Q^2 + |u-u_s(\beta)|_R^2 \\
        &\qquad + \sum_{i=1}^{n_c} w_i\max\set{0,\overline{c}_i(u,h(x,u,d))} \\
        V_f(x,\beta) &= |x-x_s(\beta)|_{P_f(\beta)}^2
      \end{align}
    \end{subequations}
  }%
  for each \((x,u)\in\bbX\times\bbU\) and \(\beta = (\ssp,d) \in \calB\).
\end{assumption}

\Cref{assum:cons,assum:cont,assum:sstp:exist,assum:quad} guarantee the existence
of solutions to \cref{eq:mpc} for all
\((x,\beta)\in\calS_N\)~\cite[Prop.~2.4]{rawlings:mayne:diehl:2020}. We denote
any such solution by \(\bfu^0(x,\beta) = (u^0(0;x,\beta), \ldots,
u^0(N-1;x,\beta))\), and define the corresponding optimal state \(x^0(k;x,\beta)
\defas \phi(k;x,\bfu^0(x,\beta),d)\), optimal state sequence by
\(\bfx^0(x,\beta) \defas (x^0(0;x,\beta), \ldots, x^0(N;x,\beta))\), and FHOCP
control law by \(\kappa_N(x,\beta) \defas u^0(0;x,\beta)\). Terminal ingredients
satisfying \cref{assum:stabilizability,assum:quad} are constructed in
\Cref{app:terminal}.

\ifthenelse{\boolean{LongVersion}}{%
  Finally, some remarks are in order.

  \begin{remark}
    Soft constraint penalties of the form~\cref{eq:quad} were also used
    in~\cite{santos:biegler:castro:2008} for regulation under plant-model
    mismatch.
  \end{remark}

  \begin{remark}
    We use a parameter-varying terminal region~\cref{eq:mpc:terminal} rather than
    an offset penalty (cf.~\cite{%
      falugi:2015,% nonlinear reference tracking MPC
      limon:ferramosca:alvarado:alamo:2018,% nonlinear tracking MPC
      galuppini:magni:ferramosca:2023% semidefinite stage cost case terminal ingredients
    }), so it is unnecessary to assume the existence of an invariant set for
    tracking.
  \end{remark}

  \begin{remark}
    With \(\beta=(\ssp,d)\in\calB\), \Cref{assum:stabilizability} and the terminal
    set definition \cref{eq:mpc:terminal} imply
    \(V_f(f(x,\kappa_f(x,\beta),d),\beta) \leq V_f(x,\beta) \leq c_f\) for all
    \(x\in\bbX_f(\beta)\) and therefore \(\bbX_f(\beta)\) is positive invariant
    for \(x^+=f(x,\kappa_f(x,\beta),d)\).
  \end{remark}
}{}

\ifthenelse{\boolean{LongVersion}}{}{%
  \begin{remark}\label{rem:cons}
    Constraint back-offs are important for regulator stability. Suppose
    \Cref{assum:cont,assum:cons,assum:sstp:exist,assum:quad} hold, except we let
    \(\overline{b}=0\). Consider a scalar system \(f(x,u,d) \defas x+u+d\),
    \(h(x,u,d) \defas x\), \(g(u,y) \defas y\), and \(\overline{c}(u,y) \defas
    y-1\). With fixed \(\beta = (1,0,1,0)\), we have \((x_s(\beta),u_s(\beta)) =
    (1,0)\), and the constraint \(\overline{c}(u,y)\leq 0\) is active at steady
    state. Suppose \(\tilde{q} \defas Q - P_f^2/(2P_f+2R) < 0\). With
    \(\mathcal{F}(x,u) \defas V_f(f(x,u,d),\beta) - V_f(x,\beta) +
    \ell(x,u,\beta)\), we have \(\min_{u\in\real} \mathcal{F}(x,u) =
    \tilde{q}(x-1)^2 + w_1\max\set{0,x-1}\). But this means \(\min_{u\in\real}
    \mathcal{F}(x,u) > 0\) for all \(0<x-1<\sqrt{w_1/\tilde{q}}\), so
    \Cref{assum:stabilizability} is not satisfied.
  \end{remark}
}%

\ifthenelse{\boolean{LongVersion}}{
  \begin{remark}\label{rem:cons}
    Given \Cref{assum:cont,assum:cons,assum:sstp:exist,assum:quad}, it may be
    impossible to satisfy \Cref{assum:stabilizability} without constraint
    back-offs, i.e., \(b=0\). This is because the terminal cost difference
    \(V_f(f(x,\kappa_f(x,\beta),d)) - V_f(x)\) is, at best, negative definite
    with quadratic scaling, whereas the stage cost
    \(\ell(x,\kappa_f(x,\beta),\beta)\) has quadratic scaling when the soft
    constraint is satisfied but linear scaling when the soft constraint is
    violated. Thus, with constraints active at the targets, the stage cost
    exceeds the terminal cost decrease in a neighborhood of the origin.
  \end{remark}

  \begin{example}
    To illustrate \Cref{rem:cons}, consider the scalar linear system
    \(x^+=x+u+d\), \(y=x\), and \(r=y\) with stage costs of the form
    \Cref{assum:quad} and the soft constraint function
    \(\overline{c}(u,y)=y-1\). Let \(\overline{b}=0\) and \(\beta=(1,0,1,0)\).
    Clearly the target is reachable, and we can take the SSTP \cref{eq:sstp}
    solution \((x_s(\beta),u_s(\beta))=(1,0)\). Then we have stage costs of the
    form \(\ell(x,u,\beta) = q(x-1)^2+ru^2 + w\max\set{0,x-1}\) and
    \(V_f(x,\beta) = p_fx^2\), where \(q,r,w,p_f>0\).
    \Cref{assum:stabilizability} is not satisfied if there exists \(x\in\real\)
    such that \ifthenelse{\boolean{OneColumn}}{%
      \begin{equation*}
        \mathcal{F}(x,u) \defas p_f(x+u-1)^2 - p_f(x-1)^2
        + q(x-1)^2 + ru^2 + w\max\set{0,x-1} > 0
      \end{equation*}
    }{%
      \begin{multline*}
        \mathcal{F}(x,u) \defas p_f(x+u-1)^2 - p_f(x-1)^2 \\
        + q(x-1)^2 + ru^2 + w\max\set{0,x-1} > 0
      \end{multline*}
    }%
    for all \(u\in\real\). Completing the squares gives
    \begin{align*}
      \mathcal{F}(x,u)
      &= (\tilde au + \tilde b(x-1))^2 + \tilde c(x-1)^2 + w\max\set{0,x-1} \\
      &\geq \tilde c(x-1)^2 + w\max\set{0,x-1}
    \end{align*}
    for all \(x\in\real\) and \(u\in\real\), where \(\tilde a\defas\sqrt{r+p_f}\),
    \(\tilde b \defas \frac{p_f}{2\tilde a}\), and \(\tilde c \defas q - \tilde
    b^2\). Ideally, we would have chosen \((q,r,p_f)\) so that \(\tilde c<0\). But
    this means we can still take \(0<x-1<\sqrt{\frac{w}{\tilde c}}\) to give
    \[
      \mathcal{F}(x,u) \geq \tilde c(x-1)^2 + w(x-1) > 0
    \]
    for all \(u\in\real\), no matter the chosen weights \(w>0\).

    On the other hand, let \(\overline{b}=1\) and \(\beta=(0,0,0,0)\). Again,
    the target is reachable and we can take the SSTP solution
    \((x_s(0),u_s(0))=(0,0)\). Notice that for both problems the backed-off
    constraint \(\overline{c}(u,y)+\overline{b}\) is active at the solution.
    This time, however, we have
    \begin{align*}
      \mathcal{F}(x,u)
      &\defas p_f(x+u)^2 - p_fx^2 + qx^2 + ru^2
        \ifthenelse{\boolean{OneColumn}}{}{\\
      &\qquad}
        + w\max\set{0,x-1} \\
      &= (\tilde au + \tilde bx)^2 + \tilde cx^2 + w\max\set{0,x-1}
    \end{align*}
    and with \(\kappa_f(x,0)\defas-\frac{\tilde b}{\tilde a}x\), we have
    \[
      \mathcal{F}(x,\kappa_f(x,0)) = \tilde cx^2 + w\max\set{0,x-1}
    \]
    for all \(x\in\real\). Let \(c_f=p_f\) and suppose \(\tilde c<0\). Then, for
    each \(x\in\bbX_f(0)\), we have \(|x|\leq 1\) and therefore
    \[
      \mathcal{F}(x,\kappa_f(x,0)) = \tilde cx^2 \leq 0.
    \]
  \end{example}
}{}

\begin{remark}\label{rem:quad}
  \Cref{assum:quad} is used for guaranteeing offset-free performance under
  mismatch. For general stage costs, arbitrarily small mismatch may cause offset
  in standard MPC, even with known steady-state
  targets~\cite{kuntz:rawlings:2024d}.
\end{remark}

\subsubsection{State and disturbance estimation}\label{ssec:est}
Consider the following estimator, to be designed according to the model
\cref{eq:model}.
\begin{definition}\label{defn:est}
  We define a \emph{joint state and disturbance estimator} as a sequence of
  functions \(\Phi_k : \bbX \times \bbD \times \bbU^k \times \bbY^k \rightarrow
  \bbX \times \bbD,k\in\nnegint\), and the \emph{state and disturbance
    estimates} as
  \begin{equation}\label{eq:est}
    (\hat{x}(k),\hat{d}(k)) \defas \Phi_k(\overline{x}, \overline{d},
    \bfu_{0:k-1}, \bfy_{0:k-1})
  \end{equation}
  where \((\overline{x}, \overline{d}) \in \bbX \times \bbD\) is the initial
  guess at time \(k=0\), \(\bfu\in\bbU^\infty\) is the input data, and
  \(\bfy\in\bbY^\infty\) is the output data.
\end{definition}
\ifthenelse{\boolean{LongVersion}}{%
  %% NOTE: The definition above is straight from Doug's paper and doesn't really
  %% need to be explained. Only for LongVersion/thesis.
  \begin{remark}\label{rem:est:1}
    Since the regulator requires a state estimate to compute, and the input
    directly affects the output, the current state and disturbance estimates
    \((\hat{x}(k),\hat{d}(k))\) must be functions of past data, not including
    the current measurement \(y(k)\). Therefore, at time \(k=0\), there is no
    data available to update the prior guess, and most estimator designs will
    take \(\Phi_0\) as the identity map, i.e.,
    \[
      (\hat{x}(0),\hat{d}(0)) \defas \Phi_0(\overline{x}, \overline{d}) =
      (\overline{x}, \overline{d}).
    \]
    However, we can also consider models without direct feedthrough effects
    (i.e., \(y = h(x,d)\)) in which case \Cref{defn:est} can be modified so the
    estimator functions also take \(y(k)\) as an argument.
  \end{remark}
}{}

To analyze the estimator performance in terms of the model
equations~\cref{eq:model}, we consider the following noisy model:
\begin{subequations}\label{eq:model:noise}
  \begin{align}
    x^+ &= f(x,u,d) + w \label{eq:model:noise:a} \\
    d^+ &= d + w_d \label{eq:model:noise:b} \\
    y &= h(x,u,d) + v \label{eq:model:noise:c} %\\
    % \tilde{w} &\defas (w,w_d,v) \in \tilde{\bbW}(x,u,d)
    %% NOTE This alignment was carefully constructed to make it look like a
    %% multline environment was appended to an align environment, without the
    %% pesky double-spacing added between. If these equations get edited, make
    %% sure to check that the alignment looks right by switching back and forth
    %% from the multline environment below.
    % \tilde{w} \defas (w,w_d,v) \in \tilde{\bbW}
    % &(x,u,d) \defas \{\; (w,w_d,v) \;| \nonumber \\
    % (x^+&, d^+, y) \in\bbX\times\bbD\times\bbY,\, \cref{eq:model:noise} \;\}.
  \end{align}
  % \begin{multline}
  %   \tilde{w} \defas (w,w_d,v) \in \tilde{\bbW}
  %   (x,u,d) \defas \{\; (w,w_d,v) \;| \\
  %   (x^+, d^+,y) \in\bbX\times\bbD\times\bbY,\, \cref{eq:model:noise} \;\}.
  % \end{multline}
\end{subequations}
where \(\tilde{w} \defas (w,w_d,v)\) denote process, disturbance, and
measurement noises. We restrict the noise as
\ifthenelse{\boolean{LongVersion}}{%
  \[
    \tilde{w} \in \tilde{\bbW}(x,u,d) \defas \{\; (w,w_d,v) \;|\; (x^+, d^+, y)
    \in\bbX\times\bbD\times\bbY,\, \cref{eq:model:noise} \;\}
  \]
}{%
  \(\tilde{w} \in \tilde{\bbW}(x,u,d) \defas \{\, (w,w_d,v) \,|\, (x^+, d^+, y)
  \in\bbX\times\bbD\times\bbY,\, \cref{eq:model:noise} \,\}\) %
}%
to satisfy physical constraints. The estimation errors are defined by
\begin{subequations}\label{eq:est:err}
  \begin{align}
    e_x(k) &\defas x(k) - \hat{x}(k),
    & e_d(k) &\defas d(k) - \hat{d}(k), \\
    e(k) &\defas \begin{bmatrix} e_x(k) \\ e_d(k) \end{bmatrix},
    & \overline e &\defas \begin{bmatrix} x(0) - \overline x \\
                            d(0) - \overline d \end{bmatrix}.
  \end{align}
\end{subequations}
We define robust stability of the estimator \cref{eq:est} as follows.
\begin{definition}\label{defn:rges}
  The estimator \cref{eq:est} is \emph{robustly globally exponentially stable}
  (RGES) for the system \cref{eq:model:noise} if there exist constants
  \(c_{e,1},c_{e,2}>0\) and \(\lambda_e\in(0,1)\) such that
  \begin{equation*}
    |e(k)| \leq c_{e,1}\lambda_e^k|\overline e| + c_{e,2}\sum_{j=1}^k
    \lambda_e^{j-1}|\tilde{w}(k-j)|
  \end{equation*}
  for all \(k\in\nnegint\), \((\overline{x},\overline{d}) \in \bbX\times\bbD\),
  and trajectories \((\bfx,\bfu,\bfd,\bfy,\tilde{\bfw})\) satisfying
  \cref{eq:model:noise} and \(\tilde{w} \defas (w,w_d,v) \in \tilde{\bbW}(x,u,d)
  \), given~\cref{eq:est:err}.
\end{definition}

For the case with plant-model mismatch, the estimator \cref{eq:est} is not only
assumed to be RGES for the system \cref{eq:model:noise}, but is also assumed to
admit a robust global Lyapunov function.
\begin{assumption}[Estimator stability]\label{assum:est}
  The initial estimator \(\Phi_0\) is the identity map. There exists a function
  \(V_e : \bbX \times \bbD \times \bbX \times \bbD \rightarrow \nnegreal\) and
  constants \(c_1,c_2,c_3,c_4>0\) such that
  \begin{subequations}\label{eq:est:lyap}
    \begin{align}
      c_1|e(k)|^2 &\leq V_e(k) \leq c_2|e(k)|^2 \label{eq:est:lyap:a} \\
      V_e(k+1) &\leq V_e(k) - c_3|e(k)|^2 + c_4|\tilde{w}(k)|^2
                 \label{eq:est:lyap:b}
    \end{align}
  \end{subequations}
  for all \(k\in\nnegint\), \((\overline{x},\overline{d}) \in \bbX\times\bbD\),
  and trajectories \((\bfx,\bfu,\bfd,\bfy,\tilde{\bfw})\) satisfying
  \cref{eq:model:noise} and \(\tilde{w} \defas (w,w_d,v) \in \tilde{\bbW}(x,u,d)
  \), given \cref{eq:est}, \cref{eq:est:err}, and \(V_e(k) \defas
  V_e(x(k),d(k),\hat{x}(k),\hat{d}(k))\).
\end{assumption}

\ifthenelse{\boolean{LongVersion}}{%
  The following theorem establishes that \Cref{assum:est} implies RGES of the
  estimator~\cref{eq:est} for the system~\cref{eq:model:noise}
  (see~\Cref{app:est:rges} for proof).
  \begin{theorem}\label{thm:est:rges}
    Suppose the estimator~\cref{eq:est} for the system~\cref{eq:model:noise}
    satisfies \Cref{assum:est}. Then the estimator is RGES under
    \Cref{defn:rges}.
  \end{theorem}

  \begin{remark}
    In \Cref{assum:est}, we assume \(\Phi_0\) is the identity map, and therefore
    \(e(0)=\overline e\). However, as mentioned in \Cref{rem:est:1}, if we
    consider models without direct input-output effects (i.e.,
    \(y=\hat{h}(x,d)\)), then the estimator functions \(\Phi_k\) may become a
    function of the current output \(y(k)\) and it is no longer reasonable to
    assume \(\Phi_0\) is the identity map. Then \(e(0)\neq\overline e\) in
    general. However, we can modify \Cref{defn:est} to include robustness to the
    current noise \(\tilde n(k)\), and we can modify \Cref{assum:est} to include
    a linear bound of the form \(|e(0)| \leq \overline a_1|\overline e| +
    \overline a_2|\tilde{w}(0)|\), for some \(\overline a_1,\overline a_2>0\),
    to again imply RGES of the estimator.
  \end{remark}
}{}

While \Cref{assum:est} is satisfied for stable full-order observers
of~\cref{eq:model:noise},\footnote{A full-order state observer
  of~\cref{eq:model:noise} is a dynamical system, evolving in the same state
  space as~\cref{eq:model:noise}, stabilized with respect to \(x\) by output
  feedback.} we know of no nonlinear results that guarantee a Lyapunov function
characterization of stability (i.e., \Cref{assum:est}) for the full information
estimation (FIE) or moving horizon estimation (MHE) algorithms. FIE and MHE were
shown to be RGES for exponentially detectable and stabilizable systems
by~\cite{allan:rawlings:2021a}, but they use a \(Q\)-function to demonstrate
stability. To the best of our knowledge, the closest construction is the
\(N\)-step Lyapunov function
of~\cite{schiller:muntwiler:kohler:zeilinger:muller:2023}. %
\ifthenelse{\boolean{LongVersion}}{%
  If we treat the disturbance as a parameter, rather than an uncontrollable
  integrator, there are FIE and MHE algorithms for combined state and parameter
  estimation that could also be used to estimate the states and
  disturbances~\citep{muntwiler:kohler:zeilinger:2023a,schiller:muller:2023}.%
  \footnote{The estimation algorithms of~\cite{muntwiler:kohler:zeilinger:2023a}
    produce RGES state estimates, but it is not shown the parameter estimates
    are RGES.~The estimation algorithm of~\cite{schiller:muller:2023} produces
    RGES state and parameter estimates, but only under a persistence of
    excitation condition.} %
}{}

\section{Robust stability for reference tracking}\label{sec:stability}
In this section, we present stability theory relevant to the setpoint-tracking
problem. We consider the system,\footnote{To ensure unphysical states are not
  produced by additive disturbances, we let the disturbance set be a function of
  the state and input.%
  \ifthenelse{\boolean{LongVersion}}{ %
    However, we can convert~\cref{eq:sys:ctrl} to a standard form by taking
    \(\xi^+ = \tilde{F}(\xi,u,\omega)\), \(\omega\in\Omega\) where
    \(\tilde{F}(\xi,u,\omega) = F(\xi,\textrm{proj}_{\Omega(\xi,u)}(\omega))\),
    \(\Omega \defas \bigcup_{(\xi,u)\in\Xi\times\bbU} \Omega(\xi,u)\), and
    \(\textrm{proj}_{\Omega(\xi,u)}(\omega) = \argmin_{\omega'\in\Omega(\xi,u)}
    |\omega-\omega'|\).%
  }{}}
\begin{align}\label{eq:sys:ctrl}
  \xi^+ &= F(\xi,u,\omega), & \omega &\in \Omega(\xi,u).
\end{align}
The system \cref{eq:sys:ctrl} represents the evolution of an \emph{extended}
plant state \(\xi\in\Xi\subseteq\real^{n_\xi}\) subject to the input
\(u\in\bbU\) and \emph{extended} disturbance
\(\omega\in\Omega(\xi,u)\subseteq\real^{n_\omega}\) (to be defined). Greek
letters are used for the extended state and disturbance \((\xi,\omega)\) to
avoid confusion with the states and disturbances
of~\cref{eq:plant,eq:model,eq:model:noise}. Throughout, we assume \(\Xi\) is
closed and \(0\in \Omega(\xi,u)\) and \(F(\xi,u,\omega)\in\Xi\) for all
\((\xi,u)\in\Xi\times\bbU\) and \(\omega\in\Omega(\xi,u)\).

\subsection{Robust stability with respect to two outputs}\label{ssec:stability}
We first consider stabilization of \cref{eq:sys:ctrl} under state feedback,
\begin{align}\label{eq:sys:cl}
  \xi^+ &= F_c(\xi,\omega), & \omega &\in \Omega_c(\xi)
\end{align}
where \(\kappa:\Xi\rightarrow\bbU\) is the control law, \(F_c(\xi,\omega) \defas
F(\xi,\kappa(\xi),\omega)\), and \(\Omega_c(\xi) \defas
\Omega(\xi,\kappa(\xi))\). We define robust positive invariance for the
system~\cref{eq:sys:cl} as follows.
\begin{definition}[Robust positive invariance]\label{defn:rpi}
  A closed set \(X\subseteq\Xi\) is \emph{robustly positive invariant} (RPI) for
  the system~\cref{eq:sys:cl} if \(\xi\in X\) and \(\omega\in\Omega_c(\xi)\)
  imply \(F_c(\xi,\omega)\in X\).
\end{definition}

To address robust setpoint-tracking stability, we extend the definition of
input-to-state stability (ISS) with respect to two measurement
functions~\citep{tran:kellett:dower:2015}. Consider the outputs
\begin{align}\label{eq:output}
  \zeta_1 &= G_1(\xi,\omega), & \zeta_2 &= G_2(\xi,\omega)
\end{align}
where \(\zeta_1\in \real^{n_{\zeta_1}}\) and \(\zeta_2\in \real^{n_{\zeta_2}}\).
In this context, ``output'' refers to any function of the extended state and
disturbance, not only the output \(y\) used for state estimation. From
\cref{eq:output}, the measurement functions of~\cite{tran:kellett:dower:2015}
are the special case where \(G_1\) and \(G_2\) are scalar-valued, positive
semidefinite functions of \(\xi\).

\begin{definition}[Robust stability w.r.t.~two
  outputs]\label{defn:ras}\label{defn:res}
  We say the system~\cref{eq:sys:cl} (with outputs~\cref{eq:output}) is
  \emph{robustly asymptotically stable} (RAS) (\emph{on a RPI set
    \(X\subseteq\Xi\)}) \emph{with respect to \((\zeta_1,\zeta_2)\)} if there
  exist \(\beta_\zeta\in\calKL\) and \(\gamma_\zeta\in\calK\) such that
  \begin{equation}\label{eq:ras}
    |\zeta_1(k)| \leq \beta_\zeta(|\zeta_2(0)|,k) +
    \gamma_\zeta(\|\bfomega\|_{0:k})
  \end{equation}
  for each \(k\in\nnegint\) and trajectories
  \((\bfxi,\bfomega,\bfzeta_1,\bfzeta_2)\) satisfying \cref{eq:sys:cl},
  \cref{eq:output}, and \(\xi(0)\in X\). We say~\cref{eq:sys:cl} is
  \emph{robustly exponentially stable} (RES) w.r.t.~\((\zeta_1,\zeta_2)\) if it
  is RAS w.r.t.~\((\zeta_1,\zeta_2)\) with \(\beta_\zeta(s,k)\defas
  c_\zeta\lambda_\zeta^ks\) for some \(c_\zeta>0\) and
  \(\lambda_\zeta\in(0,1)\).
\end{definition}

For the nominal case (i.e., \(\Omega(\xi,u)\equiv\set{0}\)), we drop the word
\emph{robust} from \Cref{defn:rpi,defn:ras} and simply write \emph{positive
  invariant}, \emph{asymptotically stable} (AS), and \emph{exponentially stable}
(ES). Moreover, if the system~\cref{eq:sys:cl} is RAS (RES)
w.r.t.~\((\zeta,\zeta)\), where \(\zeta=G(\xi,\omega)\), we simply say it is RAS
(RES) w.r.t.~\(\zeta\).

\begin{remark}\label{rem:ras}
  If~\cref{eq:sys:cl} is RAS (on \(X\) w.r.t.~\((\zeta_1,\zeta_2)\)), then the
  disturbance \(\omega\) vanishing implies the output \(\zeta_1\) vanishes,
  i.e., \(\omega(k)\rightarrow 0\) (and \(\xi(0)\in X\)) implies
  \(\zeta_1(k)\rightarrow 0\)~\cite[Lem.~2]{tran:kellett:dower:2015}.
\end{remark}

\begin{remark}%\label{rem:ras:applications}
  In \Cref{sec:nominal,sec:robust}, we demonstrate nominal stability and
  robustness to estimate error, noise, and SSTP parameter changes. The following
  cases of the system~\cref{eq:sys:ctrl}, control law \(u=\kappa(\xi)\), and
  outputs \cref{eq:output} are considered.
  \begin{enumerate}
  \item \emph{Nominal stability}: \ifthenelse{\boolean{LongVersion}}{Let}{}
    \(\xi\defas x\), \(u = \kappa(\xi) \defas \kappa_N(x,\beta)\), \(\omega\defas
    0\), \(\zeta_1 \defas g(u,h(x,u,d)) - \rsp\), and \(\zeta_2 \defas x -
    x_s(\beta)\). %
    \ifthenelse{\boolean{LongVersion}}{%
      Then, for each \emph{fixed} \(\beta=(\rsp,\usp,\ysp,d)\in\calB\), the
      closed-loop system has dynamics~\cref{eq:sys:cl} and
      outputs~\cref{eq:output} with
      \begin{align*}
        F(\xi,\omega) &\defas f(x,\kappa_N(x,\beta),\beta) \\
        G_1(\xi) &\defas g(x,h(x,\kappa_N(x,\beta),d)) - \rsp \\
        G_2(\xi) &\defas x - x_s(\beta)
      \end{align*}
      for each \(\xi\in\calX_N^\rho \defas \lev_\rho V_N^0\) and \(\omega=0\). AS
      (ES) w.r.t.~\(\zeta_2\) corresponds to (exponential) target-tracking
      stability, and AS (ES) w.r.t.~\((\zeta_1,\zeta_2)\) corresponds to
      (exponential) setpoint-tracking stability. %
    }{}
  \item \emph{Robust stability (w.r.t.~estimate error, noise, SSTP parameter
      changes)}: \ifthenelse{\boolean{LongVersion}}{Let}{}
    \(\xi\defas(\hat{x},\hat{\beta})\), \(\kappa(\xi)\defas\kappa_N(\xi)\),
    \(\omega\defas(e,e^+,\Delta\ssp,\tilde{w})\), \(\zeta_1 \defas r - \rsp\),
    \(\zeta_2 \defas \hat{x} - x_s(\hat{\beta})\), where \(r\defas g(u,
    h(\hat{x}+e_x,u,\hat{d}+e_d) + v)\) and \(\hat{\beta} \defas (\ssp,
    \hat{d})\). %
    \ifthenelse{\boolean{LongVersion}}{%
      Then the closed-loop system has dynamics~\cref{eq:sys:cl} and
      outputs~\cref{eq:output} with
      \begin{align*}
        F(\xi,\omega)
        &\defas \begin{bmatrix} f(\hat{x}+e_x,\kappa_N(\hat{x},\hat{\beta}),\hat{d}+e_d) + w - e_x^+ \\
                  \ssp + \Delta\ssp \\ \hat{d} + e_d + w_d - e_d^+ \end{bmatrix} \\
        G_1(\xi) &\defas g(x,h(\hat{x}+e_x,\kappa_N(\hat{x},\hat{\beta}),\hat{d}+e_d) + v) - \rsp, \\
        G_2(\xi) &\defas \hat{x} - x_s(\hat{\beta})
      \end{align*}
      for each \(\xi=(\hat{x},\hat{\beta})\) in a to-be-defined RPI set
      \(\hat{\calS}_N^\rho\) and \(\omega\in\Omega_c(\xi)\) (to be defined). RAS
      (RES) of \cref{eq:sys:cl} w.r.t.~\(\zeta_2\) alone corresponds to robust
      (exponential) target-tracking stability, and RAS (RES)
      w.r.t.~\((\zeta_1,\zeta_2)\) corresponds to robust (exponential)
      setpoint-tracking stability. %
    }{}
  \end{enumerate}
  \ifthenelse{\boolean{LongVersion}}{}{%
    For either case, robust stability of \cref{eq:sys:cl} w.r.t.~\(\zeta_2\) alone
    corresponds to robust target-tracking stability, and robust stability
    w.r.t.~\((\zeta_1,\zeta_2)\) corresponds to robust setpoint-tracking
    stability. %
  }
\end{remark}

\begin{remark}\label{rem:iss}
  While \Cref{defn:ras} generalizes many ISS and input-to-output stability (IOS)
  definitions\ifthenelse{\boolean{LongVersion}}{ originally posed for
    continuous-time systems
    by}{}~\cite{sontag:wang:1995,sontag:wang:1999,sontag:wang:2000}, these
  special cases are not suitable for analyzing %
  \ifthenelse{\boolean{LongVersion}}{%
    both target- and setpoint-tracking performance of %
  }{}%
  offset-free MPC.\@ ISS is not appropriate as the SSTP parameters \(\beta\) are
  often part of the extended state \(\xi\). IOS allows the tracking performance
  to degrade with the magnitude of the SSTP parameters. While state-independent
  IOS (SIIOS) coincides with the special case of \(\zeta = G_1(\xi)\equiv
  G_2(\xi)\) (e.g., for target-tracking), we find it is not general enough for
  setpoint tracking.
\end{remark}

Next, we define an (exponential) ISS Lyapunov function with respect to the
disturbance-free outputs
\begin{align}\label{eq:output:nonoise}
  \zeta_1 &= G_1(\xi), & \zeta_2 &= G_2(\xi)
\end{align}
and show its existence implies RAS (RES) of \cref{eq:sys:cl} with respect to
\((\zeta_1,\zeta_2)\) (see \Cref{app:lyap} for proof).
\begin{definition}[ISS Lyapunov function]\label{defn:lyap}
  Consider the system~\cref{eq:sys:cl} with outputs \cref{eq:output:nonoise}. We
  call \(V:\Xi\rightarrow\nnegreal\) an \emph{ISS Lyapunov function} (\emph{on a
    RPI set \(X\subseteq\Xi\)}) \emph{with respect to \((\zeta_1,\zeta_2)\)} if
  there exist \(\alpha_i\in\calKinf,i\in\intinterval{1}{3}\) and
  \(\sigma\in\calK\) such that, for each \(\xi\in X\) and
  \(\omega\in\Omega_c(\xi)\),
  \begin{subequations}\label{eq:lyap}
    \begin{align}
      \alpha_1(|G_1(\xi)|) &\leq V(\xi) \leq \alpha_2(|G_2(\xi)|) \label{eq:lyap:a} \\
      V(F_c(\xi,\omega)) &\leq V(\xi) - \alpha_3(V(\xi)) + \sigma(|\omega|). \label{eq:lyap:b}
    \end{align}
  \end{subequations}
  We say \(V\) is an \emph{exponential ISS Lyapunov function}
  with respect to \((\zeta_1,\zeta_2)\) if it is an ISS Lyapunov function
  with respect to \((\zeta_1,\zeta_2)\) with \(\alpha_i = a_i\id^b\) for some
  \(a_i,b>0,i\in\intinterval{1}{3}\).
\end{definition}

\begin{theorem}[ISS Lyapunov theorem]\label{thm:lyap}
  If the system \cref{eq:sys:cl} with outputs \cref{eq:output:nonoise} admits an
  (exponential) ISS Lyapunov function \(V:\Xi\rightarrow\nnegreal\) on an RPI
  set \(X\subseteq\Xi\) with respect to \((\zeta_1,\zeta_2)\), then it is RAS
  (RES) on \(X\) with respect to \((\zeta_1,\zeta_2)\).
\end{theorem}

As in \Cref{defn:rpi,defn:ras}, we call \(V\) a \emph{Lyapunov function} or
\emph{exponential Lyapunov function} w.r.t.~\((\zeta_1,\zeta_2)\) if it
satisfies \Cref{defn:lyap} in the nominal case (i.e.,
\(\Omega(\xi,u)\equiv\set{0}\)). %
\ifthenelse{\boolean{LongVersion}}{%
  Moreover, we note that the proof of \Cref{thm:lyap} easily extends to the
  nominal case by setting \(\omega = 0\) throughout. %
}{}

\begin{remark}\label{rem:lyap}
  If \(G_1 \equiv G_2\), then we can replace \cref{eq:lyap:b} with
  \(V(F_c(\xi,\omega)) \leq V(\xi) - \tilde{\alpha}_3(|G_1(\xi)|) +
  \sigma(|\omega|)\) in \Cref{defn:lyap}, where \(\tilde{\alpha}_3\in\calKinf\).
  Then \cref{eq:lyap:b} holds with
  \(\alpha_3\defas\tilde{\alpha}_3\circ\alpha_2^{-1}\).
\end{remark}

\subsection{Joint controller-estimator robust
  stability}\label{ssec:stability:est}
Without plant-model mismatch, RES of each subsystem implies RES of the joint
system. This is because the controller and estimator error systems are connected
\emph{sequentially}%
\ifthenelse{\boolean{LongVersion}}{%
  , with the tracking errors having no influence on the estimation errors%
}{}%
. However, plant-model mismatch makes this a \emph{feedback interconnection},
with the tracking errors influencing the state estimate errors and vice versa.
Therefore it is necessary to analyze stability of the joint system.

We define the \emph{extended} sensor output
\(\upsilon\in\Upsilon\subseteq\real^{n_\upsilon}\) by
\begin{equation}\label{eq:meas}
  \upsilon = H(\xi,u,\omega).
\end{equation}
Assume \(\Upsilon\) is closed and \(H(\xi,u,\omega)\in\Upsilon\) for all
\((\xi,u)\in\Xi\times\bbU\) and \(\omega\in\Omega(\xi,u)\). We consider the
extended state estimator
\begin{equation}\label{eq:state:est}
  \hat{\xi}(k) \defas \Phi^\xi_k(\overline{\xi},\bfu_{0:k-1},\bfupsilon_{0:k-1})
\end{equation}
and stabilization via state estimate feedback,
\begin{equation}\label{eq:ctrl}
  u = \hat{\kappa}(\hat{\xi})
\end{equation}
where \(\overline{\xi}\in\hat{\Xi}\subseteq\real^{n_{\hat{\xi}}}\) is the prior
guess, \(\Phi^\xi_k : \hat{\Xi}\times\bbU^k\times\Upsilon^k \rightarrow
\hat{\Xi},k\in\nnegint\) is the estimator, and
\(\hat{\kappa}:\hat{\Xi}\rightarrow\bbU\) is the control law. The set
\(\hat{\Xi}\) is closed but is not necessarily the same, let alone of the same
dimension, as \(\Xi\). In other words, the \emph{extended} plant and model
states may evolve on different spaces. Thus, we define the estimator error
\(\varepsilon\in\real^{n_{\hat{\xi}}}\) as the deviation of the estimate
\(\hat{\xi}\) from a function \(G_\varepsilon:\Xi\rightarrow\hat{\Xi}\) of the
state \(\xi\),
\begin{align}\label{eq:state:est:err}
  \varepsilon(k) &= G_\varepsilon(\xi(k)) - \hat{\xi}(k),
  & \overline{\varepsilon} &\defas G_\varepsilon(\xi(0)) - \overline{\xi}.
\end{align}
Finally, with the outputs
\begin{align}\label{eq:output:est}
  \zeta_1 &= G_1(\xi,\hat{\xi},u,\omega),
  & \zeta_2 &= G_2(\xi,\hat{\xi},u,\omega)
\end{align}
we define a RPI set and robust stability as follows.

\begin{definition}[Joint RPI]
  \label{defn:rpi:est}
  A closed set \(S\subseteq\Xi\times\hat{\Xi}\) is RPI for the
  system~\cref{eq:sys:ctrl,eq:meas,eq:state:est,eq:ctrl} if
  \((\xi(k),\hat{\xi}(k))\in\calS,k\in\nnegint\) for all
  \((\bfxi,\bfu,\bfomega,\bfupsilon)\) satisfying~\cref{eq:sys:ctrl},
  \cref{eq:meas,eq:ctrl,eq:state:est}, and \((\xi(0),\overline{\xi})\in\calS\).
\end{definition}

\begin{definition}[Joint robust stability]\label{defn:ras:ctrl:est}
  The system~\cref{eq:sys:ctrl}, \cref{eq:meas,eq:state:est,eq:ctrl} (with
  outputs \cref{eq:output:est}) is RAS in a RPI set
  \(\calS\subseteq\bbX\times\hat{\bbX}\) w.r.t. \((\zeta_1,\zeta_2)\) if there
  exist \(\beta_\zeta,\gamma_\zeta\in\calKL\) such that
  \begin{equation}\label{eq:ras:ctrl:est}
    |(\zeta_1(k),\varepsilon(k))| \leq
    \beta_\zeta(|(\zeta_2(0),\overline{\varepsilon})|,k) + \sum_{i=0}^k
    \gamma_\zeta(|\omega(k-i)|,i)
  \end{equation}
  for all \(k\in\nnegint\) and all trajectories
  \((\bfxi,\bfu,\bfomega,\bfupsilon,\bfepsilon,\bfzeta_1,\bfzeta_2)\)
  satisfying~\cref{eq:sys:ctrl},
  \cref{eq:meas,eq:ctrl,eq:state:est,eq:output:est,eq:state:est:err}, and
  \((\xi(0),\overline{\xi})\in\calS\). We
  say~\cref{eq:sys:ctrl,eq:meas,eq:state:est,eq:ctrl} is RES
  w.r.t.~\((\zeta_1,\zeta_2)\) if it is RAS w.r.t.~\((\zeta_1,\zeta_2)\) with
  \(\beta_\zeta(s,k)\defas c_\zeta\lambda_\zeta^ks\) and
  \(\gamma_\zeta(s,k)\defas \lambda_\zeta^k\sigma_\zeta(s)\) for some
  \(c_\zeta>0\), \(\lambda_\zeta\in(0,1)\), and \(\sigma_\zeta\in\calK\).
\end{definition}

As in \Cref{ssec:stability}, we
say~\cref{eq:sys:ctrl,eq:meas,eq:state:est,eq:ctrl} is RAS (RES)
w.r.t.~\(\zeta=G(\xi,\omega)\) if it is RAS (RES) w.r.t.~\((\zeta,\zeta)\).

%% TODO Not sure ctrl+est, w/o mismatch works at all. We don't have a bound like
%% \(V_N^0(\hat{x},\hat{\beta}) - V_N^0(\hat{x},\hat{\beta}) \leq
%% -a_3|\delta\hat{x}|^2 + a_4|(e,e^+)|^2 + \sigma_w(|(\Delta\ssp,\tilde{w})|)\)
%% for the regulator until we assume \(z_s\) is Lipschitz, and I don't want to
%% add that assumption to Section 5.
\ifthenelse{\boolean{LongVersion}}{%
  In \Cref{sec:mismatch}, we establish robustness of offset-free MPC with
  plant-model mismatch in terms of \Cref{defn:ras:ctrl:est}, using the following
  definition of the system~\cref{eq:sys:ctrl,eq:meas,eq:state:est,eq:ctrl},
  estimate errors~\cref{eq:state:est:err}, and outputs \cref{eq:output:est}:
  % In \Cref{sec:robust,sec:mismatch}, we establish robustness of offset-free MPC,
  % with and without plant-model mismatch, in terms of \Cref{defn:ras:ctrl:est}.
  % The following cases of the
  % system~\cref{eq:sys:ctrl,eq:meas,eq:state:est,eq:ctrl}, estimate
  % errors~\cref{eq:state:est:err}, and outputs \cref{eq:output:est} are
  % considered.
  \begin{enumerate}
  \item[3.] \emph{With mismatch}: \ifthenelse{\boolean{LongVersion}}{Let}{}
    \(\xi\defas(\xP,\alpha)\), \(\hat{\xi}\defas(\hat{x},\hat{\beta})\),
    \(u\defas\kappa_N(\hat{\xi})\), \(\omega\defas(\Delta\ssp,\Delta\wP)\),
    \(\upsilon\defas(y,\Delta\ssp)\), \(\varepsilon\defas (\xP + \Delta
    x_s(\alpha), \ssp, d_s(\alpha)) - \hat{\xi}\), \(\zeta_1 \defas r - \rsp\),
    \(\zeta_2 \defas \hat{x} - x_s(\hat{\beta})\), where \(r\defas
    g(u,\hP(x,u,\wP))\), \(\alpha \defas (\ssp, \wP)\), \(\hat{\beta} \defas
    (\ssp, \hat{d})\), and \((\Delta x_s(\alpha),d_s(\alpha))\) are to be
    defined. %
    \ifthenelse{\boolean{LongVersion}}{%
      Then the closed-loop system has
      dynamics~\cref{eq:sys:ctrl,eq:meas,eq:state:est,eq:ctrl},
      errors~\cref{eq:state:est:err}, and outputs~\cref{eq:output:est} with
      \begin{align*}
        F(\xi,u,\omega)
        &\defas \begin{bmatrix} \fP(\xP,u,\wP) \\ \ssp + \Delta\ssp \\ \wP + \Delta\wP \end{bmatrix},
          \ifthenelse{\boolean{OneColumn}}{
        &}{\\} H(\xi,u,\omega) &\defas \begin{bmatrix} \hP(\xi,u,\wP) \\ \Delta\ssp \end{bmatrix}, \\
        \Phi^\xi_k(\overline{\xi},\bfu_{0:k-1},\bfupsilon_{0:k-1})
        &\defas (\hat{x}(k),\ssp(k),\hat{d}(k)),
          \ifthenelse{\boolean{OneColumn}}{
        &}{\\} G_\varepsilon(\xi) &\defas \begin{bmatrix} \xP + \Delta x_s(\alpha) \\
                                            d_s(\alpha) \end{bmatrix} \\
        G_1(\xi,u,\omega)
        &\defas g(u,\hP(\xP,u,\wP)) - \rsp,
          \ifthenelse{\boolean{OneColumn}}{
        &}{\\} G_2(\hat{\xi}) &\defas \hat{x} - x_s(\hat{\beta})
      \end{align*}
      for each \((\xi,\hat{\xi})=(x,\beta,\hat{x},\hat{\beta})\) in a
      to-be-defined RPI set \(\calS_N^{\rho,\tau}\) and
      \(\omega\in\Omega_c(\xi)\) (to be defined), where
      \((\hat{x}(k),\hat{d}(k)) \defas
      \Phi_k(\overline{x},\overline{d},\bfu_{0:k-1},\bfy_{0:k-1})\) as in
      \cref{defn:est}. %
    }{}
  \end{enumerate}
  % For either case, RAS (RES) w.r.t.~\(\zeta_2\) corresponds to robust
  % (exponential) target-tracking stability, and RAS (ES)
  % w.r.t.~\((\zeta_1,\zeta_2)\) corresponds to robust (exponential)
  % setpoint-tracking stability.
  As in \Cref{ssec:stability}, RAS (RES) w.r.t.~\(\zeta_2\) corresponds to
  robust (exponential) target-tracking stability, and RAS (ES)
  w.r.t.~\((\zeta_1,\zeta_2)\) corresponds to robust (exponential)
  setpoint-tracking stability. %
}{}

\begin{remark}\label{rem:ras:ctrl:est}
  As in \Cref{rem:ras}, if~\cref{eq:sys:ctrl,eq:meas,eq:state:est,eq:ctrl} is
  RAS (on \(\calS\) w.r.t.~\((\zeta_1,\zeta_2)\)), then the disturbance
  \(\omega\) vanishing implies both the output \(\zeta_1\) and the error
  \(\varepsilon\) vanish, i.e., \(\omega(k)\rightarrow 0\) (and
  \((\xi(0),\overline{\xi})\in\calS\)) implies
  \((\zeta_1(k),\varepsilon(k))\rightarrow 0\)
  \ifthenelse{\boolean{LongVersion}}{%
    (cf.~Proposition~3.11 of \cite{allan:rawlings:2021a}). %
  }{%
    (cf.~\cite[Prop.~3.11]{allan:rawlings:2021a}).
  }
\end{remark}

To analyze stability of the
system~\cref{eq:sys:ctrl,eq:meas,eq:ctrl,eq:state:est}, we use the following
theorem (see \Cref{app:smallgain} for proof).
\begin{theorem}[Joint Lyapunov theorem]\label{thm:smallgain}
  Consider the system~\cref{eq:sys:ctrl}, \cref{eq:meas,eq:state:est,eq:ctrl}
  with errors~\cref{eq:state:est:err} and output \(\zeta=G(\hat{\xi})\). Suppose
  \(\Phi^\xi_0\) is the identity map and there exist %constants
  \(a_i,b_i>0,i\in\intinterval{1}{4}\), a RPI set
  \(\calS\subseteq\bbX\times\hat{\bbX}\), %and functions
  \(V:\hat{\Xi}\rightarrow\nnegreal\),
  \(V_\varepsilon:\Xi\times\hat{\Xi}\rightarrow\nnegreal\), and
  \(\sigma,\sigma_\varepsilon\in\calK\) such that \(\frac{a_4c_4}{a_3c_1} < 1\),
  \(\frac{a_4c_4}{a_3c_3} < \frac{c_1}{c_1+c_2}\), and, for all trajectories
  \((\bfxi,\hat{\bfxi},\bfu,\bfomega,\bfupsilon,\bfepsilon,\bfzeta)\)
  satisfying~\cref{eq:sys:ctrl,eq:meas,eq:state:est,eq:ctrl,eq:state:est:err},
  \(\zeta=G(\hat{\xi})\), and \((\xi(0),\overline{\xi})\in\calS\), we also
  satisfy
  \begin{subequations}\label{eq:smallgain}
    \begin{align}
      a_1|\zeta|^2
      &\leq V(\hat{\xi}) \leq a_2|\zeta|^2 \label{eq:smallgain:a} \\
      V(\hat{\xi}^+)
      &\leq V(\hat{\xi}) - a_3|\zeta|^2 + a_4|(\varepsilon,\varepsilon^+)|^2 +
        \sigma(|\omega|) \label{eq:smallgain:b} \\
      c_1|\varepsilon|^2
      &\leq V_\varepsilon(\xi,\hat{\xi}) \leq c_2|\varepsilon|^2
        \label{eq:smallgain:c} \\
      V_\varepsilon(\xi^+,\hat{\xi}^+)
      &\leq V_\varepsilon(\xi,\hat{\xi}) - c_3|\varepsilon|^2 + c_4|\zeta|^2 +
        \sigma_\varepsilon(|\omega|). \label{eq:smallgain:d}
    \end{align}
  \end{subequations}
  Then the system~\cref{eq:sys:ctrl,eq:meas,eq:state:est,eq:ctrl} is RES in
  \(\calS\) w.r.t.~\(\zeta\).
\end{theorem}

\section{Nominal offset-free performance}\label{sec:nominal}
In this section, we consider the application of offset-free MPC to the model
\cref{eq:model} in the \emph{nominal} case (i.e., without estimate errors or
setpoint and disturbance changes). Contrary to the subsequent sections, we
assume the SSTP parameters \(\beta=(\ssp,d)\) are fixed, and the disturbance
\(d\) is known.

Consider the following \emph{modeled} closed-loop system:
\begin{subequations}\label{eq:model:cl}
  \begin{align}
    x^+ &= f_c(x,\beta) \defas f(x,\kappa_N(x,\beta),d)
          \label{eq:model:cl:a} \\
    y &= h_c(x,\beta) \defas h(x,\kappa_N(x,\beta),d) \\
    r &= g_c(x,\beta) \defas g(\kappa_N(x,\beta),h_c(x,\beta))
  \end{align}
\end{subequations}
where \((x,\beta)\defas(x,\ssp,d)\in\calS_N\). For each \(\rho>0\) and
\(\beta\in\calB\), we define the candidate domain of stability
\begin{equation}\label{eq:mpc:dos}
  \calX_N^\rho(\beta) \defas \lev_\rho V_N^0(\cdot,\beta).
\end{equation}
\Cref{thm:mpc:nominal} generalizes standard MPC nominal stability results %
\ifthenelse{\boolean{LongVersion}}{%
  (cf.~Section~2.4 of \cite{rawlings:mayne:diehl:2020}) %
}{%
  (cf.~\cite[Sec.~2.4]{rawlings:mayne:diehl:2020}) %
}%
to consider steady-state targets based on the SSTP~\cref{eq:sstp} (see
\Cref{app:mpc:nominal} for a proof).
\begin{theorem}[Nominal offset-free stability]\label{thm:mpc:nominal}
  Suppose \Cref{assum:cont,assum:cons,assum:sstp:exist,assum:stabilizability,%
    assum:quad} hold. Let \(\rho>0\).
  \begin{enumerate}[(a)]
  \item For each compact \(\calB_c\subseteq\calB\), there exist constants
    \(a_1,a_2,a_3>0\) such that, for all \(x\in\calX_N^\rho(\beta)\) and
    \(\beta\in\calB_c\),
    \begin{subequations}\label{eq:mpc:lyap}
      \begin{align}
        a_1|x-x_s(\beta)|^2 &\leq V_N^0(x,\beta) \leq a_2|x-x_s(\beta)|^2
                              \label{eq:mpc:lyap:a} \\
        V_N^0(f_c(x,\beta),\beta) &\leq V_N^0(x,\beta) - a_3|x-x_s(\beta)|^2.
                                    \label{eq:mpc:lyap:b}
      \end{align}
    \end{subequations}
  \item For each \(\beta\in\calB\), the closed-loop system~\cref{eq:model:cl:a}
    is ES on \(\calX_N^\rho(\beta)\) w.r.t. the target-tracking error \(\delta
    x\defas x-x_s(\beta)\).
  \item For each \(\beta=(\rsp,\zsp,d)\in\calB\), the closed-loop
    system~\cref{eq:model:cl:a} is AS on \(\calX_N^\rho(\beta)\) w.r.t.
    \((\delta r,\delta x)\), where \(\delta r\defas g_c(x,\beta)-\rsp\) is the
    setpoint-tracking error.
  \item If \(g\) and \(h\) are Lipschitz continuous on bounded sets, then part
    (c) can be upgraded to ES.\@
  \end{enumerate}
\end{theorem}

\begin{remark}\label{rem:mpc:nominal}
  Contrary to standard MPC results~\citep[Sec.~2.4]{rawlings:mayne:diehl:2020},
  but similar to tracking MPC results~\citep{%
    falugi:2015,% nonlinear reference tracking MPC
    limon:ferramosca:alvarado:alamo:2018,% nonlinear tracking MPC
    galuppini:magni:ferramosca:2023% semidefinite stage cost case
  }, the Lyapunov bounds in \Cref{thm:mpc:nominal}(a) are \emph{uniform} in
  \(\beta\). This implies a guaranteed decay rate \(\lambda\in(0,1)\) for the
  tracking error \(\delta x\) and paves the way to robustness w.r.t.
  \(\Delta\beta\), but introduces a trade-off: as the set \(\calB_c\) grows, the
  rate of decay \(\lambda\) degrades.
\end{remark}

\section{Offset-free performance without mismatch}\label{sec:robust}
In this section, we prove offset-free MPC (without plant-model mismatch) is
robustly stable with respect to estimate errors and setpoint and disturbance
changes. We assume the plant evolves according to~\cref{eq:model:noise} and the
setpoints evolve as
\begin{equation}\label{eq:sp}
  \ssp^+ = \ssp + \Delta\ssp.
\end{equation}
With \(\Delta\beta\defas(\Delta\ssp,w_d)\), we have
\(\beta^+=\beta+\Delta\beta\). Similarly %
\ifthenelse{\boolean{LongVersion}}{%
  to Section~4.6 of \cite{rawlings:mayne:diehl:2020}, %
}{%
  to~\cite[Sec.~4.6]{rawlings:mayne:diehl:2020}, %
}%
we write the estimate error system as
\begin{subequations}\label{eq:model:est}
  \begin{align}
    \hat x^+ &= f(\hat x+e_x,u,\hat d+e_d) + w - e_x^+ \\
    \hat d^+ &= \hat d + e_d + w_d - e_d^+ \\
    y &= h(\hat{x}+e_x,u,\hat{d}+e_d) + v.
  \end{align}
\end{subequations}
Let \(\tilde{d}\defas(e,e^+,\Delta\ssp,\tilde{w})\) denote the lumped
perturbation term. %
\ifthenelse{\boolean{LongVersion}}{%
  To ensure the noise does not result in unphysical states, disturbances, or
  measurements, we %
}{%
  To satisfy physical constraints \((\hat x^+,\hat
  d^+,y)\in\bbX\times\bbD\times\bbY\), we %
}%
restrict the perturbations \(\tilde{d}\) to the set
%% NOTE For some reason multline is producing a lot of space above the
%% equation, so I simulated the behavior with an align environment. If the
%% equation is changed, just be careful to adjust the alignment.
% \begin{multline*}
%   \tilde{\bbD}(\hat{x},u,\hat{d}) \defas \{\;
%   (e_x,e_d,e_x^+,e_d^+,\Delta\ssp,\tilde{w}) \;|\; \cref{eq:model:est},\\
%   (\hat{x}^+,\hat{d}^+)\in\bbX\times\bbD,
%   \tilde{w}\in\tilde{\bbW}(\hat x+e_x,u,\hat d+e_d) \;\}
% \end{multline*}
\begin{align*}
  \tilde{\bbD}(\hat{x},u,\hat{d})
  &\defas \{\; (e_x,e_d,e_x^+,e_d^+,\Delta\ssp,\tilde{w}) \;|\; \cref{eq:model:est}, \\
  (&\hat{x}^+,\hat{d}^+)\in\bbX\times\bbD,
  \tilde{w}\in\tilde{\bbW}(\hat x+e_x,u,\hat d+e_d) \;\}
\end{align*}
for each \((\hat{x},u,\hat{d})\in\bbX\times\bbU\times\bbD\). The closed-loop
estimate error system, defined by
\cref{eq:sstp,eq:mpc,eq:est,eq:sp,eq:model:est}, evolves as
\begin{subequations}\label{eq:model:est:cl}
  \begin{align}
    \hat x^+ &= \hat{f}_c(\hat{x},\hat{\beta},\tilde{d})
               \ifthenelse{\boolean{OneColumn}}{}{\nonumber \\
             &\qquad\quad} \defas f(\hat x+e_x,\kappa_N(\hat{x},\hat{\beta}),\hat{d}+e_d)
               + w - e_x^+ \label{eq:model:est:cl:a} \\
    \hat{\beta}^+ &= \hat{f}_{\beta,c}(\hat{\beta},\tilde{d}) \defas
                    \begin{bmatrix} \ssp + \Delta\ssp \\
                      \hat d + e_d + w_d - e_d^+ \end{bmatrix}
                    \label{eq:model:est:cl:b} \\
    y &= \hat{h}_c(\hat{x},\hat{\beta},\tilde{d})
        \defas h(\hat{x}+e_x,\kappa_N(\hat{x},\hat{\beta}),\hat{d}+e_d) + v \nonumber \\
    r &= \hat{g}_c(\hat{x},\hat{\beta},\tilde{d})
        \defas g(\kappa_N(\hat{x},\hat{\beta}),h_c(\hat{x},\hat{\beta},\tilde{d})) \nonumber
  \end{align}
\end{subequations}
where \(\hat{\beta} \defas (\ssp,\hat{d})\). %
\ifthenelse{\boolean{LongVersion}}{%
\subsection{Steady-state target problem assumptions}
}{}%
To guarantee the SSTP~\cref{eq:sstp} is robustly feasible at all times,
\ifthenelse{\boolean{LongVersion}}{%
  and the targets themselves are robust to disturbance estimate errors, %
}{}%
we make the following assumption.

\begin{assumption}[SSTP continuity]\label{assum:sstp}
  There exists a compact set \(\calB_c\subseteq\calB\) and constant
  \(\delta_0>0\) such that %
  % \ifthenelse{\boolean{LongVersion}}{%
    \begin{enumerate}[(a)]
    \item \(\hat{\calB}_c \defas \{\; (s,\hat{d}) \;|\; (s,d)\in\calB_c,\;
      |e_d|\leq\delta_0,\; \hat{d}\defas d-e_d\in\bbD \;\} \subseteq \calB\); and
    \item \(z_s\) is continuous on \(\hat{\calB}_c\).
    \end{enumerate}
  % }{%
  %   (a) \(\hat{\calB}_c \defas \{\; (s,\hat{d}) \;|\; (s,d)\in\calB_c,\;
  %   |e_d|\leq\delta_0,\; \hat{d}\defas d-e_d\in\bbD \;\} \subseteq \calB\) and
  %   (b) \(z_s\) is continuous on \(\hat{\calB}_c\). %
  % }%
\end{assumption}

\Cref{assum:sstp}(a) guarantees robust feasibility of the SSTP so long as
\(\bfbeta\in\calB_c^\infty\) and \(\|\bfe_d\|\leq\delta_0\), as well as
robustness of the targets \(z_s(\beta)\) to perturbations in \(\beta\).
% Whenever \Cref{assum:sstp}(a) is satisfied, it is convenient to define
Consider the set
\begin{equation*}
  \tilde{\bbD}_c(\hat{x},\hat{\beta}) \defas \set{ \tilde{d} \in
    \tilde{\bbD}(\hat{x},\kappa_N(\hat{x},\hat{\beta}),\hat{\beta}) |
    \hat{f}_{\beta,c}(\hat{\beta},\tilde{d}) \in \hat{\calB}_c }
\end{equation*}
for each \((\hat{x},\hat{\beta})\in\calS_N\). So long as \(\tilde{d} \in
\tilde{\bbD}_c(\hat{x},\hat{\beta})\), the SSTP is feasible. %
\ifthenelse{\boolean{LongVersion}}{%
  In \Cref{app:terminal}, we construct, under \Cref{assum:sstp}, terminal
  ingredients satisfying \Cref{assum:quad,assum:stabilizability}. %
}{}%
In \Cref{sec:linear}, we use properties of the linearized system to show
\Cref{assum:sstp} holds near the origin.

\ifthenelse{\boolean{LongVersion}}{%
\subsection{Robust stability of offset-free MPC}
}{}%
\Cref{thm:mpc:robust} extends results on inherent robustness of
MPC~\cite{allan:bates:risbeck:rawlings:2017,pannocchia:rawlings:wright:2011},
establishing robust stability of the closed-loop offset-free
MPC~\cref{eq:model:est:cl} (see~\Cref{app:mpc:robust} for a proof).

\begin{theorem}[Robust offset-free stability]\label{thm:mpc:robust}
  If \Cref{assum:cont,assum:cons,assum:sstp:exist,assum:stabilizability,%
    assum:quad,assum:sstp} hold and \(\rho>0\), then there exists \(\delta>0\)
  such that
  \begin{enumerate}[(a)]
  \item the following set is RPI for the closed-loop
    system~\cref{eq:model:est:cl} with disturbance
    \(\tilde{d}\in\tilde{\bbD}_c(\hat{x},\hat{\beta}) \cap
    \delta\bbB^{n_{\tilde{d}}}\):
    \begin{equation}\label{eq:mpc:robust:rpi}
      \hat{\calS}_N^\rho \defas \set{ (\hat{x},\hat{\beta}) \in \calS_N |
      \hat{x}\in\calX_N^\rho(\hat{\beta}), \hat{\beta}\in\hat{\calB}_c };
    \end{equation}
  \item there exist \(a_i>0,i\in\intinterval{1}{3}\) and \(\sigma_r\in\calKinf\)
    such that
    \begin{subequations}\label{eq:mpc:robust:lyap}
      \begin{align}
        a_1|\delta\hat{x}|^2
        &\leq V_N^0(\hat{x},\hat{\beta}) \leq a_2|\delta\hat{x}|^2
          \label{eq:mpc:robust:lyap:a} \\
        V_N^0(\hat{x}^+,\hat{\beta}^+)
        &\leq V_N^0(\hat{x},\hat{\beta}) - a_3|\delta\hat{x}|^2 +
          \sigma_r(|\tilde{d}|) \label{eq:mpc:robust:lyap:b}
      \end{align}
    \end{subequations}
    for all \((\hat{x},\hat{\beta})\in\hat{\calS}_N^\rho\) and
    \(\tilde{d}\in\tilde{\bbD}_c(\hat{x},\hat{\beta}) \cap
    \delta\bbB^{n_{\tilde{d}}}\), given \cref{eq:model:est:cl} and the
    target-tracking error \(\delta\hat{x}\defas\hat{x}-x_s(\hat{\beta})\);
  \item the closed-loop system~\cref{eq:model:est:cl} with disturbance
    \(\tilde{d}\in\tilde{\bbD}_c(\hat{x},\hat{\beta}) \cap
    \delta\bbB^{n_{\tilde{d}}}\) is RES on \(\hat{\calS}_N^\rho\) w.r.t.
    \(\delta\hat{x}\);
  \item the closed-loop system~\cref{eq:model:est:cl} with disturbance
    \(\tilde{d}\in\tilde{\bbD}_c(\hat{x},\hat{\beta}) \cap
    \delta\bbB^{n_{\tilde{d}}}\) is RAS on \(\hat{\calS}_N^\rho\) w.r.t.
    \((\delta r,\delta\hat{x})\), where \(\delta
    r\defas\hat{g}_c(\hat{x},\hat{\beta},\tilde{d})-\rsp\) is the
    setpoint-tracking error and \(\hat{\beta}=(\rsp,\zsp,\hat{d})\); and
  \item if \(g\) and \(h\) are Lipschitz continuous on bounded sets, then
  part (d) can be upgraded to RES.
  \end{enumerate}
\end{theorem}

\begin{remark}\label{rem:mpc:robust}
  \Cref{thm:mpc:robust}(c,d) implies the following tracking error convergence
  result:~\ifthenelse{\boolean{LongVersion}}{we have }{}%
  \(|(\delta\hat{x}(k),\delta r(k))|\rightarrow 0\) so long as \((\hat{x}(0),
  \hat{\beta}(0)) \in \hat{\calS}_N^\rho\), \(|\tilde{d}(k)|\rightarrow 0\), and
  \(\|\tilde{\bfd}\|\leq\delta\) (cf.~\Cref{rem:ras}).
\end{remark}

\begin{remark}\label{rem:mpc:robust:tradeoffs}
  There is a trade-off between \(\rho\) and \(\delta\) implied by
  \Cref{thm:mpc:robust}(a): to be robust everywhere is to not be robust at all.
  As the size of the domain of stability \(\hat{\calS}_N^\rho\) grows to
  \(\calS\), the allowed disturbance magnitude \(\delta\) shrinks to 0.
\end{remark}

\section{Offset-free MPC under mismatch}\label{sec:mismatch}
In this section, we show offset-free MPC, \emph{despite (sufficiently small)
  plant-model mismatch}, is robust to setpoint and disturbance changes. We
consider the plant~\cref{eq:plant}, setpoint dynamics~\cref{eq:sp}, and plant
disturbance dynamics
\begin{equation}\label{eq:wp}
  \wP^+ = \wP + \Delta\wP.
\end{equation}
With \(\alpha\defas(\ssp,\wP)\) and
\(\Delta\alpha\defas(\Delta\ssp,\Delta\wP)\), we have the relationship
\(\alpha^+=\alpha+\Delta\alpha\). The SSTP and regulator are designed with the
model~\cref{eq:model}, and the estimator is designed with the noisy
model~\cref{eq:model:noise}.

\ifthenelse{\boolean{LongVersion}}{%
\subsection{Target selection under mismatch}
}{}%
With plant-model mismatch, the connection between the steady-state targets and
plant steady states becomes more complicated. To guarantee there is a plant
steady state providing offset-free performance and that we can align the plant
and model steady states using the disturbance estimate, we make the following
assumptions about the SSTP.
\begin{assumption}[Existence of mismatch corrections]\label{assum:sstp:mismatch}
  There exist compact sets
  \(\calA_c\subseteq\real^{n_r}\times\overline{\bbZ}_y\times\bbW\) and
  \(\calB_c\subseteq\calB\) containing the origin, continuous functions
  \((\xPs,d_s) : \calA_c \rightarrow \bbX \times \bbD\), and a constant
  \(\delta_0>0\) for which
  % \ifthenelse{\boolean{LongVersion}}{%
    \begin{enumerate}[(a)]
    \item \(\hat{\calB}_c\) (as defined in \Cref{assum:sstp}) is contained in
      \(\calB\);
    \item \(z_s\) is Lipschitz continuous on \(\hat{\calB}_c\);
    \item for each \(\alpha=(\ssp,\wP)\in\calA_c\), the pair
      \((\xPs,d_s)=(\xPs(\alpha),d_s(\alpha))\) is the
      unique solution to
      \begin{subequations}\label{eq:ssop}
        \begin{align}
          \xPs
          &= \fP(\xPs,u_s(\ssp,d_s),\wP) \\
          y_s(\ssp,d_s) &= \hP(\xPs,u_s(\ssp,d_s),\wP)
        \end{align}
      \end{subequations}
      where \(y_s(\ssp,d_s) \defas h(x_s(\ssp,d_s),u_s(\ssp,d_s),d_s)\);
    \item \((\ssp,d_s(\ssp,\wP))\in\calB_c\) for all \((\ssp,\wP)\in\calA_c\); and
    \item \((\ssp,0)\in\calA_c\) for all \((\ssp,\wP)\in\calA_c\).
    \end{enumerate}
  % }{%
  %   (a) \Cref{assum:sstp} holds; (b) \(z_s\) is Lipschitz continuous on
  %   \(\hat{\calB}_c\); (c) for each \(\alpha=(\ssp,\wP)\in\calA_c\), the pair
  %   \((\xPs,d_s)=(\xPs(\alpha),d_s(\alpha))\) is the unique solution to
  %   \begin{subequations}\label{eq:ssop}
  %     \begin{align}
  %       \xPs
  %       &= \fP(\xPs,u_s(\ssp,d_s),\wP) \\
  %       y_s(\ssp,d_s) &= \hP(\xPs,u_s(\ssp,d_s),\wP)
  %     \end{align}
  %   \end{subequations}
  %   where \(y_s(\ssp,d_s) \defas h(x_s(\ssp,d_s),u_s(\ssp,d_s),d_s)\); (d)
  %   \((\ssp,d_s(\ssp,\wP))\in\calB_c\) for all \((\ssp,\wP)\in\calA_c\); and (e)
  %   \((\ssp,0)\in\calA_c\) for all \((\ssp,\wP)\in\calA_c\). %
  % }%
\end{assumption}

Intuitively, \Cref{assum:sstp:mismatch} guarantees, for each
\(\alpha\in\calA_c\), there is unique point at which both systems achieve steady
state and output matching, and the point is robust to perturbations in
\(\alpha\). Given \Cref{assum:sstp:mismatch}, we let
\begin{align*}
  \calA_c(\delta_w)
  &\defas \set{ (\ssp,\wP) \in \calA_c | |\wP|\leq\delta_w } \\
  \bbA_c(\alpha,\delta_w)
  &\defas \set{ \Delta\alpha \in \real^{n_\alpha} |
    \alpha + \Delta\alpha \in \calA_c(\delta_w) }.
\end{align*}
Then \(\calA_c(\delta_w)\) is RPI for the system \(\alpha^+ = \alpha +
\Delta\alpha, \Delta\alpha\in\bbA(\alpha,\delta_w)\), and if
\(\|\bfe_d\|\leq\delta_0\), then \(\hat{\beta} = (\ssp,d_s(\alpha)-e_d) \in
\hat{\calB}_c\) and the SSTP~\cref{eq:sstp} is feasible at all times.

\ifthenelse{\boolean{LongVersion}}{%
\subsection{Correcting the model state under mismatch}
}{}%
We can define the ``corrected'' model state as \(x\defas\xP-\Delta x_s(\alpha)\)
where \(\Delta x_s\defas \xPs(\alpha) - x_s(\ssp,d_s(\alpha))\) and
\(\alpha=(\ssp,\wP)\). %
In terms of the corrected model state \(x\) and parameters \(\alpha\), the
closed-loop plant is
\begin{subequations}\label{eq:model:mismatch:cl}
  \begin{align}
    x^+ &= \fP(x + \Delta x_s(\alpha), \kappa_N(\hat{x},\hat{\beta}), \wP) -
          \Delta x_s(\alpha^+) \\
    \alpha^+ &= \alpha + \Delta\alpha \\
    y &= \hP(x + \Delta x_s(\alpha), \kappa_N(\hat{x},\hat{\beta}), \wP).
        \label{eq:model:mismatch:cl:c}
  \end{align}
\end{subequations}
To analyze the estimator, we consider the noisy model~\cref{eq:model:noise} with
the following noises:
\begin{subequations}\label{eq:noise}
  \begin{align}
    w &\defas \fP(x + \Delta x_s(\alpha),u,\wP) - f(x,u,d_s(\alpha)) -
        \Delta x_s(\alpha^+) \\
    w_d &\defas d_s(\alpha^+) - d_s(\alpha) \\
    v &\defas \hP(x + \Delta x_s(\alpha),u,\wP) - h(x,u,d_s(\alpha)).
  \end{align}
\end{subequations}
Clearly \(\tilde{w} \defas (w,w_d,v) \in \bbW(x,u,d)\) by construction, so under
\Cref{assum:est}, the estimator~\cref{eq:est} produces RGES estimates of the
corrected model state \(x\) and disturbance \(d_s(\alpha)\).
However, the noise \(\tilde{w}\) is still a function of the corrected model
state \(x\), input \(u\), and steady-state parameters \(\alpha\). In the proof
of the following result, we take the approach of \cite{kuntz:rawlings:2024d} and
use a differentiability assumption to relate the magnitude of \(\tilde{w}\) to
more convenient quantities: the tracking error \(z-z_s(\beta)\), plant
disturbance \(\wP\), and parameter changes \(\Delta\alpha\).
% NOTE See Assumption 2 (Constraints) for links on eliminating the
% \((\bbX,\bbU)\) compact assumption.
\begin{assumption}[Differentiability]\label{assum:diff}
  % The derivatives \(\partial_{(x,u)} \fP\) and \(\partial_{(x,u)} \hP\) exist
  % and are continuous on \(\bbX\times\bbU\times\bbW\). The functions \(f,h\) and
  % \(g\) are continuously differentiable on \(\bbX\times\bbU\times\bbD\) and
  % \(\bbU\times\bbY\).
  The derivatives \(\partial_{(u,y)} g\), \(\partial_{(x,u,d)} (f,h)\), and
  \(\partial_{(x,u)} (\fP,\hP)\) exist and are continuous on \(\bbU\times\bbY\),
  \(\bbX\times\bbU\times\bbD\), and \(\bbX\times\bbU\times\bbW\), respectively.
\end{assumption}

\ifthenelse{\boolean{LongVersion}}{%
\subsection{Main result}
}{}%
Finally, \Cref{thm:mpc:mismatch} establishes the main result of this work:
robust stability of offset-free MPC, despite plant-model mismatch
(see~\Cref{app:mpc:mismatch} for proof).
\begin{theorem}[Offset-free stability]\label{thm:mpc:mismatch} If
  \Cref{assum:cons,assum:cont,assum:sstp:exist,assum:stabilizability,%
    assum:quad,assum:est,assum:sstp,assum:sstp:mismatch,assum:diff} hold and
  \(\rho>0\), then there exists \(\tau,\delta_w,\delta_\alpha>0\) such that
  \begin{enumerate}[(a)]
  \item the following set is RPI for the closed-loop
    system~\cref{eq:est,eq:model:mismatch:cl} with disturbance
    \(\Delta\alpha\in\bbA_c(\alpha,\delta_w)\cap\delta_\alpha\bbB^{n_\alpha}\):
    \ifthenelse{\boolean{OneColumn}}{%
    % \[
    %   \calS_N^{\rho,\tau} \defas \set{ (x,\alpha,\hat{x},\hat{\beta}) \in
    %     \bbX\times\calA_c\times\hat{\calS}_N^\rho |
    %     V_e(x,d_s(\alpha),\hat{x},\hat{d}) \leq \tau, \, \alpha = (\ssp,\wP), \,
    %     \hat{\beta} = (\ssp,\hat{d}) };
    %   \]
      \begin{multline*}
        \calS_N^{\rho,\tau} \defas \{\, (x,\alpha,\hat{x},\hat{\beta}) \in
        \bbX\times\calA_c\times\hat{\calS}_N^\rho \,|\, \alpha = (\ssp,\wP), \,
        \hat{\beta} = (\ssp,\hat{d}),  \\
        V_e(x,d_s(\alpha),\hat{x},\hat{d}) \leq \tau \,\};
      \end{multline*}
  }{%
    %% NOTE For some reason multline is producing a lot of space above the
    %% equation, so I simulated the behavior with an align environment. If the
    %% equation is changed, just be careful to adjust the alignment.
    % \begin{multline*}
    %   \calS_N^{\rho,\tau} \defas \{\, (x,\alpha,\hat{x},\hat{\beta}) \in
    %   \bbX\times\calA_c\times\hat{\calS}_N^\rho \,|\, \alpha = (\ssp,\wP), \\
    %   \hat{\beta} = (\ssp,\hat{d}),  \,
    %   V_e(x,d_s(\alpha),\hat{x},\hat{d}) \leq \tau \,\};
    % \end{multline*}
    \begin{align*}
      \calS_N^{\rho,\tau} \defas \{\, (x,\alpha
      &,\hat{x},\hat{\beta}) \in \bbX\times\calA_c\times\hat{\calS}_N^\rho \,|\,
        \alpha = (\ssp,\wP),\\
      &\hat{\beta} = (\ssp,\hat{d}),\, V_e(x,d_s(\alpha),\hat{x},\hat{d}) \leq
        \tau \,\};
    \end{align*}
  }%
\item the closed-loop system~\cref{eq:est,eq:model:mismatch:cl} with disturbance
  \(\Delta\alpha\in\bbA_c(\alpha,\delta_w)\cap\delta_\alpha\bbB^{n_\alpha}\) is
  RES on \(\calS_N^{\rho,\tau}\) w.r.t. the target-tracking error
  \(\delta\hat{x} \defas \hat{x}-x_s(\hat{\beta})\); and
  \item the closed-loop system~\cref{eq:est,eq:model:mismatch:cl} with
    disturbance \(\Delta\alpha \in \bbA_c(\alpha,\delta_w) \cap
    \delta_\alpha\bbB^{n_\alpha}\) is RES on \(\calS_N^{\rho,\tau}\) w.r.t.
    \((\delta r,\delta\hat{x})\), where \(\delta r\defas r-\rsp\) is the
    setpoint-tracking error, \(\alpha=(\rsp,\zsp,\wP)\), \(r =
    g(\kappa_N(\hat{x},\hat{\beta}), y)\), and \cref{eq:model:mismatch:cl:c}.
  \end{enumerate}
\end{theorem}

\begin{remark}\label{rem:mpc:mismatch}
  %% TODO minor adjustments to fix diff spacing. Undo this.
  % \Cref{thm:mpc:mismatch} implies the following error convergence result: % reg
  \Cref{thm:mpc:mismatch} implies the error convergence result: % diff
  \((\delta\hat{x}(k),\delta r(k),\varepsilon(k)) \rightarrow 0\) so long as
  \(\Delta\alpha(k) \rightarrow 0\), \((x(0), \alpha(0), \hat{x}(0),
  \hat{\beta}(0)) \in \calS_N^{\rho,\tau}\), %
  % \(\Delta\alpha(k) \in \bbA_c(\alpha,\delta_w),k\in\nnegint\), % reg
  \(\Delta\alpha(k) \in \bbA_c(\alpha,\delta_w)\), \(k\in\nnegint\), % diff
  and \(\|\boldsymbol{\Delta\alpha}\|\leq\delta_\alpha\)
  (cf.~\Cref{rem:ras:ctrl:est}).
\end{remark}

\begin{remark}\label{rem:mpc:mismatch:tradeoffs}
  As with~\cref{rem:mpc:robust:tradeoffs}, increasing \(\rho\) decreases the
  other constants \(\tau,\delta_w,\delta_\alpha\). With \(\rho\) fixed,
  increasing one of the error allowance \(\tau\), mismatch allowance
  \(\delta_w\), or parameter drift allowance \(\delta_\alpha\) necessarily
  decreases the other two. These trade-offs are fairly intuitive. For example,
  as we allow greater estimate errors (\(\tau\) increases) the tolerance for
  mismatch and drift is reduced (\(\delta_w,\delta_\alpha\) decrease).
\end{remark}

\section{Linear systems connections}\label{sec:linear}
Consider the linearization of~\cref{eq:model:noise,eq:ref} about the origin,
\begin{subequations}\label{eq:lin}
  \begin{align}
    x^+ &= Ax + Bu + B_dd + w \label{eq:lin:a} \\
    d^+ &= d + w_d \label{eq:lin:b} \\
    y &= Cx + Du + C_dd + v \label{eq:lin:c} \\
    r &= H_uu + H_yy \label{eq:lin:d}
  \end{align}
\end{subequations}
where %
\ifthenelse{\boolean{LongVersion}}{%
  \begin{align*}
    A &\defas \partial_x f(0,0,0), & B &\defas \partial_u f(0,0,0),
    & B_d &\defas \partial_d f(0,0,0), \\
    C &\defas \partial_x h(0,0,0), & D &\defas \partial_u h(0,0,0),
    & C_d &\defas \partial_d h(0,0,0), \\
    H_u &\defas \partial_u g(0,0), & H_y &\defas \partial_y g(0,0).
  \end{align*}
}{%
  \(A \defas \partial_x f(0,0,0)\), \(B \defas \partial_u f(0,0,0)\), \(C\defas
  \partial_x h(0,0,0)\), \(D \defas \partial_u h(0,0,0)\), \(B_d\defas
  \partial_d f(0,0,0)\), \(C_d\defas \partial_d h(0,0,0)\), \(H_u \defas
  \partial_u g(0,0)\), and \(H_y \defas \partial_y g(0,0)\). %
}%
In \Cref{lem:sstp,lem:sstp:mismatch}, we provide sufficient % reg
conditions under which the SSTP assumptions
(\Cref{assum:sstp,assum:sstp:mismatch}, respectively) are guaranteed to hold
(see \Cref{app:sstp,app:sstp:mismatch} for proofs).

\begin{lemma}\label{lem:sstp}
  %% NOTE This ifthenelse statement just cancels out any latexdiff markup in
  %% this section. Otherwise, the spacing gets wonky and text spills onto the
  %% next page.
  % \ifthenelse{\boolean{true}}{%
    Suppose~\Cref{assum:cont,assum:cons,assum:sstp:exist} hold and let
    \begin{equation}\label{eq:linearization:1}
      M_1 \defas \begin{bmatrix} A-I & B \\ H_yC & H_yD+H_u \end{bmatrix}.
    \end{equation}
  % }{}%
  If %
  \ifthenelse{\boolean{LongVersion}}{%
    \begin{enumerate}[(a)]
      %% Continuous differentiability (for regularity)
    \item \(f,g,h,\overline{c}\) are continuously differentiable;
      %% Regularity
    \item \(M_1\) is full row rank;
      %% Locally inactive constraints
    \item \(\bbX,\bbU,\bbY,\bbD\) contain neighborhoods of the origin; %
    \item there exist continuously differentiable functions %
      \(c_x,c_u,c_y\) for which
      \begin{align*}
        \bbX &= \set{ x\in\real^n | c_x(x) \leq 0 }, \\
        \bbU &= \set{ u\in\real^{n_u} | c_u(u) \leq 0 }, \\
        \bbY &= \set{ y\in\real^{n_y} | c_y(y) \leq 0 };
      \end{align*}
      % \(\bbX = \set{ x | c_x(x) \leq 0 }\), \(\bbU =
      % \set{ u | c_u(u) \leq 0 }\), and \(\bbY = \set{ y | c_y(y) \leq 0 }\);
      %% Local uniqueness
    \item \(h(x,0,0) \neq 0\) for all \((x,0) \in \calZ_O(0) \setminus
      \set{(0,0)}\); and
    \item \(\ell_s\) is positive definite, i.e., \(\ell_s(\tilde u,\tilde y)>0\)
      for all \((\tilde{u}, \tilde{y}) \in \real^{n_u+n_y} \setminus
      \set{(0,0)}\);
    \end{enumerate}
  }{%
    %% Continuous differentiability (for regularity)
    (a) \(f,g,h,\overline{c}\) are continuously differentiable;
    %% Regularity
    (b) \(M_1\) is full row rank;
    %% Locally inactive constraints
    (c) \(\bbX,\bbU,\bbY,\bbD\) contain neighborhoods of the origin; %
    (d) there exist continuously differentiable functions %
    \(c_x,c_u,c_y\) for which \(\bbX = \set{ x | c_x(x) \leq 0 }\), \(\bbU = \set{
      u | c_u(u) \leq 0 }\), and \(\bbY = \set{ y | c_y(y) \leq 0 }\);
    %% Local uniqueness
    (e) \(h(x,0,0) \neq 0\) for all \((x,0)\in\calZ_O(0)\setminus\set{0}\); and
    % (f) \(\ell_s(\tilde u,\tilde y)>0\) for all \((\tilde{u}, \tilde{y}) \in
    % \real^{n_u+n_y}\);
    (f) \(\ell_s\) is positive definite;
  }%
  %% Continuity and uniqueness of solns
  then there exists a neighborhood of the origin \(\calB_c\subseteq\calB\),
  constant \(\delta_0>0\), and function \(z_s:\calB\rightarrow\bbX\times\bbU\)
  satisfying \Cref{assum:sstp}. Moreover, \(z_s(\hat\beta)\) uniquely
  solves~\cref{eq:sstp} for all \(\hat\beta\in\hat{\calB}_c\).
\end{lemma}

\begin{lemma}\label{lem:sstp:mismatch}
  %% NOTE This ifthenelse statement just cancels out any latexdiff markup in
  %% this section. Otherwise, the spacing gets wonky and text spills onto the
  %% next page.
  \ifthenelse{\boolean{true}}{%
    Suppose the conditions of \Cref{lem:sstp} hold and let
    \begin{equation}\label{eq:sstp:mismatch}
      M_2 \defas \begin{bmatrix} A - I & B_d \\ C & C_d \end{bmatrix}.
    \end{equation}
  }{}%
  If %
  \ifthenelse{\boolean{LongVersion}}{%
    \begin{enumerate}[(a)]
    \item \(f,g,h,\ell_s,\fP,\hP\) are twice continuously differentiable;
    \item \(M_2\) is invertible,
    \item \(\partial_{(u,y)} \ell_s(0,0) = 0\); and
    \item \(\partial_{(u,y)}^2\ell_s(0,0)\) is positive definite;
    \end{enumerate}
  }{%
    (a) \(f,g,h,\ell_s,\fP,\hP\) are twice continuously differentiable, (b)
    \(M_2\) is invertible, (c) \(\partial_{(u,y)} \ell_s(0,0) = 0\), and (d)
    \(\partial_{(u,y)}^2\ell_s(0,0)\) is positive definite, %
  }%
  %% NOTE This ifthenelse statement just cancels out any latexdiff markup in
  %% this section. Otherwise, the spacing gets wonky and text spills onto the
  %% next page.
  \ifthenelse{\boolean{true}}{%
    then there exist compact sets
    \(\calA_c\subseteq\real^{n_r}\times\overline{\bbZ}_y\times\bbW\) and
    \(\calB_c\subseteq\calB\) containing neighborhoods of the origin and functions
    \(z_s:\calB\rightarrow\bbX\times\bbU\) and
    \((\xPs,d_s):\calA_c\rightarrow\bbX\times\bbD\) satisfying all parts
    of~\Cref{assum:sstp:mismatch}. Moreover, \(z_s(\beta)\) and
    \((\xPs(\alpha),d_s(\alpha))\) are the unique solutions
    to~\cref{eq:sstp,eq:ssop} for all \(\alpha=(\ssp,\wP)\in\calA_c\), where
    \(\beta\defas(\ssp,d_s(\alpha))\). %
  }{}
\end{lemma}

\ifthenelse{\boolean{LongVersion}}{%
  To conclude this section, we connect rank conditions in
  \Cref{lem:sstp,lem:sstp:mismatch} to steady-state versions of the reachability
  and observability of parts of the linearized system~\cref{eq:lin}. %
}{}

\begin{remark}\label{rem:sstp}
  The rank condition \Cref{lem:sstp}(b) can be interpreted as the following %
  %% NOTE This ifthenelse statement just cancels out any latexdiff markup in
  %% this section. Otherwise, the spacing gets wonky and text spills onto the
  %% next page.
  \ifthenelse{\boolean{true}}{%
    \emph{steady-state reachability} condition: each disturbance \(d\), each
    reference \(r\) can be reached by some \(u\) at steady-state. %
  }{}%
  A similar reachability assumption is also used in
  \ifthenelse{\boolean{LongVersion}}{%
    Assumption~1 and Remark~1 of \cite{limon:ferramosca:alvarado:alamo:2018}, %
  }{%
    \cite[Assm.~1,~Remark~1]{limon:ferramosca:alvarado:alamo:2018}, %
  }%
  but it is enforced on the entire domain \(\bbX\times\bbU\), and the functions
  \((x_s,u_s)\) are simply assumed to exist, rather than produced by the
  SSTP~\cref{eq:sstp}.
\end{remark}

\begin{remark}\label{rem:sstp:mismatch}
  Invertibility of \(M_2\) is a key assumption in linear offset-free
  MPC~\citep{muske:badgwell:2002,pannocchia:rawlings:2003}. In fact, it is known
  that the system~\cref{eq:lin:a,eq:lin:b,eq:lin:c} is detectable if and only if
  \(M_2\) is full column rank and \((A,C)\) is
  detectable~\cite[Lem.~1]{pannocchia:rawlings:2003}. Moreover, \(M_2\) full row
  rank can be interpreted as a \emph{steady-state observability} condition: at
  steady-state, the disturbance \(d\) can be uniquely recovered from the input
  \(u\) and output \(y\). On the other hand, \(M_2\) full row rank can be
  interpreted as the following \emph{steady-state reachability} condition: for
  each the input \(u\) and output \(y\), a disturbance \(d\) exists that
  achieves the output \(y\) at steady state. In other words, invertibility of
  \(M_2\) guarantees the existence and uniqueness of a disturbance providing
  steady-state output matching with the plant.
\end{remark}

\ifthenelse{\boolean{LongVersion}}{%
\section{Examples}\label{sec:example}
In this section, we illustrate the main results using the example systems
depicted in \Cref{fig:sys}. We compare two MPCs in our experiments.

\begin{figure}[t]
  \centering
  \ifthenelse{\boolean{OneColumn}}{%
    \subfloat[Simple pendulum\label{fig:pendulum}]{
      \centering
      \includegraphics[width=0.45\textwidth]{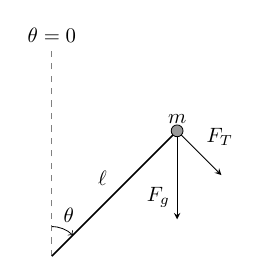}
    }\subfloat[Continuous stirred-tank reactor\label{fig:cstr}]{
      \centering
      \def\svgwidth{0.45\textwidth}
      {\input{cstr.tex}}
    }
  }{%
    \subfloat[Simple pendulum\label{fig:pendulum}]{
      \centering
      \includegraphics[width=0.225\textwidth]{pendulum}
    }\subfloat[Continuous stirred-tank reactor\label{fig:cstr}]{
      \centering
      \def\svgwidth{0.225\textwidth}
      {\tiny\input{cstr.tex}}
    }
  }%
  \caption{Example systems.}\label{fig:sys}
\end{figure}
}{%
\section{Example}\label{sec:example}
In this section, we illustrate the main results by comparing two MPCs. %
}%
First, the offset-free MPC (OFMPC) uses \cref{eq:sstp}, \cref{eq:mpc}, and the
following state-disturbance MHE:
\begin{equation}\label{eq:mhe:aug}
  \min_{(\bfx,\bfd)\in\bbX^{T_k+1}\times\bbD^{T_k+1}}
  V_T^{\textrm{MHE}}(k;\bfx,\bfd,\bfu,\bfy)
\end{equation}
where \(T_k\defas\min\set{k,T}\), \(T\in\posint\),
\ifthenelse{\boolean{LongVersion}}{%
  \(w \defas x^+ - f(x,u,d)\), \(w_d \defas d^+ - d\), \(v \defas y -
  h(x,u,d)\), and
  \[
    V_T^{\textrm{MHE}}(k;\bfx,\bfd,\bfu,\bfy) \defas \sum_{j=0}^{T_k-1}
    |w(j)|_{Q_w^{-1}}^2 + |w_d(j)|_{Q_d^{-1}}^2 + |v(j)|_{R_v^{-1}}^2.
  \]
}{%
  \(V_T^{\textrm{MHE}}(k;\bfx,\bfd,\bfu,\bfy) \defas \sum_{j=0}^{T_k-1}
  |w(j)|_{Q_w^{-1}}^2 + |w_d(j)|_{Q_d^{-1}}^2 + |v(j)|_{R_v^{-1}}^2\), \(w
  \defas x^+ - f(x,u,d)\), \(w_d \defas d^+ - d\), and \(v \defas y -
  h(x,u,d)\). }%
\ifthenelse{\boolean{LongVersion}}{%
  For simplicity, a prior term is not used. %
}{}%
Let \(\hat{x}(j;\bfu,\bfy)\) and \(\hat{d}(j;\bfu,\bfy)\) denote solutions to
the above problem, and define the estimates by %
\ifthenelse{\boolean{OneColumn}}{%
  \begin{align*}
    \hat{x}(k)
    &\defas \hat{x}(k;\bfu_{k-T_k:k-1},\bfy_{k-T_k:k-1}),
    & \hat{d}(k) &\defas \hat{d}(k;\bfu_{k-T_k:k-1},\bfy_{k-T_k:k-1}).
  \end{align*}
}{%
  \(\hat{x}(k) \defas \hat{x}(k;\bfu_{k-T_k:k-1},\bfy_{k-T_k:k-1})\) and
  \(\hat{d}(k) \defas \hat{d}(k;\bfu_{k-T_k:k-1},\bfy_{k-T_k:k-1})\). %
}%
Second, the nominal tracking MPC (TMPC) uses \cref{eq:sstp}, \cref{eq:mpc}, and
a state-only MHE,
\begin{equation}\label{eq:mhe}
  \min_{\bfx\in\bbX^{T_k+1}}
  V_T^{\textrm{MHE}}(k;\bfx,0,\bfu,\bfy).
\end{equation}
With solutions denoted by \(\hat{x}(j;\bfu,\bfy)\), we define the estimates
by %
\ifthenelse{\boolean{OneColumn}}{%
  \begin{align*}
    \hat{x}(k)
    &\defas \hat{x}(k;\bfu_{k-T_k:k-1},\bfy_{k-T_k:k-1}),
    & \hat{d}(k) &\defas 0.
  \end{align*}
}{%
  \(\hat{x}(k) \defas \hat{x}(k;\bfu_{k-T_k:k-1},\bfy_{k-T_k:k-1})\) and
  \(\hat{d}(k) \defas 0\).
}%

\ifthenelse{\boolean{LongVersion}}{%
\subsection{Simple pendulum}
Consider the following nondimensionalized pendulum system (\Cref{fig:pendulum}):
\begin{subequations}\label{eq:pendulum}
  \begin{align}
    \dot x
    &= \FP(x,u,\wP) \defas
      \ifthenelse{\boolean{OneColumn}}{%
      \begin{bmatrix} x_2 \\ \sin x_1 - (\wP)_1^2x_2 +
        (\hat k+(\wP)_2)u + (\wP)_3 \end{bmatrix}
      }{%
      \begin{bsmallmatrix} x_2 \\ \sin x_1 - (\wP)_1^2x_2 +
        (\hat k+(\wP)_2)u + (\wP)_3 \end{bsmallmatrix}
      } \\
    y &= \hP(x,u,\wP) \defas x_1 + (\wP)_4 \\
    r &= g(u,y) \defas y
  \end{align}
\end{subequations}
where \((x_1,x_2)\in \bbX \defas \real^2\) are the angle and angular velocity,
\(u\in\bbU \defas [-1,1]\) is the (dimensionless) motor voltage, \(\hat
k=5\textnormal{ rad} / \textnormal{s}^2\) is the estimated motor gain,
\((\wP)_1\) is an air resistance factor, \((\wP)_2\) is the error in the motor
gain estimate, \((\wP)_3\) is an externally applied torque, and \((\wP)_4\) is the
measurement noise. Let \(\psi(t;x,u,\wP)\) denote the solution
to~\cref{eq:pendulum} at time \(t\) given \(x(0)=x\), \(u(t)=u\), and
\(\wP(t)=\wP\). We model the discretization of~\cref{eq:pendulum} by
\begin{subequations}\label{eq:plant:pendulum}
  \begin{equation}
    x^+ = \fP(x,u,\wP) \defas x + \Delta \FP(x,u,\wP) + (\wP)_5
    r_d(x,u,\wP)
  \end{equation}
  where \((\wP)_5\) scales the discretization error, \(r_d\) is a residual
  function given by
  \begin{equation}
    r_d(x,u,\wP) \defas \int_0^\Delta [\FP(x(t),u,\wP) - \FP(x,u,\wP)]dt
  \end{equation}
\end{subequations}
and \(x(t) = \psi(t;x,u,\wP)\). Assuming a zero-order hold on the input \(u\)
and disturbance \(\wP\), the system \cref{eq:pendulum} is discretized (exactly)
as \cref{eq:plant:pendulum} with \((\wP)_5 \equiv 1\). We model the system with
\(\wP = w(d) \defas (0,0,d,0,0)\), i.e.,
\begin{subequations}\label{eq:model:pendulum}
  \begin{align}
    x^+
    &= f(x,u,d) \defas
      \ifthenelse{\boolean{LongVersion}}{%
      \fP(x,u,w(d)) = %
      }{}%
      x + \Delta
      \ifthenelse{\boolean{OneColumn}}{%
      \begin{bmatrix} x_2 \\ \sin x_1 + \hat ku + d \end{bmatrix}
      }{%
      \begin{bsmallmatrix} x_2 \\ \sin x_1 + \hat ku + d \end{bsmallmatrix}
      } \\
    y &= h(x,u,d) \defas %
        \ifthenelse{\boolean{LongVersion}}{%
        \hP(x,u,w(d)) = %
        }{}%
        x_1
  \end{align}
\end{subequations}
and therefore we do not need access to the residual function \(r_d\) to design
the offset-free MPC.\@

For the following simulations, %
\ifthenelse{\boolean{LongVersion}}{%
  assume \(\wP \in \bbW \defas [-3,3]^3\times[-0.05,0.05]\times\set{0,1}\),
  and %
}{}%
let the sample time be \(\Delta = 0.1 \textnormal{ s}\). %
\ifthenelse{\boolean{LongVersion}}{%
  Regardless of objective \(\ell_s\), the SSTP~\cref{eq:sstp} is uniquely solved
  by
  \begin{align*}
    x_s(\beta) &\defas \begin{bmatrix} \rsp \\ 0 \end{bmatrix},
    &u_s(\beta) &\defas -\frac{1}{\hat k}(\sin\rsp + d)
  \end{align*}
  for each \(\beta = (\rsp,\usp,\ysp,d) \in \calB_c\), where
  \[
    \calB_c \defas \set{ (r,u,y,d)\in\real^4 | |\sin r+d|,|\sin y+d|\leq
      \hat{k}, |u|\leq 1}
  \]
  and \(\delta_0>0\). Likewise, the solution to \cref{eq:ssop} is
  \begin{align}
    \xPs(\alpha) &\defas \begin{bmatrix} \rsp \\ 0 \end{bmatrix}
    &d_s(\alpha)
    &\defas (\wP)_3 - \frac{(\wP)_2}{\hat{k}+(\wP)_2}(\sin\rsp + (\wP)_3)
  \end{align}
  for each \(\alpha = (\rsp,\usp,\ysp,\wP) \in \calA_c\), where
  \[
    \calA_c \defas \set{ (r,u,y,w) \in \real^3 \times \bbW | |\sin r+(\wP)_3|,
      |\sin y+(\wP)_3|\leq \hat{k} + (\wP)_2, |u|\leq 1}.
  \]
  Notice that \(\calA_c\) and \(\calB_c\) are compact and satisfy
  \Cref{assum:sstp:mismatch}. %
}{%
  We define a SSTP with \(\ell(u,y)=u^2+y^2\) and no output constraints. %
}%
We define a regulator with \(N\defas 20\), \(\mathbb{U}\defas[-1,1]\),
\(\ell_s(u,y) = |u|^2 + |y|^2\), \(\ell(x,u,\Delta u,\beta)\defas
|x-x_s(\beta)|^2 + 10^{-2}(u-u_s(\beta))^2 + 10^2(\Delta u)^2\),\footnote{The
  \(\Delta u(k)\defas u(k) - u(k-1)\) penalty is a standard generalization used
  by practitioners to ``smooth'' the closed-loop response in a tuneable
  fashion.} \(V_f(x,\beta)\defas |x-x_s(\beta)|_{P_f(\beta)}^2\), and
\(\mathbb{X}_f\defas\lev_{c_f} V_f\), where \(P_f(\beta)\) and \(c_f\approx
0.4364\) are constructed according to \Cref{app:terminal} to satisfy
\Cref{assum:stabilizability,assum:quad}. \Cref{assum:cons} is clearly satisfied, 
and \Cref{assum:cont,assum:sstp:mismatch,assum:diff} are satisfied
since smoothness of \(F\) implies that \(\psi\), \(r\), and \(f\) are
smooth~\cite[Thm.~3.3]{hale:1980}. Finally, we use MHE
designs~\cref{eq:mhe:aug,eq:mhe} for the offset-free MPC and tracking MPC,
respectively, where \(T=5\), \(Q_w\defas
\begin{bsmallmatrix} 10^{-3} \\ & 10^{-6} \end{bsmallmatrix}\), and \(Q_d\defas
R_v\defas 1\). \ifthenelse{\boolean{LongVersion}}{%
  While the estimators defined by \cref{eq:mhe:aug,eq:mhe} should be
  RGES~\citep{allan:rawlings:2021a}, it is not known if they satisfy
  \Cref{assum:est}. %
}{}%
If \Cref{assum:est} is satisfied, then \Cref{thm:mpc:mismatch} gives robust
stability with respect to the tracking errors.

\begin{figure*} %
  \centering %
  \ifthenelse{\boolean{LongVersion}}{%
    \subfloat[No mismatch: \((\wP)_1 \equiv 0\) and \((\wP)_2 \equiv
    0\).\label{fig:pendulum:traj:a}]{%
      \includegraphics[width=0.49\linewidth,page=1]{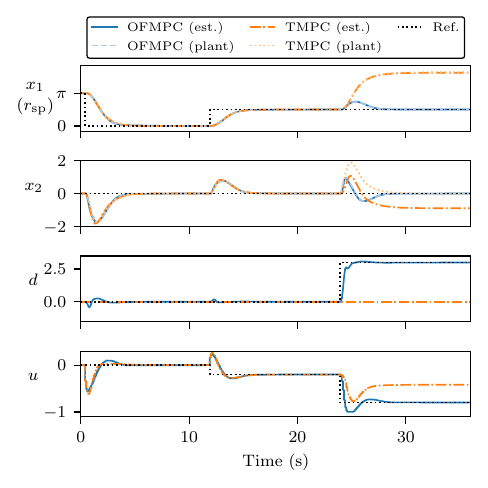}%
    }\hfill %
    \subfloat[Mismatch: \((\wP)_1 \equiv 1\) and \((\wP)_2 \equiv
    2\).\label{fig:pendulum:traj:b}]{%
      \includegraphics[width=0.49\linewidth,page=2]{pendulum_plot.pdf}%
    }

    \subfloat[Noise and mismatch: \((\wP)_3^+ = (\wP)_3 + (\Delta\wP)_3\),
    \((\Delta\wP)_3 \sim \norm(0,10^{-2})\), and \((\wP)_4 \sim
    \norm(0,10^{-4})\).\label{fig:pendulum:traj:c}]{%
      \includegraphics[width=0.49\linewidth,page=3]{pendulum_plot.pdf}%
    }\hfill %
    \subfloat[Oscillating disturbance and mismatch: \((\wP)_3(k) =
    1-\cos(\frac{2\pi k}{50})\) and
    \(\rsp(k)\equiv\pi\).\label{fig:pendulum:traj:d}]{%
      \includegraphics[width=0.49\linewidth,page=4]{pendulum_plot.pdf}%
    } %
  }{%
    \subfloat[No mismatch.\label{fig:pendulum:traj:a}]{%
      \includegraphics[width=0.33\linewidth,page=1]{pendulum_plot.pdf}%
    }\hfill %
    \subfloat[Mismatch: \((\wP)_1 \equiv 1\) and \((\wP)_2 \equiv
    2\).\label{fig:pendulum:traj:b}]{%
      \includegraphics[width=0.33\linewidth,page=2]{pendulum_plot.pdf}%
    }\hfill %
    % \subfloat[Noise: \((\wP)_3^+ = (\wP)_3 + (\Delta\wP)_3\), \((\Delta\wP)_3
    % \sim \norm(0,10^{-2})\), and \((\wP)_5 \sim
    % \norm(0,10^{-4})\).\label{fig:pendulum:traj:c}]{%
    %   \includegraphics[width=0.33\linewidth,page=3]{pendulum_plot.pdf}%
    % } %
    \subfloat[Oscillating disturbance and mismatch: \((\wP)_3(k) =
    1-\cos(\frac{2\pi k}{50})\) and \(\rsp(k)\equiv\pi\).\label{fig:pendulum:traj:d}]{%
      \includegraphics[width=0.33\linewidth,page=4]{pendulum_plot.pdf}%
    } %
  }%
  \caption[Closed-loop trajectories for the offset-free MPC and tracking MPC of
  the simple pendulum.]{Simulated closed-loop trajectories for the offset-free
    MPC and tracking MPC of \cref{eq:pendulum}. Solid blue and dot-dashed orange
    lines represent the closed-loop estimates and inputs \((\hat{x},\hat{d},u)\)
    for the offset-free MPC and tracking MPC simulations, respectively. Dashed
    blue and dotted orange lines represent the closed-loop plant states \(\xP\)
    for the offset-free MPC and tracking MPC simulations, respectively. Dotted
    black lines represent the intended steady-state targets and disturbance
    values \((\xPs, d_s, u_s)\) found by solving~\cref{eq:sstp,eq:ssop}. We set
    \((\wP)_1\equiv 1\), \((\wP)_2\equiv 2\), \((\wP)_3(k)=3H(k-240)\),
    \((\wP)_4\equiv 1\), \((\wP)_5\equiv 0\), and \(\rsp(k) = \pi H(5-k) +
    \frac{\pi}{2}H(k-120)\), unless otherwise specified.}%
  \label{fig:pendulum:traj}
\end{figure*}

We present the results of numerical experiments in \Cref{fig:pendulum:traj}. To
ensure numerical accuracy, the plant~\cref{eq:pendulum} is simulated by four
4th-order Runga-Kutta steps per sample time. Unless otherwise specified, we
consider, in each simulation, unmodeled air resistance \((\wP)_1\equiv 1\),
motor gain error \((\wP)_2\equiv 2\), an exogenous torque
\((\wP)_3(k)=3H(k-240)\), the discretization parameter \((\wP)_4\equiv 1\), no
measurement noise \((\wP)_5\equiv 0\), and a reference signal \(\rsp(k) = \pi
H(5-k) + \frac{\pi}{2}H(k-120)\), where \(H\) denotes the unit step function.
The setpoint brings the pendulum from the resting state \(x_1=\pi\), to the
upright position \(x_1=0\), to the half-way position \(x_1=\frac{\pi}{2}\).

In the first experiment, we consider the case without plant-model mismatch,
i.e., \((\wP)_1\equiv 0\) and \((\wP)_2\equiv 0\) (\Cref{fig:pendulum:traj:a}).
Both offset-free and tracking MPC remove offset after the setpoint changes.
However, only offset-free MPC removes offset after the disturbance is injected.
Without a disturbance model, the tracking MPC cannot produce useful steady-state
targets, and the pendulum drifts far from the setpoint. Moreover, the tracking
MPC produces pathological state estimates, with nonzero velocity at steady
state.

The second experiment considers plant-model mismatch \((\wP)_1 \equiv 1\) and
\((\wP)_2 \equiv 2\) (\Cref{fig:pendulum:traj:b}). As in the first experiment,
both the tracking MPC and offset-free MPC bring the pendulum to the upright
position \(x_1=0\), without offset. However, only the offset-free MPC brings the
pendulum to the half-way position \(x_1=\frac{\pi}{2}\). The tracking MPC, not
accounting for motor gain errors, provides an insufficient force and does not
remove offset. Note the intended disturbance estimate \(d_s=\frac{13}{7}\) is a
smaller value that the actual injected disturbance \((\wP)_3=3\), as
underestimation of the motor gain necessitates a smaller disturbance value to be
corrected. Again, the tracking MPC produces pathological state estimates.

\ifthenelse{\boolean{LongVersion}}{%
  The third experiment follows the second, except the exogenous torque is an
  integrating disturbance \((\wP)_3^+ = (\wP)_3 + (\Delta\wP)_3\) where \((\wP)_3
  \sim \norm(0,10^{-2})\), and we have measurement noise \((\wP)_5 \sim
  \norm(0,10^{-4})\) (\Cref{fig:pendulum:traj:c}). In this experiment, we see the
  remarkable ability of offset-free MPC to track a reference subject to random
  disturbances. While the tracking MPC is robust to the disturbance \((\wP)_3\),
  it is not robust to the disturbance changes \((\Delta\wP)_3\) and wanders far
  from the setpoint as a result. On the other hand, offset-free MPC is robust to
  both and exhibits practically offset-free performance. %
}{}
\ifthenelse{\boolean{LongVersion}}{%
  We remark that, while the example is mechanical in nature, we are illustrating
  a behavior that is often desired in chemical process control, where process
  specifications must be met despite constantly, but slowly varying upstream
  conditions.
}{}

\ifthenelse{\boolean{LongVersion}}{%
  In the fourth and final experiment, %
}{%
  In the third experiment, %
}%
the pendulum maintains the resting position \(\rsp=\pi\) subject to an
oscillating torque \((\wP)_3(k) = 1 - \cos(\frac{2\pi k}{50})\)
(\Cref{fig:pendulum:traj:d}). Tracking MPC wanders away from the setpoint,
whereas offset-free MPC oscillates around it with small amplitude. We note the
disturbance estimate \(\hat{d}\) does not ever ``catch'' the intended value
\(d_s\) as the disturbance model has no ability to match its \emph{velocity} or
\emph{acceleration}. More general integrator schemes (e.g., double or triple
integrators) could provide more dynamic tracking performance at the cost of a
higher disturbance dimension (c.f.,~\cite{maeder:morari:2010}
\ifthenelse{\boolean{LongVersion}}{%
  or Chapter~5 of \cite{zagrobelny:2014}). %
}{%
  or~\cite[Ch.~5]{zagrobelny:2014}). %
}

\subsection{Continuous stirred-tank reactor}
  \begin{figure}
    \centering %
    \ifthenelse{\boolean{OneColumn}}{%
      \includegraphics[width=0.75\linewidth,page=1]{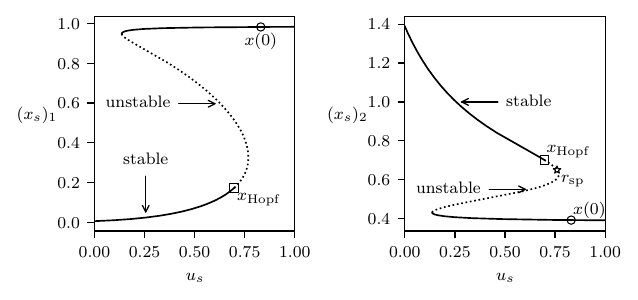}
    }{%
      \includegraphics[width=\linewidth,page=1]{cstr_ss.pdf}
    }%
    \caption{Nominal steady states for the CSTR~\cref{eq:cstr}.}
    \label{fig:cstr:ss}
  \end{figure}
}{}

We consider the following continuous stirred-tank reactor (CSTR) model, adapted
from \cite{falugi:2015}, %
\ifthenelse{\boolean{LongVersion}}{%
  Example~1.11 of \cite{rawlings:mayne:diehl:2020} (\Cref{fig:cstr}): %
}{%
  \cite[Ex.~1.11]{rawlings:mayne:diehl:2020}: %
}%
\begin{subequations}\label{eq:cstr}
  \begin{align}
    \dot{x}
    &= \FP(x,u,\wP) \nonumber \\
    &\defas
      \ifthenelse{\boolean{OneColumn}}{
      \begin{bmatrix} \theta^{-1}(1 + (\wP)_1 - x_1) - k\exp\left(
        \frac{(\wP)_2-M}{x_2} \right)x_1 \\
        \theta^{-1}(x_f - x_2) + k\exp\left( \frac{(\wP)_2-M}{x_2} \right)x_1 -
               \gamma u(x_2 - x_c - (\wP)_3) \end{bmatrix}
      }{
      \begin{bsmallmatrix} \theta^{-1}(1 + (\wP)_1 - x_1) - ke^{((\wP)_2-M)/x_2}x_1 \\
        \theta^{-1}(x_f - x_2) + ke^{(\wP)_2-M/x_2}x_1 -
        \gamma u(x_2 - x_c - (\wP)_3) \end{bsmallmatrix}
      } \\
    y &= \hP(x,u,\wP) \defas x_2 + (\wP)_4 \\
    r &= g(u,y) \defas y
  \end{align}
\end{subequations}
\ifthenelse{\boolean{LongVersion}}{%
  where \((x_1,x_2)\in\bbX\defas\nnegreal^2\) are the concentration and
  temperature, \(u\in\bbU\defas[0,2]\) is the coolant flowrate, \(\theta=20
  \textrm{ s}\) is the residence time, \(k=300 \textrm{ s}^{-1}\) is the rate
  coefficient, \(M=5\) is the dimensionless activation energy, \(x_f = 0.3947\)
  and \(x_c = 0.3816\) are dimensionless feed and coolant temperatures,
  \(\gamma=0.117 \textrm{ s}^{-1}\) is the heat transfer coefficient, \((\wP)_1\)
  is a kinetic modeling error, \((\wP)_2\) is a change to the coolant temperature,
  and \((\wP)_4\) is the measurement noise. %
  Again, we discretize the system \cref{eq:cstr} via the equations
  \cref{eq:plant:pendulum}, where the continuous system is recovered with
  \((\wP)_5=1\) and zero-order holds on \(u\) and \(\wP\). %
}{%
  where \((x_1,x_2)\in\bbX\defas\nnegreal^2\) are the concentration and
  temperature, and \(u\in\bbU\defas[0,2]\) is the coolant flowrate. The model
  parameters are \(\theta=20\), \(k=300\), \(M=5\), \(x_f = 0.3947\), \(x_c =
  0.3816\), and \(\gamma=0.117\). %
  We discretize~\cref{eq:cstr} as
  \begin{equation*}
    x^+ = \fP(x,u,\wP) \defas x + \Delta \FP(x,u,\wP) + (\wP)_5
    r_d(x,u,\wP)
  \end{equation*}
  assuming a zero-order hold on \(u\) and \(\wP\), where \(r_d\) is a residual
  function and \((\wP)_5\) scales the discretization error. %
}%
The system is modeled with \(\wP=w(d)\defas(0,d,0,0,0)\), i.e.,
\begin{subequations}\label{eq:cstr:model}
  \begin{align}
    x^+ &= f(x,u,d) \ifthenelse{\boolean{OneColumn}}{}{\nonumber \\
        &}\defas x + \Delta
          \ifthenelse{\boolean{OneColumn}}{
        \begin{bmatrix} \theta^{-1}(1-x_1) - k\exp\left( -M/x_2 \right) x_1 \\
          \theta^{-1}(x_f-x_2) + k\exp\left( -M/x_2 \right)x_1 -
            \gamma u(x_2 - x_c - d) \end{bmatrix}
    }{%
    \begin{bsmallmatrix} \theta^{-1}(1-x_1) - ke^{-M/x_2}x_1 \\
      \theta^{-1}(x_f-x_2) + ke^{-M/x_2}x_1 -
      \gamma u(x_2 - x_c - d) \end{bsmallmatrix}
    } \\
    y &= h(x,u,d) \defas x_2.
  \end{align}
\end{subequations}

The control objective is to steer the CSTR~\cref{eq:cstr} from a nominal steady
state %
\ifthenelse{\boolean{LongVersion}}{ %
  \[
    (x(0), u(-1)) \approx (0.9831, 0.3918, 0.8305)
  \]
}{%
  \((x(0), u(-1)) \approx (0.9831, 0.3918, 0.8305)\) %
}%
to a temperature setpoint \(\rsp\in[0.6, 0.7]\). In this range the nominal
steady states are unstable, with a nearby Hopf bifurcation
at%
\ifthenelse{\boolean{LongVersion}}{%
  ~\citep{falugi:2015}:
  \[
    (x_{\mathrm{Hopf}},u_{\mathrm{Hopf}}) \approx (0.1728, 0.7009, 0.6973).
  \]
}{ %
  \((x_{\mathrm{Hopf}},u_{\mathrm{Hopf}}) \approx (0.1728, 0.7009,
  0.6973)\)~\cite{falugi:2015}. %
}%
\ifthenelse{\boolean{LongVersion}}{%
  We plot the nominal steady states (i.e., \(\wP=0\)) along with the initial
  steady state \(x(0)\) and the Hopf bifurcation \(x_{\mathrm{Hopf}}\) in
  \Cref{fig:cstr:ss}. %
}{}

For the following simulations, the plant~\cref{eq:cstr} is simulated by ten
4th-order Runga-Kutta steps per sample time %
\ifthenelse{\boolean{LongVersion}}{%
  \(\Delta = 1 \textnormal{ s}\). Assume disturbance set is \(\wP \in \bbW
  \defas [-0.05,0.05]^4 \times \set{0,1}\). %
}{%
  \(\Delta = 1\). %
}\ifthenelse{\boolean{LongVersion}}{%
  Regardless of objective \(\ell_s\), the SSTP~\cref{eq:sstp} is uniquely solved
  by
  \begin{align*}
    x_s(\beta)
    &\defas \begin{bmatrix} \frac{1}{1 + \theta k\exp\left( -M/\rsp \right)} \\
              \rsp \end{bmatrix},
    &u_s(\beta)
    &\defas \frac{x_f - \rsp + 1 - (x_s(\beta))_1}{\theta\gamma(\rsp-x_c-d)}
  \end{align*}
  for each \(\beta = (\rsp,\usp,\ysp,d) \in \calB_c\), where
  \[
    \calB_c \defas [0.6,0.7]\times\bbU\times[0.6,0.7]\times[-0.1,0.1]
  \]
  and we have used the identity \(\frac{a}{1+a} = 1 - \frac{1}{1+a}\) for all
  \(a\neq 1\). Likewise, the solution to \cref{eq:ssop} is
  %% TODO broken
  \begin{align}
    \xPs(\alpha)
    &\defas \begin{bmatrix} \frac{1 + (\wP)_1}{1 + \theta k \exp\left(
              \frac{(\wP)_2-M}{\rsp-(\wP)_4} \right)} \\ \rsp - (\wP)_4 \end{bmatrix}, \\
    d_s(\alpha)
    &\defas (\wP)_3 + (\wP)_4 + \frac{((\wP)_1 + (\wP)_4 -
      (\Delta x_s(\alpha))_1)((\xPs(\alpha))_2 - x_c - (\wP)_3)}{x_f -
      (\xPs(\alpha))_2 + 1 + (\wP)_1 - (\xPs(\alpha))_1}
  \end{align}
  for each \(\alpha = (\rsp,\usp,\ysp,\wP) \in \calA_c\), where
  \[
    (\Delta x_s(\alpha))_1 \defas (\xPs(\alpha))_1 - (x_s(\beta))_1 = \frac{1 +
      (\wP)_1}{1 + \theta k \exp\left( \frac{(\wP)_2-M}{\rsp-(\wP)_4} \right)} -
    \frac{1}{1 + \theta k\exp\left( -M/\rsp \right)},
  \]
  \(\beta \defas (\rsp,\usp,\ysp,d_s(\alpha))\), and
  \[
    \calA_c \defas [0.6,0.7]\times\bbU\times[0.6,0.7]\times\bbW.
  \]
  It is straightforward to verify \(\calA_c\) and \(\calB_c\) are compact and
  satisfy \Cref{assum:sstp:mismatch}. %

}{%
  We define a SSTP with \(\ell_s(u,y)=u^2+y^2\) and no output constraints. %
}%
We define a regulator with \(N\defas 150\), \(\ell(x,u,\Delta u,\beta)\defas
|x-x_s(\beta)|_Q^2 + 10^{-3}(u-u_s(\beta))^2 + (\Delta u)^2\),%
%% TODO is a cross term really necessary?
\footnote{The rate-of-change penalty \(\Delta u\) is easily implemented in
  the FHOCP via state augmentation~\cite[Ex.~1.25]{rawlings:mayne:diehl:2020}.
  While this introduces a cross term to the stage cost~\cref{eq:quad}, i.e.,
  \(\ell(x,u,\beta) \defas |(x,u)-(x_s(\beta),u_s(\beta))|_S^2\), the proofs are
  also easily extended by replacing
  \(\underline{\sigma}(Q),\underline{\sigma}(R)\) with \(\underline{\sigma}(S)\)
  throughout.} %
\(Q=\begin{bsmallmatrix} 10^{-3} \\ & 1 \end{bsmallmatrix}\),
\(V_f(x,\beta)\defas |x-x_s(\beta)|_{P_f(\beta)}^2\), and
\(\mathbb{X}_f\defas\lev_{c_f} V_f\), where \(P_f(\beta)\) and \(c_f\approx
6.5154\times 10^{-16}\) are constructed according to \Cref{app:terminal} to
satisfy \Cref{assum:quad,assum:stabilizability}.\footnote{While \(c_f\) was
  chosen near machine precision, the CSTR tends to evolve to the nearest stable
  steady state, and the horizon is chosen long enough to easily achieve this
  steady state to a high degree of precision. Thus, the system remains robust
  despite the tight terminal constraint.} Finally, we use MHE
designs~\cref{eq:mhe:aug,eq:mhe} for the offset-free MPC and tracking MPC,
respectively, where \(T\defas N\), \(Q_w\defas 10^{-4}I\), \(Q_d\defas
10^{-2}\), and \(R_v\defas 1\). %
\ifthenelse{\boolean{LongVersion}}{%
  As in the simple pendulum example, if %
}{%
  If %
}%
\Cref{assum:est} is satisfied, then \Cref{thm:mpc:mismatch} implies the
offset-free MPC can robustly track setpoints despite plant-model mismatch.

\begin{figure*}
  \centering %
  \ifthenelse{\boolean{LongVersion}}{%
    \subfloat[No mismatch: \((\wP)_1\equiv 0\) and \((\wP)_2\equiv
    0\).\label{fig:cstr:traj:a}]{%
      \includegraphics[width=0.49\linewidth,page=1]{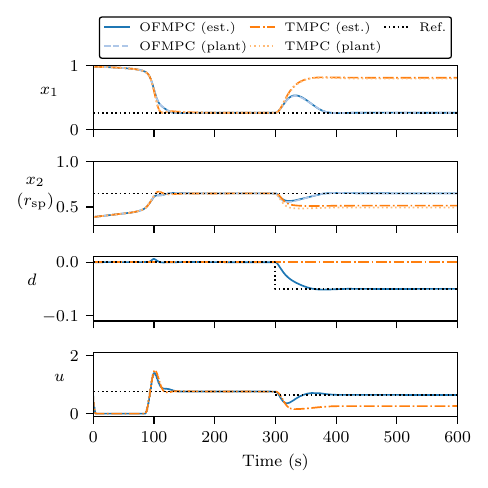}%
    }\hfill %
    \subfloat[Mismatch: \((\wP)_1\equiv -0.05\) and \((\wP)_2\equiv
    -0.05\).\label{fig:cstr:traj:b}]{%
      \includegraphics[width=0.49\linewidth,page=2]{cstr_plot.pdf}%
    }

    \subfloat[Noise and mismatch: \((\wP)_3^+ = (\wP)_3 + (\Delta\wP)_3\),
    \((\Delta\wP)_3 \sim \norm(0,10^{-6})\), and \((\wP)_4 \sim
    \norm(0,10^{-6})\).\label{fig:cstr:traj:c}]{%
      \includegraphics[width=0.49\linewidth,page=3]{cstr_plot.pdf}%
    }\hfill %
    \subfloat[Oscillating setpoint: \(\rsp(k) = 0.05\sin(\frac{2\pi k}{90}) +
    0.65\).\label{fig:cstr:traj:d}]{%
      \includegraphics[width=0.49\linewidth,page=4]{cstr_plot.pdf}%
    } %
  }{%
    \subfloat[No mismatch: \((\wP)_1\equiv 0\) and \((\wP)_2\equiv
    0\).\label{fig:cstr:traj:a}]{%
      \includegraphics[width=0.33\linewidth,page=1]{cstr_plot.pdf}%
    }\hfill %
    \subfloat[Mismatch: \((\wP)_1\equiv -0.05\) and \((\wP)_2\equiv
    -0.05\).\label{fig:cstr:traj:b}]{%
      \includegraphics[width=0.33\linewidth,page=2]{cstr_plot.pdf}%
    }\hfill %
    \subfloat[Noise: \((\wP)_3^+ = (\wP)_3 + (\Delta\wP)_3\), \((\Delta\wP)_3
    \sim \norm(0,10^{-6})\), and \((\wP)_4 \sim
    \norm(0,10^{-6})\).\label{fig:cstr:traj:c}]{%
      \includegraphics[width=0.33\linewidth,page=3]{cstr_plot.pdf}%
    } %
  }%
  \caption[Closed-loop trajectories for the offset-free MPC and tracking MPC of
  the CSTR.]{Simulated closed-loop trajectories for the offset-free MPC and
    tracking MPC of the CSTR~\cref{eq:cstr}. Solid blue and dot-dashed orange
    lines represent the closed-loop estimates and inputs \((\hat{x},\hat{d},u)\)
    for the offset-free MPC and tracking MPC simulations, respectively. Dashed
    blue and dotted orange lines represent the closed-loop plant states \(\xP\)
    for the offset-free MPC and tracking MPC simulations, respectively. Dotted
    black lines represent the intended steady-state targets and disturbance
    values \((\xPs, d_s, u_s)\) found by solving~\cref{eq:sstp,eq:ssop}. We set
    \((\wP)_1\equiv -0.05\), \((\wP)_2\equiv -0.05\),
    \((\wP)_3(k)=-0.05H(k-300)\), \((\wP)_4\equiv 0\), \((\wP)_5\equiv 1\), and
    \(\rsp\equiv 0.65\) unless otherwise specified.}%
  \label{fig:cstr:traj}
\end{figure*}

The results of the CSTR experiments are presented in \Cref{fig:cstr:traj}.
Unless otherwise specified, each simulation is carried out with error in the
feed concentration \((\wP)_1\equiv -0.05\), error in the activation energy
\((\wP)_2\equiv -0.05\), a step in the coolant temperature
\((\wP)_3(k)=-0.05H(k-300)\), no measurement noise \((\wP)_4\equiv 0\), the
discretization parameter \((\wP)_5\equiv 1\), and a constant reference signal
\(\rsp \equiv 0.65\).

In the first experiment, we consider the case without plant-model mismatch,
i.e., \((\wP)_1\equiv 0\) and \((\wP)_2\equiv 0\) (\Cref{fig:cstr:traj:a}).
\ifthenelse{\boolean{LongVersion}}{%
  As in the pendulum experiment, both %
}{%
  Both %
}%
offset-free and tracking MPC remove offset after the setpoint changes, but only
offset-free MPC removes offset after the disturbance is injected. We also note
that, after the disturbance is injected, the tracking MPC state estimates are
slightly different than the plant states.

We consider plant-model mismatch \((\wP)_1\equiv -0.05\) and \((\wP)_2\equiv
-0.05\) in the second experiment (\Cref{fig:cstr:traj:b}). The offset-free
MPC is able to track the reference and reject the disturbance despite mismatch,
this time at the cost of a significant temperature spike around \(k=170\). On
the other hand, the tracking MPC fails to bring the temperature above
\(x_2=0.5\), far from the setpoint \(\rsp=0.65\).

In the third experiment, the coolant temperature is an integrating disturbance
\((\wP)_3^+ = (\wP)_3 + (\Delta\wP)_3\), \((\Delta\wP)_3 \sim
\norm(0,10^{-6})\), and we have measurement noise \((\wP)_4 \sim
\norm(0,10^{-6})\) (\Cref{fig:cstr:traj:c}). %
\ifthenelse{\boolean{LongVersion}}{%
  As in the corresponding pendulum experiment, offset-free MPC %
}{%
  Offset-free MPC %
}%
tracks the reference despite the randomly drifting disturbance. Here we are
illustrating a behavior that is often desired in chemical process control, where
process specifications must be met despite constantly, but slowly varying
upstream conditions. %
\ifthenelse{\boolean{LongVersion}}{%
  We remark that, while the pendulum example is mechanical in nature, it
  illustrated the same property. %
}{}%
The tracking MPC, on the other hand, still cannot handle the plant-model
mismatch and fails to bring the temperature up to the setpoint.

\ifthenelse{\boolean{LongVersion}}{%
  In the fourth and final experiment, the setpoint follows an oscillating
  pattern \(\rsp(k) = 0.05\sin(\frac{2\pi k}{90}) + 0.65\). Tracking MPC again
  fails bring the temperature up to the setpoint. Offset-free MPC closely
  follows the setpoint, substantially deviating from it only at the start-up
  phase and when the coolant temperature disturbance is injected. Again, we note
  that a precise tracking of this disturbance and reference signal could be
  accomplished by more general integrator schemes.
  (c.f.,~\cite{maeder:morari:2010} or Sections~5.3 and~5.4 of
  \cite{zagrobelny:2014}). %
}{}%

\section{Conclusions}\label{sec:conclusion}
In this paper, we presented a nonlinear offset-free MPC design that is robustly
stable with respect to setpoint- and target-tracking errors, despite persistent
disturbances and plant-model mismatch. We assume neither stability of the
closed-loop system (as
in~\cite{muske:badgwell:2002,pannocchia:rawlings:2003,morari:maeder:2012}), nor
the existence of an invariant set for tracking (as in~\cite{%
  falugi:2015,% nonlinear reference tracking MPC
  limon:ferramosca:alvarado:alamo:2018,% nonlinear tracking MPC
  galuppini:magni:ferramosca:2023% semidefinite stage cost case terminal
  % ingredients
}). However, using an offset constraint (in the SSTP~\cref{eq:sstp}) rather than
an offset penalty limits the tracking domain to \(\calX_N(\beta)\) rather than
its union over \(\beta\in\hat{\calB}_c\).

These results form a foundation on which offset-free performance guarantees can
be established on a wider class of MPC designs and applications. By
incorporating offset penalties (cf.~\cite{%
  falugi:2015,% nonlinear reference tracking MPC
  limon:ferramosca:alvarado:alamo:2018,% nonlinear tracking MPC
  galuppini:magni:ferramosca:2023% semidefinite stage cost case terminal
  % ingredients
}) the tracking domain may be extended. Relaxing some of the restrictions of
this work, notably the requirement of a Lyapunov function for the estimator
(\Cref{assum:est}), and the necessity of quadratic costs (\Cref{assum:quad}),
are also possible
%% NOTE This ifthenelse statement just cancels out any latexdiff markup in
%% this section. Otherwise, the spacing gets wonky and text spills onto the
%% next page.
\ifthenelse{\boolean{true}}{%
  areas of future research. %
}{}%
  Throughout this work, %
\ifthenelse{\boolean{true}}{%
``sufficiently small mismatch''}{} is never quantified. Quantification of the
bounding constants (e.g., as done for linear systems
in %
\ifthenelse{\boolean{LongVersion}}{%
  Chapter~6 of \cite{kuntz:2024}) %
}{%
  \cite[Ch.~6]{kuntz:2024}) %
}%
is another possible area of future research.

%% file: cstr.tex
%% Creator: Inkscape 1.2.2 (b0a8486541, 2022-12-01), www.inkscape.org
%% Accompanies image file 'cstr.pdf' (pdf, eps, ps)
\begingroup%
  \makeatletter%
  \providecommand\color[2][]{%
    \errmessage{(Inkscape) Color is used for the text in Inkscape, but the package 'color.sty' is not loaded}%
    \renewcommand\color[2][]{}%
  }%
  \providecommand\transparent[1]{%
    \errmessage{(Inkscape) Transparency is used (non-zero) for the text in Inkscape, but the package 'transparent.sty' is not loaded}%
    \renewcommand\transparent[1]{}%
  }%
  \providecommand\rotatebox[2]{#2}%
  \newcommand*\fsize{\dimexpr\f@size pt\relax}%
  \newcommand*\lineheight[1]{\fontsize{\fsize}{#1\fsize}\selectfont}%
  \ifx\svgwidth\undefined%
    \setlength{\unitlength}{386.101872bp}%
    \ifx\svgscale\undefined%
      \relax%
    \else%
      \setlength{\unitlength}{\unitlength * \real{\svgscale}}%
    \fi%
  \else%
    \setlength{\unitlength}{\svgwidth}%
  \fi%
  \global\let\svgwidth\undefined%
  \global\let\svgscale\undefined%
  \makeatother%
  \begin{picture}(1,0.86543339)%
    \lineheight{1}%
    \setlength\tabcolsep{0pt}%
    \put(0,0){\includegraphics[width=\unitlength,page=1]{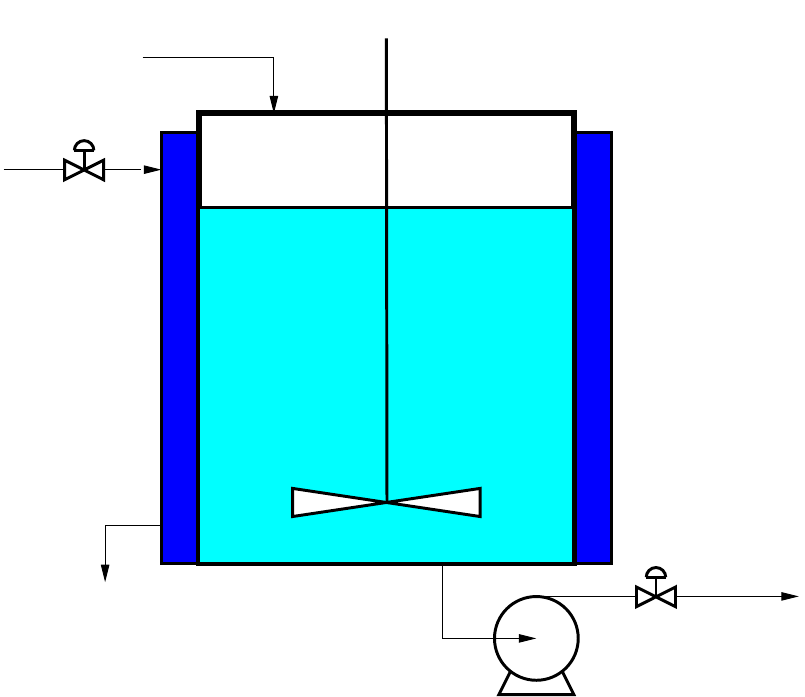}}%
    \put(0.17146658,0.83791214){\color[rgb]{0,0,0}\makebox(0,0)[lt]{\lineheight{1.25}\smash{\begin{tabular}[t]{l}$F_0,T_0,c_0$\end{tabular}}}}%
    \put(-0.00224811,0.57401095){\color[rgb]{0,0,0}\makebox(0,0)[lt]{\lineheight{1.25}\smash{\begin{tabular}[t]{l}$F_c,T_c$\end{tabular}}}}%
    \put(0.77114195,0.0462079){\color[rgb]{0,0,0}\makebox(0,0)[lt]{\lineheight{1.25}\smash{\begin{tabular}[t]{l}$F_0$\end{tabular}}}}%
    \put(0.91544201,0.04628105){\color[rgb]{0,0,0}\makebox(0,0)[lt]{\lineheight{1.25}\smash{\begin{tabular}[t]{l}$T,c$\end{tabular}}}}%
  \end{picture}%
\endgroup%

%% file: paper_appendix.tex
\ifthenelse{\boolean{LongVersion}}{%
\section{Proofs of robust estimation and tracking stability}\label[appendix]{app:stability}
\subsection{Proof of \Cref{thm:est:rges}}\label[appendix]{app:est:rges}
First, note that \(c_3\leq c_2\), as otherwise, this would imply \(V_e(k+1)
\leq 0\) whenever \(\tilde{w}(k)=0\). We combine the upper bound
\cref{eq:est:lyap:a} and bound on the difference \cref{eq:est:lyap:b} to
give
\[
  V_e(k+1) \leq \lambda V_e(k) + c_4|\tilde{w}(k)|^2
\]
where \(\lambda\defas 1 - \frac{c_3}{c_2}\in(0,1)\). Recursively applying
the above inequality gives
\begin{align*}
  V_e(k)
  &\leq \lambda^k V_e(0) + \sum_{j=1}^k
    c_4\lambda^{j-1}|\tilde{w}(k-j)|^2 \\
  &\leq c_2\lambda^{k+1}|\overline e|^2
    + \sum_{j=1}^k c_4\lambda^{j-1}|\tilde{w}(k-j)|^2
\end{align*}
noting that \(e(0)=\overline e\) because \(\Phi_0\) is the identity map.
Finally,
\[
  |e(k)| \leq \sqrt{\frac{V_e(k)}{c_1}} \leq c_{e,1}\lambda_e^k|\overline
  e| + c_{e,2}\sum_{j=1}^{k+1} \lambda_e^{j-1}|\tilde{w}(k-j)|
\]
where \(c_{e,1}\defas\sqrt{\frac{c_2}{c_1}}\),
\(c_{e,2}\defas\sqrt{\frac{c_4}{c_1}}\), and
\(\lambda_e\defas\sqrt{\lambda}\).
\hspace*{\fill}~\QED
}{}

\subsection{Proof of \Cref{thm:lyap}}\label[appendix]{app:lyap}
\ifthenelse{\boolean{LongVersion}}{%
  Suppose \(X\subseteq\Xi\) is RPI for \cref{eq:sys:cl}. Let the functions
  \(V:\Xi\rightarrow\nnegreal\) and
  \(\alpha_i,\sigma\in\calKinf,i\in\intinterval{1}{3}\) satisfy \cref{eq:lyap}
  for all \(\xi\in X\) and \(\omega\in\Omega_c(\xi)\). Let
  \((\bfxi,\bfomega,\bfzeta_1,\bfzeta_2)\) satisfy \cref{eq:sys:cl} and
  \(\xi(0)\in X\). %
}{}%

\paragraph*{Asymptotic case}
\ifthenelse{\boolean{LongVersion}}{%
  The proof of this part follows similarly to Lemma~3.5 of
  \cite{jiang:wang:2001} and Theorem~1 of \cite{tran:kellett:dower:2015}. We
  start by noting \cref{eq:lyap:b} can be rewritten
  \begin{equation}\label{eq:ras:1}
    V(F_c(\xi,\omega)) \leq (\id - \alpha_4)(V(\xi)) + \sigma(|\omega|)
  \end{equation}
  where \(\alpha_4\defas\alpha_3\circ\alpha_2^{-1}\in\calKinf\). Without loss of
  generality, we can assume
  \(\id-\alpha_4\in\calK\)~\cite[Lem.~B.1]{jiang:wang:2001}. Let
  \(\rho\in\calKinf\) such that \(\id-\rho\in\calKinf\).

  Let \(b \defas \alpha_4^{-1}(\rho^{-1}(\sigma(\|\bfomega\|)))\) and
  \(D\defas \set{\xi\in \Xi | V(\xi) \leq b}\). The following intermediate
  result is required.
  \begin{lemma}
    If there exists \(k_0\in\nnegint\) such that \(\xi(k_0)\in D\), then \(\xi(k)\in
    D\) for all \(k\geq k_0\).
  \end{lemma}
  \begin{proof}
    Suppose \(k\geq k_0\) and \(\xi(k)\in D\). Then \(V(\xi(k))\leq b\) and
    % \(\rho(\alpha_4(V(\xi(k)))) \leq \sigma(\|\bfomega\|)\). Then
    by \cref{eq:ras:1},
    \begin{align*}
      V(\xi(k+1))
      &\leq (\id - \alpha_4)(V(\xi(k))) + \sigma(\|\bfomega\|) \\
      &\leq (\id - \alpha_4)(b) + \sigma(\|\bfomega\|) \\
      &= \underbrace{-(\id - \rho)(\alpha_4(b))}_{\leq 0} + b
        \underbrace{- \rho(\alpha_4(b)) + \sigma(\|\bfomega\|)}_{=0} \leq b.
    \end{align*}
    The result follows by induction.
  \end{proof}

  Next, let \(j_0 \defas \min\set{ k\in\nnegint | \xi(k)\in D}\). The above
  lemma gives \(V(\xi(k)) \leq \gamma(\|\bfomega\|)\) for all \(k\geq j_0\),
  where \(\gamma\defas\alpha_4^{-1}\circ\rho^{-1}\circ\sigma\). On the other
  hand, if \(k<j_0\), then we have \(\rho(\alpha_4(V(\xi(k)))) >
  \sigma(\|\bfomega\|)\) and therefore
  \begin{align*}
    V(\xi(k+1)) - V(\xi(k))
    &\leq -\alpha_4(V(\xi(k))) + \sigma(\|\bfomega\|) \\
    &= -\alpha_4(V(\xi(k))) + \rho(\alpha_4(V(\xi(k))))
      -\rho(\alpha_4(V(\xi(k)))) + \sigma(\|\bfomega\|) \\
    &\leq -\alpha_4(V(\xi(k))) + \rho(\alpha_4(V(\xi(k)))).
  \end{align*}
  By Lemma~4.3 of \cite{jiang:wang:2001}, there exists \(\beta\in\calKL\) such
  that
  \[
    \alpha_1(|\zeta_1(k)|) \leq V(\xi(k)) \leq \beta(V(\xi(0)),k) \leq
    \beta(\alpha_2(|\zeta_2(0)|),k).
  \]
  Combining the above inequalities gives
  \[
    |\zeta_1(k)| \leq \max\{\beta_\zeta(|\zeta_2(0)|,k), \;
    \gamma_\zeta(\|\bfomega\|)\} \leq \beta_\zeta(|\zeta_2(0)|,k) +
    \gamma_\zeta(\|\bfomega\|)
  \]
  where \(\beta_\zeta(s,k) \defas \alpha_1^{-1}(\beta(\alpha_2(s),k))\) and
  \(\gamma_\zeta \defas \alpha_1^{-1}\circ\gamma\). Finally, causality lets us
  drop future terms of \(\omega\) from the signal norm in the above inequality
  and simply write~\cref{eq:ras}. %
}{%
  See~\cite[Thm.~1]{tran:kellett:dower:2015}, noting that the proof does not
  depend on continuity of \(F_c\), \(G_1\), or \(G_2\), and does not require the
  measurement functions to be zero at zero. %
}%

\paragraph*{Exponential case}
\ifthenelse{\boolean{LongVersion}}{%
  Suppose, additionally, that \(\alpha_i
  \defas a_i\id^b,i\in\intinterval{1}{3}\). %
}{}%
Without loss of generality, we can assume \(\lambda\defas 1-a_3\in(0,1)\).
Recursively applying \cref{eq:lyap:b} gives %
\ifthenelse{\boolean{LongVersion}}{%
  \begin{align*}
    V(\xi(k))
    &\leq \lambda^k V(\xi(0)) + \sum_{i=1}^k \lambda^{i-1}\sigma(|\omega(k-i)|) \\
    &\leq \lambda^k a_2|\zeta_2(0)|^b + \frac{\sigma(\|\bfomega\|_{0:k-1})}{1-\lambda}.
  \end{align*}
}{%
  \(V(\xi(k)) \leq \lambda^k a_2|\zeta_2(0)|^b +
  \frac{\sigma(\|\bfomega\|_{0:k-1})}{1-\lambda}\). %
}%
Applying \cref{eq:lyap:a}, we have %
\ifthenelse{\boolean{LongVersion}}{%
  \begin{equation*}
    |\zeta_1(k)| \leq \left(\frac{a_2}{a_1}\lambda^k|\zeta_2(0)|^b +
      \frac{\sigma(\|\bfomega\|_{0:k-1})}{a_1(1-\lambda)}\right)^{1/b}.
  \end{equation*}
}{%
  \(|\zeta_1(k)| \leq \left(\frac{a_2}{a_1}\lambda^k|\zeta_2(0)|^b +
    \frac{\sigma(\|\bfomega\|_{0:k-1})}{a_1(1-\lambda)}\right)^{1/b}\). %
}%
\ifthenelse{\boolean{LongVersion}}{%
  If \(b\geq 1\), the triangle inequality gives %
  \begin{equation}\label{eq:lyap:2}
    |\zeta_1(k)| \leq c_\zeta\lambda_\zeta^k|\zeta_2(0)| + \gamma_\zeta(\|\bfomega\|_{0:k-1})
  \end{equation}
  with \(c_\zeta\defas\left(\frac{a_2}{a_1}\right)^{1/b}\),
  \(\lambda_\zeta\defas\lambda^{1/b}\), and \(\gamma_\zeta(\cdot)\defas \left(
    \frac{\sigma(\cdot)}{a_1(1-\lambda)} \right)^{1/b}\). Otherwise, if \(b<1\),
  then convexity gives~\cref{eq:lyap:2} with
  \(c_\zeta\defas\frac{1}{2}\left(\frac{2a_2}{a_1}\right)^{1/b}\),
  \(\lambda_\zeta\defas\lambda^{1/b}\), and \(\gamma_\zeta(\cdot) \defas
  \frac{1}{2}\left( \frac{2\sigma(\cdot)}{a_1(1-\lambda)}\right)^{1/b}\).
  %% Convexity argument:
  %% f convex => f(x/2+y/2) \leq f(x)/2 + f(y)/2
  %% a\geq 1 => (x/2+y/2)^a \leq x^a/2 + y^a/2
  %% a\geq 1 => (x+y)^a \leq (2x)^a/2 + (2y)^a/2
  %% OR more generally,
  %% f convex => f(\lambda x+(1-\lambda)y) \leq \lambda f(x) + (1-\lambda)f(y)
  %% a\geq 1 => (\lambda x+(1-\lambda)y)^a \leq \lambda x^a + (1-\lambda)y^a
  %% a\geq 1 => (x+y)^a \leq \lambda(x/\lambda)^a + (1-\lambda)(y/(1-\lambda))^a
}{%
  Finally, applying the triangle inequality when \(b\geq 1\) or a convexity
  argument when \(b<1\) gives \cref{eq:ras} with \(\beta_\zeta(s,k)\defas
  c_\zeta\lambda_\zeta^ks\) for appropriately defined \(c_\zeta>0\),
  \(\lambda_\zeta\in(0,1)\), and \(\gamma_\zeta\in\calK\). %
}%
\hspace*{\fill}~\QED

\subsection{Proof of \Cref{thm:smallgain}}\label[appendix]{app:smallgain}
\ifthenelse{\boolean{LongVersion}}{%
  Throughout, we fix \(k\in\nnegint\) and drop dependence on \(k\) when it is
  understood from context. Let the trajectories
  \((\bfxi,\hat{\bfxi},\bfu,\bfomega,\bfupsilon,\bfepsilon,\bfzeta)\)
  satisfy~\cref{eq:sys:ctrl,eq:meas,eq:state:est,eq:ctrl,eq:state:est:err},
  \(\zeta=G(\hat{\xi})\), and \((\xi(0),\overline{\xi})\in\calS\), where
  \(\calS\) is RPI.\@ Suppose \(\Phi^\xi_0\) is the identity map. Let
  \(a_i,b_i>0,i\in\intinterval{1}{4}\), \(V:\hat{\Xi}\rightarrow\nnegreal\),
  \(V_\varepsilon:\Xi\times\hat{\Xi}\rightarrow\nnegreal\), and
  \(\sigma,\sigma_\varepsilon\in\calK\) satisfy \(\frac{a_4c_4}{a_3c_1} < 1\),
  \(\frac{a_4c_4}{a_3c_3} < \frac{c_1}{c_1+c_2}\), and \cref{eq:smallgain}.

}{}%
\paragraph*{Joint Lyapunov function}
\ifthenelse{\boolean{LongVersion}}{%
  Our first goal is to construct a Lyapunov function for the joint
  regulator-estimator system. Combining the fact
  \(|(\varepsilon,\varepsilon^+)|^2 = |\varepsilon|^2 + |\varepsilon^+|^2\) %
  with the inequalities~\cref{eq:smallgain:b,eq:smallgain:c,eq:smallgain:d}, we
  have %
  \begin{align*}
    V \ifthenelse{\boolean{OneColumn}}{}{
    &}(\hat{\xi}^+) - V(\hat{\xi}) \ifthenelse{\boolean{OneColumn}}{}{\\}
    &\overset{\cref{eq:smallgain:b}}{\leq} -a_3|\zeta|^2 + a_4|\varepsilon|^2
      + a_4|\varepsilon^+|^2 + \sigma(|\omega|) \\
    &\overset{\cref{eq:smallgain:c}}{\leq} -a_3|\zeta|^2 + a_4|\varepsilon|^2
      + \frac{a_4}{c_1}V_\varepsilon(\xi^+,\hat{\xi}^+) + \sigma(|\omega|) \\
    &\overset{\cref{eq:smallgain:d}}{\leq} -\tilde{a}_3|\zeta|^2 +
      a_4\left( 1 - \frac{c_3}{c_1} \right)|\varepsilon|^2
      + \frac{a_4}{c_1} V_\varepsilon(\xi,\hat{\xi}) + \tilde\sigma(|\omega|) \\
    &\overset{\cref{eq:smallgain:c}}{\leq} -\tilde{a}_3|\zeta|^2 +
      \tilde{a}_4|\varepsilon|^2 + \tilde\sigma(|\omega|)
  \end{align*}
}{%
  Combining~\cref{eq:smallgain:b,eq:smallgain:c,eq:smallgain:d} gives %
  \[
    V(\hat{\xi}^+) - V(\hat{\xi}) \leq -\tilde{a}_3|\zeta|^2 +
    \tilde{a}_4|\varepsilon|^2 + \tilde\sigma(|\omega|)
  \]
}%
where \(\tilde{a}_3\defas a_3 - \frac{a_4c_4}{c_1}\), %
\ifthenelse{\boolean{LongVersion}}{%
  \(\tilde{a}_4\defas a_4\left( 1 + \frac{c_2-c_3}{c_1} \right)\), %
}{%
  \(\tilde{a}_4\defas a_4(1 + \frac{c_2-c_3}{c_1})\), %
}%
and \(\tilde{\sigma}\defas\frac{a_4}{c_1}\sigma_\varepsilon + \sigma \in
\calK\). Note that %
\ifthenelse{\boolean{LongVersion}}{%
  \(\tilde{a}_3 = a_3\left( 1 - \frac{a_4c_4}{a_3c_1} \right) > 0\) %
}{%
  \(\tilde{a}_3 = a_3(1 - \frac{a_4c_4}{a_3c_1}) > 0\) %
}%
by assumption, and \(\tilde{a}_4 > 0\) since \(c_2>c_3\). Let
\(W(\xi,\hat{\xi})\defas V(\hat{\xi})+qV_\varepsilon(\xi,\hat{\xi})\) where
\(q>0\). With \(b_1\defas\min\set{a_1,qc_1}\), we have the lower bound, %
\ifthenelse{\boolean{OneColumn}}{%
  \begin{equation}\label{eq:smallgain:1}
    b_1|(\zeta,\varepsilon)|^2 = b_1|\zeta|^2 + b_1|\varepsilon|^2
    \leq a_1|\zeta|^2 + qc_1|\varepsilon|^2
    \leq V(\hat{\xi}) + qV_\varepsilon(\xi,\hat{\xi}) \asdef W(\xi,\hat{\xi}).
  \end{equation}
}{%
  %% NOTE For some reason multline is producing a lot of space above the
  %% equation, so I simulated the behavior with an align environment. If the
  %% equation is changed, just be careful to adjust the alignment.
  % \begin{multline}\label{eq:smallgain:1}
  %   b_1|(\zeta,\varepsilon)|^2 = b_1|\zeta|^2 + b_1|\varepsilon|^2
  %   \leq a_1|\zeta|^2 + qc_1|\varepsilon|^2 \\
  %   \leq V(\hat{\xi}) + qV_\varepsilon(\xi,\hat{\xi}) \asdef W(\xi,\hat{\xi}).
  % \end{multline}
  \begin{align}
    b_1|(\zeta,\varepsilon)|^2 = b_1|\zeta|^2
    &+ b_1|\varepsilon|^2 \leq a_1|\zeta|^2 + qc_1|\varepsilon|^2 \nonumber \\
    &\leq V(\hat{\xi}) + qV_\varepsilon(\xi,\hat{\xi}) \asdef W(\xi,\hat{\xi}). \label{eq:smallgain:1}
  \end{align}
}%
With \(b_2\defas\max\set{a_2,qc_2}\), we have the upper bound %
\ifthenelse{\boolean{OneColumn}}{%
  \begin{equation}\label{eq:smallgain:2}
    W(\xi,\hat{\xi}) \defas V(\hat{\xi}) + qV_\varepsilon(\xi,\hat{\xi})
    \leq a_2|\zeta|^2 + qc_2|\varepsilon|^2
    \leq b_2|\zeta|^2 + b_2|\varepsilon|^2 = b_2|(\zeta,\varepsilon)|^2.
  \end{equation}
}{%
  %% NOTE For some reason multline is producing a lot of space above the
  %% equation, so I simulated the behavior with an align environment. If the
  %% equation is changed, just be careful to adjust the alignment.
  % \begin{multline}\label{eq:smallgain:2}
  %   W(\xi,\hat{\xi}) \defas V(\hat{\xi}) + qV_\varepsilon(\xi,\hat{\xi})
  %   \leq a_2|\zeta|^2 + qc_2|\varepsilon|^2 \\
  %   \leq b_2|\zeta|^2 + b_2|\varepsilon|^2 = b_2|(\zeta,\varepsilon)|^2.
  % \end{multline}
  \begin{align}
    W(\xi,\hat{\xi}) \defas V(\hat{\xi}) + q
    &V_\varepsilon(\xi,\hat{\xi}) \leq a_2|\zeta|^2 + qc_2|\varepsilon|^2
      \nonumber \\
    &\leq b_2|\zeta|^2 + b_2|\varepsilon|^2 = b_2|(\zeta,\varepsilon)|^2.
      \label{eq:smallgain:2}
  \end{align}
}%
For the cost decrease, we first note that \(\frac{a_4c_4}{a_3c_3} <
\frac{c_1}{c_1 + c_2}\) implies %
\ifthenelse{\boolean{LongVersion}}{%
  \ifthenelse{\boolean{OneColumn}}{%
    \begin{equation*}
      \tilde{a}_4c_4 = a_4\left( \frac{c_1 + c_2}{c_1} - \frac{c_3}{c_1}
      \right)c_4 < a_4\left( \frac{a_3c_3}{a_4c_4} - \frac{c_3}{c_1} \right)c_4
      = a_3c_3 - \frac{a_4c_3c_4}{c_1} = \tilde{a}_3c_3
    \end{equation*}
  }{%
    \begin{multline*}
      \tilde{a}_4c_4 = a_4\left( \frac{c_1 + c_2}{c_1} - \frac{c_3}{c_1}
      \right)c_4 < a_4\left( \frac{a_3c_3}{a_4c_4} - \frac{c_3}{c_1} \right)c_4 \\
      = a_3c_3 - \frac{a_4c_3c_4}{c_1} = \tilde{a}_3c_3
    \end{multline*}
  }%
}{%
  \(\tilde{a}_4c_4 = a_4( \frac{c_1 + c_2}{c_1} - \frac{c_3}{c_1} )c_4 < a_4(
  \frac{a_3c_3}{a_4c_4} - \frac{c_3}{c_1} )c_4 = \tilde{a}_3c_3\) %
}%
and therefore \(\frac{\tilde{a}_4}{c_3} < \frac{\tilde{a}_3}{c_4}\). With
\ifthenelse{\boolean{LongVersion}}{%
  \(q\in\left(\frac{\tilde{a}_4}{c_3}, \frac{\tilde{a}_3}{c_4}\right)\), %
}{%
  \(q\in(\frac{\tilde{a}_4}{c_3}, \frac{\tilde{a}_3}{c_4})\), %
}%
we have
\(b_3\defas\min\set{\tilde{a}_3 - qc_4,qc_3 - \tilde{a}_4} > 0\), \(\sigma_W
\defas \tilde\sigma + q\sigma_\varepsilon \in \calK\), and %
\ifthenelse{\boolean{OneColumn}}{%
  \begin{equation}\label{eq:smallgain:3}
    W(\xi^+,\hat{\xi}^+) \leq V(\hat{\xi}^+) + qV_\varepsilon(\xi^+,\hat{\xi}^+) \\
    \leq W(\xi,\hat{\xi}) - b_3|(\zeta,\varepsilon)|^2 + \sigma_W(|\omega|).
  \end{equation}
}{%
  %% NOTE For some reason multline is producing a lot of space above the
  %% equation, so I simulated the behavior with an align environment. If the
  %% equation is changed, just be careful to adjust the alignment.
  % \begin{multline}\label{eq:smallgain:3}
  %   W(\xi^+,\hat{\xi}^+) \leq V(\hat{\xi}^+) + qV_\varepsilon(\xi^+,\hat{\xi}^+) \\
  %   \leq W(\xi,\hat{\xi}) - b_3|(\zeta,\varepsilon)|^2 + \sigma_W(|\omega|).
  % \end{multline}
  \begin{align}
    W(\xi^+,\hat{\xi}^+) \leq V
    &(\hat{\xi}^+) + qV_\varepsilon(\xi^+,\hat{\xi}^+) \nonumber \\
    &\leq W(\xi,\hat{\xi}) - b_3|(\zeta,\varepsilon)|^2 + \sigma_W(|\omega|).
      \label{eq:smallgain:3}
  \end{align}
}%

\paragraph*{Robust exponential stability}
\ifthenelse{\boolean{LongVersion}}{%
  Substituting the upper bound~\cref{eq:smallgain:2} into the cost
  decrease~\cref{eq:smallgain:3} gives
  \begin{equation}\label{eq:smallgain:4}
    W(\xi^+,\hat{\xi}^+) \leq \lambda W(\xi,\hat{\xi}) -
    b_3|(\zeta,\varepsilon)|^2 + \sigma_W(|\omega|)
  \end{equation}
  where \(\lambda\defas 1-\frac{b_3}{b_2}\) and we can assume
  \(\lambda\in(0,1)\) since
  \[
    b_2 \geq qc_2 > qc_3 > qc_3 - \tilde{a}_4 \geq b_3.
  \]
  Recursively applying~\cref{eq:smallgain:4} gives
  \begin{align*}
    W(\xi(k),\hat{\xi}(k))
    &\leq \lambda^kW(\xi(0),\hat{\xi}(0)) + \sum_{i=1}^k
      \lambda^{i-1}\sigma(|\omega(k-i)|) \\
    &\leq b_2\lambda^k|(\zeta(0),\varepsilon(0))|^2 + \sum_{i=1}^k
      \lambda^{i-1}\sigma(|\omega(k-i)|)
  \end{align*}
  where the second inequality follows from~\cref{eq:smallgain:2}. Finally,
  by~\cref{eq:smallgain:1} and the triangle inequality, we have
  \begin{equation*}
    |(\zeta(k),e(k))|
    \leq c_\zeta\lambda_\zeta^k|(\zeta(0),\varepsilon(0))| + \sum_{i=1}^k
    \gamma_\zeta(|\omega(k-i)|,i)
  \end{equation*}
  where \(c_\zeta\defas \sqrt{\frac{b_2}{b_1}}\),
  \(\lambda_\zeta\defas\sqrt{\lambda}\), and \(\gamma_\zeta(s,k)\defas
  \lambda_\zeta^{k-1}\sqrt{\frac{\sigma(s)}{b_1}}\). %
}{%
  Note that \(\lambda \defas 1-\frac{b_3}{b_2}\in(0,1)\) since \(b_2 \geq qc_2 >
  qc_3 > qc_3 - \tilde{a}_4 \geq b_3\). Then
  \cref{eq:smallgain:2,eq:smallgain:3} give \(W(\xi^+,\hat{\xi}^+) \leq \lambda
  W(\xi,\hat{\xi}) - b_3|(\zeta,\varepsilon)|^2 + \sigma_W(|\omega|)\) at all
  times. Moreover, by \cref{eq:smallgain:2}, \(W(\xi(k),\hat{\xi}(k)) \leq
  b_2\lambda^k|(\zeta(0),\varepsilon(0))|^2 + \sum_{i=1}^k
  \lambda^{i-1}\sigma(|\omega(k-i)|)\) for all \(k\in\nnegint\).
  By~\cref{eq:smallgain:1} and the triangle inequality, we have
  \cref{eq:ras:ctrl:est} with \(\beta_\zeta(s,k) \defas
  c_\zeta\lambda_\zeta^ks\), \(c_\zeta \defas \sqrt{\frac{b_2}{b_1}}\),
  \(\lambda_\zeta \defas \sqrt{\lambda}\), and \(\gamma_\zeta(s,k) \defas
  \lambda_\zeta^{k-1}\sqrt{\frac{\sigma(s)}{b_1}}\). %
}%
\hspace*{\fill}~\QED

\ifthenelse{\boolean{LongVersion}}{%
\section{Proofs of offset-free MPC stability}\label[appendix]{app:nominal}
}{}
\subsection{Proof of \Cref{thm:mpc:nominal}}\label[appendix]{app:mpc:nominal}
In this proof and the subsequent proofs, we require some facts from the MPC
literature. From %
\ifthenelse{\boolean{LongVersion}}{%
  Proposition~2.4 of \cite{rawlings:mayne:diehl:2020}, %
}{%
  \cite[Prop.~2.4]{rawlings:mayne:diehl:2020}, %
}%
we have
\begin{equation}\label{eq:mpc:descent}
  V_N(x^+,\tilde{\bfu}(x,\beta),\beta) \leq
  V_N^0(x,\beta) - \ell(x,\kappa_N(x,\beta),\beta)
\end{equation}
for all \((x,\beta)\in\calS_N\), where \(x^+ \defas f_c(x,\beta)\) and %
\ifthenelse{\boolean{OneColumn}}{%
  \begin{equation}\label{eq:mpc:suboptimal}
    \tilde{\bfu}(x,\beta) \defas (u^0(1;x,\beta), \ldots,
    u^0(N-1;x,\beta), \kappa_f(x^0(N;x,\beta),\beta))
  \end{equation}
}{%
  \(\tilde{\bfu}(x,\beta) \defas (u^0(1;x,\beta), \ldots, u^0(N-1;x,\beta),
  \kappa_f(x^0(N;x,\beta),\beta))\) %
}%
is a suboptimal (yet feasible) sequence for \(x^+\) as the initial state.
Moreover, for each \((x,\beta)\in\calS_N\), the suboptimal sequence
\(\tilde{\bfu}(x,\beta)\) steers the system from \(f_c(x,\beta)\) to the
terminal constraint \(\bbX_f(\beta)\) and keeps it there (by
\Cref{assum:stabilizability}). Therefore
\(\tilde{\bfu}(x,\beta)\in\calU_N(f_c(x,\beta),\beta)\) and
\(f_c(x,\beta)\in\calX_N(\beta)\).

Throughout, fix \(x\in\calX_N^\rho(\beta)\) and \(\beta=(\rsp,\zsp,d)\in\calB\),
let \(\calB_c\subseteq\calB\) be compact, containing \(\beta\), and define
\(\delta r\defas g_c(x,\beta) - \rsp\) and \(\delta x\defas x-x_s(\beta)\).

\paragraph*{Part (a)}
Since \(\tilde{\bfu}(x,\beta)\) is feasible, %
\ifthenelse{\boolean{LongVersion}}{%
  \[
    V_N^0(f_c(x,\beta),\beta) \leq V_N(f_c(x,\beta),\tilde{\bfu}(x,\beta),\beta)
  \]
  and, applying the inequality \cref{eq:mpc:descent}, we have
  \[
    V_N^0(f_c(x,\beta),\beta) \leq V_N^0(x,\beta) -
    \ell(x,\kappa_N(x,\beta),\beta).
  \]
}{%
  \cref{eq:mpc:descent} implies
  \begin{align*}
    V_N^0(f_c(x,\beta),\beta)
    &\leq V_N(f_c(x,\beta),\tilde{\bfu}(x,\beta),\beta) \\
    &\leq V_N^0(x,\beta) - \ell(x,\kappa_N(x,\beta),\beta).
  \end{align*}
}%
But
\ifthenelse{\boolean{LongVersion}}{%
  \[
    \underline{\sigma}(Q)|x-x_s(\beta)|^2 \leq
    \ell(x,\kappa_N(x,\beta),\beta) \leq V_N^0(x,\beta)
  \]
}{%
  \(\underline{\sigma}(Q)|x-x_s(\beta)|^2 \leq \ell(x,\kappa_N(x,\beta),\beta)
  \leq V_N^0(x,\beta)\) %
}%
so the lower bound~\cref{eq:mpc:lyap:a} and the cost
decrease~\cref{eq:mpc:lyap:b} both hold with \(a_1 = a_3 =
\underline{\sigma}(Q)\). Only the upper bound of \cref{eq:mpc:lyap:a} remains.
Since \(P_f(\cdot)\) is continuous and positive definite, and \(\calB_c\) is
compact, the maximum \(\gamma \defas \max_{\beta\in\calB_c}
\overline{\sigma}(P_f(\beta)) > 0\) exists. Then \(|x-x_s(\beta)| \leq
\varepsilon \defas \sqrt{\frac{c_f}{\gamma}}\) implies %
\ifthenelse{\boolean{LongVersion}}{%
  \[
    V_f(x,\beta) \leq \overline{\sigma}(P_f(\beta))|x-x_s(\beta)|^2 \leq \gamma
    |x-x_s(\beta)|^2 \leq c_f
  \]
}{%
  \(V_f(x,\beta) \leq c_f\), %
}%
and therefore \(x\in\bbX_f(\beta)\). By monotonicity of the value
function~\cite[Prop.~2.18]{rawlings:mayne:diehl:2020}, we have \(V_N^0(x,\beta)
\leq V_f(x,\beta)\) whenever \(x\in\bbX_f(\beta)\), and therefore
\ifthenelse{\boolean{LongVersion}}{%
  \[
    V_N^0(x,\beta) \leq V_f(x,\beta) \leq \gamma|x-x_s(\beta)|^2
  \]
}{%
  \(V_N^0(x,\beta) \leq V_f(x,\beta) \leq \gamma|x-x_s(\beta)|^2\) %
}%
whenever \(|x-x_s(\beta)|\leq\varepsilon\). On the other hand, if
\(|x-x_s(\beta)|>\varepsilon\), then %
\ifthenelse{\boolean{LongVersion}}{%
  \[
    V_N^0(x,\beta) \leq \rho \leq \frac{\rho}{\varepsilon^2}|x-x_s(\beta)|^2.
  \]
}{%
  \(V_N^0(x,\beta) \leq \rho \leq
  \frac{\rho}{\varepsilon^2}|x-x_s(\beta)|^2\). %
}%
Finally, we have the upper bound~\cref{eq:mpc:lyap:a} with
\(a_2\defas\max\set{ \gamma, \frac{\rho}{\varepsilon^2} }\).

%% Exponential stability
\paragraph*{Part (b)}
We already have %
\ifthenelse{\boolean{LongVersion}}{%
  that \(V_N^0(\cdot,\beta)\) is a Lyapunov function (for the system
  \cref{eq:model:cl}, on \(\calX_N^\rho(\beta)\)) with respect to
  \(x-x_s(\beta)\), and %
}{}%
\(f_c(x,\beta)\in\calX_N(\beta)\) for all \(x\in\calX_N^\rho(\beta)\) by
recursive feasibility. We can choose any compact set \(\calB_c\subseteq\calB\)
containing \(\beta\) to achieve the descent property \cref{eq:mpc:lyap:b}. Then,
for each \(x\in\calX_N^\rho(\beta)\), we have %
\ifthenelse{\boolean{LongVersion}}{%
  \[
    V_N^0(f_c(x,\beta),\beta) \leq V_N^0(x,\beta) - a_1|x-x_s(\beta)|^2 \leq
    \rho
  \]
}{%
  \(V_N^0(f_c(x,\beta),\beta) \leq V_N^0(x,\beta) - a_1|x-x_s(\beta)|^2 \leq
  \rho\) %
}%
and therefore \(f_c(x,\beta)\in\calX_N^\rho(\beta)\). %
\ifthenelse{\boolean{LongVersion}}{%
  In other words, \(\calX_N^\rho(\beta)\) is positive invariant for the
  system~\cref{eq:model:cl:a}. %
}{}%
Finally, ES in \(\calX_N^\rho(\beta)\) w.r.t.~\(x-x_s(\beta)\) follows
from~\Cref{thm:lyap}\ifthenelse{\boolean{LongVersion}}{}{ and part (a)}.

\paragraph*{Intermediate results}
Consider the following propositions.

\begin{proposition}%
  [\protect{\cite[Prop.~20]{allan:bates:risbeck:rawlings:2017}}]%
  \label{prop:cont}%
  Let \(C\subseteq D\subseteq\real^m\), with \(C\) compact, \(D\) closed, and
  \(V : D \rightarrow \real^p\) continuous. Then there exists
  \(\alpha\in\calKinf\) such that \(|V(x)-V(y)|\leq\alpha(|x-y|)\) for all
  \(x\in C\) and \(y\in D\).
\end{proposition}

\begin{proposition}\label{prop:bound:xu}
  Suppose \Cref{assum:cont,assum:cons,assum:sstp:exist,assum:stabilizability,%
    assum:quad} hold. Let \(\rho>0\) and \(\calB_c\subseteq\calB\) be compact.
  There exist \(c_x,c_u>0\) such that
  \begin{subequations}\label{eq:bound:xu}
    \begin{align}
      |x^0(j;x,\beta) - x_s(\beta)| &\leq c_x|x - x_s(\beta)| \label{eq:bound:x} \\
      |u^0(k;x,\beta) - u_s(\beta)| &\leq c_u|x - x_s(\beta)| \label{eq:bound:u}
    \end{align}
  \end{subequations}
  for each \(x\in\calX_N^\rho(\beta)\), \(\beta\in\calB_c\),
  \(j\in\intinterval{1}{N}\), and \(k\in\intinterval{1}{N-1}\).
\end{proposition}
\begin{proof}
  Throughout, we fix \(x\in\calX_N^\rho(\beta)\) and \(\beta\in\calB_c\). %
  \ifthenelse{\boolean{LongVersion}}{%
    Unless otherwise specified, the constructed constants and functions are
    independent of \((x,\beta)\). %
  }{} %
  By \Cref{thm:mpc:nominal}(a), there exists \(a_2>0\) satisfying the upper
  bound \cref{eq:mpc:robust:lyap:a}. Since \(P_f\) is continuous and positive
  definite and \(\calB_c\) is compact, the minimum
  \(\gamma\defas\min_{\beta\in\calB_c} \underline{\sigma}(P_f(\beta))\) exists
  and is positive. Moreover, since \(Q,R\) are positive definite, we have
  \(\underline{\sigma}(Q),\underline{\sigma}(R)>0\). For each
  \(k\in\intinterval{0}{N-1}\),
  \begin{align*}
    \underline{\sigma}(Q)|x^0(k;x,\beta)-x_s(\beta)|^2
    &\leq |x^0(k;x,\beta)-x_s(\beta)|_Q^2 \\
    &\leq V_N^0(x,\beta) \leq a_2|x-x_s(\beta)|^2 \\
    \gamma|x^0(N;x,\beta)-x_s(\beta)|^2
    &\leq |x^0(N;x,\beta)-x_s(\beta)|_{P_f(\beta)}^2 \\
    &\leq V_N^0(x,\beta) \leq a_2|x-x_s(\beta)|^2 \\
    \underline{\sigma}(R)|u^0(k;x,\beta)-u_s(\beta)|^2
    &\leq |u^0(k;x,\beta)-u_s(\beta)|_R^2 \\
    &\leq V_N^0(x,\beta) \leq a_2|x-x_s(\beta)|^2.
  \end{align*}
  Thus, \cref{eq:bound:xu} holds for all \(j\in\intinterval{1}{N}\) and
  \(k\in\intinterval{1}{N-1}\) with
  \(c_x\defas\max\set{\sqrt{\frac{a_2}{\underline{\sigma}(Q)}},
    \sqrt{\frac{a_2}{\gamma}}}\) and
  \(c_u\defas\sqrt{\frac{a_2}{\underline{\sigma}(R)}}\).
\end{proof}

\begin{proposition}\label{prop:bound:ref}
  Suppose \Cref{assum:cont,assum:cons,assum:sstp:exist,assum:stabilizability,%
    assum:quad} hold. Let \(\rho>0\), \(\calB_c\subseteq\calB\) be compact.
  There exists \(\sigma_r\in\calKinf\) such that
  \begin{equation}\label{eq:bound:ref}
    |g_c(x,\beta) - \rsp| \leq \sigma_r(|x-x_s(\beta)|)
  \end{equation}
  for each \(x\in\calX_N^\rho(\beta)\) and \(\beta = (\rsp,\zsp,d) \in
  \calB_c\). Moreover, if \(g\) and \(h\) are Lipschitz continuous on bounded
  sets, then \cref{eq:bound:ref} holds on the same sets with \(\sigma_r \defas
  c_r\id\) and some \(c_r>0\).
\end{proposition}
\begin{proof}
  By \Cref{prop:cont},
  there exists \(\tilde\sigma_r\in\calKinf\) such that %
  %% NOTE This ifthenelse statement just cancels out any latexdiff markup in
  %% this section. Otherwise, the spacing gets wonky and text spills onto the
  %% next page.
  \ifthenelse{\boolean{true}}{%
    \begin{equation*}
      |g(u,h(z,d)) - g(\tilde{u},h(\tilde{z},\tilde{d}))| \leq
      \tilde\sigma_r(|(z,\beta)-(\tilde{z},\tilde{\beta})|)
    \end{equation*}
    for all \(z=(x,u),\tilde{z}=(\tilde{x},\tilde{u})\in\calX_N^\rho\times\bbU\),
    and \(\beta=(s,d),\tilde\beta=(\tilde{s},\tilde{d}) \in \calB_c\). %
  }{}%
  Fix \(x\in\calX_N^\rho(\beta)\) and \(\beta\in\calB_c\). %
  \ifthenelse{\boolean{LongVersion}}{%
    The following constructions are independent of \((x,\beta)\) unless otherwise
    specified. %
  }{} %
  By \Cref{prop:bound:xu}, there exists \(c_u>0\) such that %
  \ifthenelse{\boolean{LongVersion}}{%
    \[
      |\kappa_N(x,\beta) - u_s(\beta)| \leq c_u|x-x_s(\beta)|
    \]
  }{%
    \(|\kappa_N(x,\beta) - u_s(\beta)| \leq c_u|x-x_s(\beta)|\). %
  }%
  \ifthenelse{\boolean{LongVersion}}{%
    Combining these inequalities gives %
    \begin{align*}
      |g_c(x,\beta) - \rsp|
      &\leq \tilde\sigma_r(|(x - x_s(\beta), \kappa_N(x,\beta) - u_s(\beta))|) \\
      &\leq \tilde\sigma_r((1+c_u)|x - x_s(\beta)|) \\
      &\leq \sigma_r(|x - x_s(\beta)|)
    \end{align*}
  }{%
    Moreover, \(|g_c(x,\beta) - \rsp| \leq \sigma_r(|x - x_s(\beta)|)\) %
  }%
  where \(\sigma_r\defas\tilde\sigma_r\circ(1+c_u)\id\in\calKinf\). If,
  additionally, \(g\) and \(h\) are Lipschitz on bounded sets, then we can take
  \(\sigma_r \defas c_r\id\) and \(c_r\defas L_r(1+c_u) > 0\), where \(L_r>0\)
  is the Lipschitz constant for \(g(u,h(x,u,d))\) over
  \(\calX_N^\rho\times\bbU\times\calB_c\).
\end{proof}

\paragraph*{Part (c)}
\Cref{prop:bound:ref} gives \(\sigma_r\in\calKinf\) satisfying
\cref{eq:bound:ref}. Then %
\ifthenelse{\boolean{LongVersion}}{%
  \[
    \alpha_1(|\delta r|) \leq a_1|\delta x|^2 \leq V_N^0(x,\beta)
  \]
  where \(\alpha_1(\cdot) \defas a_1[\sigma_r^{-1}(\cdot)]^2 \in \calKinf\), so
  \(V_N^0(\cdot,\beta)\) is a Lyapunov function on \(\calX_N^\rho(\beta)\)
  w.r.t.~\((\delta r,\delta x)\), %
}{%
  \(\alpha_1(|\delta r|) \leq a_1|\delta x|^2 \leq V_N^0(x,\beta)\) with
  \(\alpha_1 \defas a_1[\sigma_r^{-1}]^2 \in \calKinf\), %
}%
and AS on \(\calX_N^\rho(\beta)\) w.r.t.~\((\delta r,\delta x)\) follows by
\Cref{thm:lyap}.

\paragraph*{Part (d)}
If \(g\) and \(h\) are Lipschitz continuous on bounded sets,
then by \Cref{prop:bound:ref}, we can repeat part (c) with \(\alpha_1
\defas a_1c_r^{-2}\id^2\) and some \(c_r>0\). Then \(V_N^0(\cdot,\beta)\) is
an exponential Lyapunov function on \(\calX_N^\rho(\beta)\) w.r.t.~\((\delta
r,\delta x)\), and ES on \(\calX_N^\rho(\beta)\) w.r.t.~\((\delta r,\delta x)\)
follows by \Cref{thm:lyap}. %
\hspace*{\fill}~\QED

\subsection{Proof of \Cref{thm:mpc:robust}}\label[appendix]{app:mpc:robust}
% To prove \Cref{thm:mpc:robust},
We require two preliminary results. %
%% NOTE This ifthenelse statement just cancels out any latexdiff markup in
%% this section. Otherwise, the spacing gets wonky and text spills onto the
%% next page.
First, in \Cref{prop:mpc:robust:feas} (adapted from the proof %
\ifthenelse{\boolean{LongVersion}}{%
  of Theorem~21 of \cite{allan:bates:risbeck:rawlings:2017}), %
}{%
  of~\cite[Thm.~21]{allan:bates:risbeck:rawlings:2017}), %
}%
we establish (a) recursive feasibility of the FHOCP, (b) the cost decrease
\begin{equation}\label{eq:mpc:robust:cost:decr}
  V_N(\hat{x}^+,\tilde{\bfu}(\hat{x},\hat{\beta}),\hat{\beta}^+) \leq
  V_N^0(\hat{x},\hat{\beta}) - a_3|\delta\hat{x}|^2 + \sigma_r(|\tilde{d}|)
\end{equation}
where \(a_3>0\), \(\sigma_r\in\calKinf\), and \(\delta\hat{x} \defas
\hat{x}-x_s(\hat{\beta})\), and (c) robust positive invariance of
\(\calX_N^\rho(\hat{\beta})\), given feasibility of the SSTP and sufficiently
small \(\tilde{d}\in\tilde{\bbD}_c(\hat{x},\hat{\beta})\). Second, in
\Cref{prop:ref:robust}, we establish bounds on the reference signal errors.

\ifthenelse{\boolean{LongVersion}}{%
\subsubsection{Suboptimal cost decrease and robust positive invariance}
}{}
\begin{proposition}\label{prop:mpc:robust:feas}
  %% NOTE This ifthenelse statement just cancels out any latexdiff markup in
  %% this section. Otherwise, the spacing gets wonky and text spills onto the
  %% next page.
  Suppose \Cref{assum:cont,assum:cons,assum:sstp:exist,assum:stabilizability,%
    assum:quad,assum:sstp} hold and let \(\rho>0\). There exists
  \(\sigma_r\in\calKinf\) and \(a_3,\delta>0\) such that%
  \ifthenelse{\boolean{LongVersion}}{%
    \begin{enumerate}[(a)]
    \item \(\tilde{\bfu}(\hat{x},\hat{\beta}) \in
      \calU_N(\hat{x}^+,\hat{\beta}^+)\),
    \item \cref{eq:mpc:robust:cost:decr} holds, and
    \item \(\hat{x}^+ \in \calX_N^\rho(\hat{\beta}^+)\),
    \end{enumerate}
    for all \(\hat{\beta}\in\hat{\calB}_c\),
    \(\hat{x}\in\calX_N^\rho(\hat{\beta})\) and
    \(\tilde{d}\in\tilde{\bbD}_c(\hat{x},\hat{\beta}) \cap
    \delta\bbB^{n_{\tilde{d}}}\), %
  }{%
    , for each \(\hat{\beta}\in\hat{\calB}_c\),
    \(\hat{x}\in\calX_N^\rho(\hat{\beta})\) and
    \(\tilde{d}\in\tilde{\bbD}_c(\hat{x},\hat{\beta}) \cap
    \delta\bbB^{n_{\tilde{d}}}\), we have (a) \(\tilde{\bfu}(\hat{x},\hat{\beta}) \in
    \calU_N(\hat{x}^+,\hat{\beta}^+)\), (b) \cref{eq:mpc:robust:cost:decr}
    holds, and (c) \(\hat{x}^+ \in \calX_N^\rho(\hat{\beta}^+)\), %
  }%
  %% NOTE This ifthenelse statement just cancels out any latexdiff markup in
  %% this section. Otherwise, the spacing gets wonky and text spills onto the
  %% next page.
  % \ifthenelse{\boolean{true}}{%
    where \(\hat{x}^+ \defas \hat{f}_c(\hat{x},\hat{\beta},\tilde{d})\) and
    \(\hat{\beta}^+ \defas \hat{f}_{\beta,c}(\hat{\beta},\tilde{d})\). %
  % }{}
\end{proposition}

\begin{proof}
  % \subsection{Proof of \Cref{prop:mpc:robust:feas}}\label[appendix]{app:robust}
  First, we aim to show the set %
  \ifthenelse{\boolean{LongVersion}}{%
    \begin{equation*}
      \hat{\calX}_N^\rho \defas \bigcup_{\hat{\beta}\in\hat{\calB}_c}
      \calX_N^\rho(\hat{\beta})
    \end{equation*}
    is compact, where \(\hat{\calB}_c\) is defined as in \Cref{assum:sstp}(a). %
  }{%
    \(\hat{\calX}_N^\rho \defas \bigcup_{\hat{\beta}\in\hat{\calB}_c}
    \calX_N^\rho(\hat{\beta})\) is compact. %
  }%
  Consider the lifted set \ifthenelse{\boolean{OneColumn}}{%
    \[
      \mathcal{F} \defas \set{ (\hat{x},\bfu,\hat{\beta}) \in \bbX \times \bbU^N
        \times \hat{\calB}_c | V_f(\phi(N;\hat{x},\bfu,\hat{\beta})) \leq c_f,\;
        V_N(\hat{x},\bfu,\hat{\beta}) \leq \rho }.
    \]
  }{%
    %% NOTE For some reason multline is producing a lot of space above the
    %% equation, so I simulated the behavior with an align environment. If the
    %% equation is changed, just be careful to adjust the alignment.
    % \begin{multline*}
    %   \mathcal{F} \defas \{\; (\hat{x},\bfu,\hat{\beta}) \in \bbX \times \bbU^N \times
    %   \hat{\calB}_c \;|\\ V_f(\phi(N;\hat{x},\bfu,\hat{\beta})) \leq c_f,\;
    %   V_N(\hat{x},\bfu,\hat{\beta}) \leq \rho \;\}.
    % \end{multline*}
    \begin{align*}
      \mathcal{F} \defas \{\; (\hat{x},\bfu,\hat{\beta}
      &) \in \bbX \times \bbU^N \times \hat{\calB}_c \;|\\
      V_f&(\phi(N;\hat{x},\bfu,\hat{\beta})) \leq c_f,\;
           V_N(\hat{x},\bfu,\hat{\beta}) \leq \rho \;\}.
    \end{align*}
  }%
  Notice \(\hat{\calX}_N^\rho\) is equivalent to the projection of
  \(\mathcal{F}\) onto the first coordinate, i.e., \(\hat{\calX}_N^\rho =
  P(\mathcal{F})\) where \(P(\hat{x},\bfu,\hat{\beta})=\hat{x}\). Since \(P\) is
  continuous, the image \(\hat{\calX}_N^\rho = P(\mathcal{F})\) is compact
  whenever \(\mathcal{F}\) is compact. Thus, it suffices to show \(\mathcal{F}\)
  is compact.

  The set \(\mathcal{F}\) is closed because \((\bbX,\bbU,\hat{\calB}_c)\) are
  closed, and continuity of \((f,x_s,u_s,\ell,V_f)\) implies continuity of
  \(V_f(\phi(N;\cdot,\cdot,\cdot))\) and \(V_N(\cdot,\cdot,\cdot)\). Next, we
  show \(\mathcal{F}\) is bounded. Since \(x_s\) is continuous and
  \(\hat{\calB}_c\) is compact, the maximum
  \(\rho_s\defas\max_{\hat{\beta}\in\hat{\calB}_c} |x_s(\hat{\beta})|\) exists
  and is finite. For each \((\hat{x},\bfu,\hat{\beta})\in\mathcal{F}\), we have
  \(V_N^0(\hat{x},\hat{\beta})\leq V_N(\hat{x},\bfu,\hat{\beta}) \leq \rho\) by
  construction. But \(V_N^0(\hat{x},\hat{\beta}) \geq
  \underline{\sigma}(Q)|\hat{x}-x_s(\hat{\beta})|^2\), so this implies
  \(|\hat{x}-x_s(\hat{\beta})| \leq \sqrt{\frac{\rho}{\underline{\sigma}(Q)}}\)
  and therefore \(|\hat{x}| \leq \sqrt{\frac{\rho}{\underline{\sigma}(Q)}} +
  \rho_s\). But \(\bfu\) and \(\hat{\beta}\) always lie in compact sets, so
  \(\mathcal{F}\) is bounded and \(\hat{\calX}_N^\rho\) is compact.

  For the rest of the proof, we fix \(\hat{\beta}\in\hat{\calB}_c\),
  \(\hat{x}\in\calX_N^\rho(\hat{\beta})\), and \(|\tilde{d}|\leq\delta_0\) such
  that \(\hat{\beta}^+ \defas \hat{f}_{\beta,c}(\hat{\beta},\tilde{d}) \in
  \hat{\calB}_c\). For brevity, let %
  \ifthenelse{\boolean{OneColumn}}{%
    \begin{align*}
      \tilde{\bfu} &\defas \tilde{\bfu}(\hat{x},\hat{\beta}),
      & \overline{x}^+ &\defas f_c(\hat{x},\hat{\beta}),
      & \overline{x}^+(N) &\defas \phi(N;\overline{x}^+,\tilde{\bfu},\hat{d}), \\
      \overline{x}(N) &\defas x^0(N;\hat{x},\hat{\beta}),
      & \hat{x}^+ &\defas \hat{f}_c(\hat{x},\hat{\beta},\tilde{d}),
      & \hat{x}^+(N) &\defas \phi(N;\hat{x}^+,\tilde{\bfu},\hat{d}^+).
    \end{align*}
  }{%
    %% TODO: Decide on more space, better readability, or less space, harder to
    %% read.
    % \begin{align*}
    %   \tilde{\bfu} &\defas \tilde{\bfu}(\hat{x},\hat{\beta}),
    %   & \overline{x}(N) &\defas x^0(N;\hat{x},\hat{\beta}), \\
    %   \overline{x}^+ &\defas f_c(\hat{x},\hat{\beta}),
    %   & \overline{x}^+(N) &\defas \phi(N;\overline{x}^+,\tilde{\bfu},\hat{d}), \\
    %   \hat{x}^+ &\defas \hat{f}_c(\hat{x},\hat{\beta},\tilde{d}),
    %   & \hat{x}^+(N) &\defas \phi(N;\hat{x}^+,\tilde{\bfu},\hat{d}^+).
    % \end{align*}
    \(\tilde{\bfu} \defas \tilde{\bfu}(\hat{x},\hat{\beta})\), \(\overline{x}^+
    \defas f_c(\hat{x},\hat{\beta})\), \(\overline{x}^+(N) \defas
    \phi(N;\overline{x}^+,\tilde{\bfu},\hat{d})\), \(\hat{x}^+ \defas
    \hat{f}_c(\hat{x},\hat{\beta},\tilde{d})\), \(\hat{x}^+(N) \defas
    \phi(N;\hat{x}^+,\tilde{\bfu},\hat{d}^+)\), and \(\overline{x}(N) \defas
    x^0(N;\hat{x},\hat{\beta})\). %
  }%
  Recall \(\tilde{d} \defas (e,e^+,\Delta\beta,w,v)\), \(e\defas (e_x,e_d)\),
  \(e^+\defas (e_x^+,e_d^+)\), \(\Delta\beta \defas (\Delta\ssp,w_d)\), and
  \(\hat{\calX}_N^\rho\) is compact. Since \((f,x_s,u_s,P_f)\) are continuous,
  so are \((V_f,V_N)\). By \Cref{prop:cont}, there exist
  \(\sigma_f,\sigma_{V_f},\sigma_{V_N}\in\calKinf\) such that
  \ifthenelse{\boolean{OneColumn}}{%
    \begin{gather}
      |f(x_1,u_1,\hat{d}_1) - f(x_2,u_2,\hat{d}_2)|
      \leq \sigma_f(|(x_1,u_1,\hat{d}_1)-(x_2,u_2,\hat{d}_2)|)
      \label{eq:mpc:robust:f} \\
      |V_f(\phi(N;x_1,\bfu_1,\hat{d}_1),\hat{\beta}_1) -
      V_f(\phi(N;x_2,\bfu_2,\hat{d}_2),\hat{\beta}_2)|
      \leq \sigma_{V_f}(|(x_1-x_2, \bfu_1-\bfu_2, \hat{\beta}_1-\hat{\beta}_2)|)
      \label{eq:mpc:robust:Vf} \\
      |V_N(x_1,\bfu_1,\hat{\beta}_1) - V_N(x_2,\bfu_2,\hat{\beta}_2)| \leq
      \sigma_{V_N}(|(x_1-x_2, \bfu_1-\bfu_2, \hat{\beta}_1-\hat{\beta}_2)|)
      \label{eq:mpc:robust:VN}
    \end{gather}
  }{%
    %% NOTE For some reason multline is producing a lot of space between the
    %% equations, so I simulated the behavior with an align environment. If the
    %% equations are changed, just be careful to adjust the alignment.
    % \begin{multline}\label{eq:mpc:robust:f}
    %   |f(x_1,u_1,\hat{d}_1) - f(x_2,u_2,\hat{d}_2)| \\
    %   \leq \sigma_f(|(x_1,u_1,\hat{d}_1)-(x_2,u_2,\hat{d}_2)|)
    % \end{multline}
    % \begin{multline}\label{eq:mpc:robust:Vf}
    %   |V_f(\phi(N;x_1,\bfu_1,\hat{d}_1),\hat{\beta}_1) -
    %   V_f(\phi(N;x_2,\bfu_2,\hat{d}_2),\hat{\beta}_2)| \\
    %   \leq \sigma_{V_f}(|(x_1,\bfu_1,\hat{\beta}_1)-(x_2,\bfu_2,\hat{\beta}_2)|)
    % \end{multline}
    % \begin{multline}\label{eq:mpc:robust:VN}
    %   |V_N(x_1,\bfu_1,\hat{\beta}_1) - V_N(x_2,\bfu_2,\hat{\beta}_2)| \\
    %   \leq \sigma_{V_N}(|(x_1,\bfu_1,\hat{\beta}_1)-(x_2,\bfu_2,\hat{\beta}_2)|)
    % \end{multline}
    \begin{align}
      |f(x_1,u_1,\hat{d}_1) - &f(x_2,u_2,\hat{d}_2)| \nonumber \\
        &\leq \sigma_f(|(x_1,u_1,\hat{d}_1)-(x_2,u_2,\hat{d}_2)|)
          \label{eq:mpc:robust:f} \\
      |V_f(\phi(N;x_1,\bfu_1&,\hat{d}_1),\hat{\beta}_1) -
            V_f(\phi(N;x_2,\bfu_2,\hat{d}_2),\hat{\beta}_2)| \nonumber \\
        &\leq \sigma_{V_f}(|(x_1,\bfu_1,\hat{\beta}_1)-(x_2,\bfu_2,\hat{\beta}_2)|)
          \label{eq:mpc:robust:Vf} \\%\end{align}\begin{align}%\\
      |V_N(x_1,\bfu_1,\hat{\beta}_1) &- V_N(x_2,\bfu_2,\hat{\beta}_2)| \nonumber \\
          &\leq \sigma_{V_N}(|(x_1,\bfu_1,\hat{\beta}_1)-(x_2,\bfu_2,\hat{\beta}_2)|)
          \label{eq:mpc:robust:VN}
    \end{align}
  }%
  for all \(x_1\in\bbX\), \(x_2\in\hat{\calX}_N^\rho\), \(u_1,u_2\in\bbU\),
  \(\bfu_1,\bfu_2\in\bbU^N\), and \(\hat{\beta}_1=(s_1,\hat{d}_1),
  \hat{\beta}_2=(s_2,\hat{d}_2) \in \hat{\calB}_c\).

  Substituting \(x_1=\hat{x}+e_x\), \(x_2=\hat{x}\),
  \(u_1=u_2=\kappa_N(\hat{x},\hat{\beta})\), \(\hat{d}_1=\hat{d}+e_d\), and
  \(\hat{d}_2=\hat{d}\) into \cref{eq:mpc:robust:f}, we have %
  \ifthenelse{\boolean{LongVersion}}{%
    \[
      |\hat{x}^+ - \overline{x}^+| \leq \sigma_f(|e|) + |w| + |e_x^+|.
    \]
  }{%
    \(|\hat{x}^+ - \overline{x}^+| \leq \sigma_f(|e|) + |w| + |e_x^+|\). %
  }%
  But \(|\hat{\beta}^+-\hat{\beta}| \leq |\Delta\beta| + |e_d| + |e_d^+|\), so
  \begin{equation}\label{eq:mpc:robust:feas:1}
    |(\hat{x}^+,\hat{\beta}^+) - (\overline{x}^+,\hat{\beta})| \leq
    \sigma_f(\tilde{d}) + 5|\tilde{d}|.
  \end{equation}
  Substituting \(x_1=\hat{x}^+\), \(x_2=\hat{f}_c(\hat{x},\hat{\beta})\),
  \(\bfu_1=\bfu_2=\tilde{\bfu}\), \(\hat{\beta}_1=\hat{\beta}^+\), and
  \(\hat{\beta}_2=\hat{\beta}\) into \cref{eq:mpc:robust:Vf,eq:mpc:robust:VN}
  gives %
  \begin{align}
    |V_f(\hat{x}^+ \ifthenelse{\boolean{OneColumn}}{}{
    &} (N),\hat{\beta}^+) - V_f(\overline{x}^+(N),\hat{\beta})|
      \ifthenelse{\boolean{OneColumn}}{}{\nonumber \\}
    &\leq \sigma_{V_f}(|(\hat{x}^+,\hat{\beta}^+) - (\overline{x}^+,\hat{\beta})|)
      \ifthenelse{\boolean{LongVersion}}{\nonumber \\
    &}{}
      \leq \tilde\sigma_{V_f}(|\tilde{d}|) \label{eq:mpc:robust:feas:2}
    \\
    |V_N(\hat{x}^+ \ifthenelse{\boolean{OneColumn}}{}{
    &} ,\tilde{\bfu},\hat{\beta}^+) -
      V_N(\overline{x}^+,\tilde{\bfu},\hat{\beta})|
      \ifthenelse{\boolean{OneColumn}}{}{\nonumber \\}
    &\leq \sigma_{V_N}(|(\hat{x}^+,\hat{\beta}^+) - (\overline{x}^+,\hat{\beta})|)
      \ifthenelse{\boolean{LongVersion}}{\nonumber \\
    &}{}
      \leq \sigma_r(|\tilde{d}|) \label{eq:mpc:robust:cost:decr:1}
  \end{align}
  where \(\tilde\sigma_{V_f} \defas \sigma_{V_f}\circ(\sigma_f+5\id), \sigma_r
  \defas \sigma_{V_N}\circ(\sigma_f+5\id) \in \calKinf\), and the second and
  fourth inequalities follow from~\cref{eq:mpc:robust:feas:1}.

\paragraph*{Part (a)}
By definitions \cref{eq:mpc:terminal,eq:mpc:admit,eq:mpc:admit:inputs},
\(\tilde{\bfu} \in \calU_N(\hat{x}^+,\hat{\beta}^+)\) if and only if
\(V_f(\hat{x}^+(N),\hat{\beta}^+) \leq c_f\). Thus, it suffices to construct
\(\delta_1>0\) (independently of \(\hat{\beta}\) and \(\tilde{d}\)) for which
\(\hat{x}\in\calX_N(\hat{\beta})\) implies \(V_f(\hat{x}^+(N),\hat{\beta}^+)
\leq c_f\). Since \(\hat{x}\in\calX_N(\hat{\beta})\), we already have
\(V_f(\overline{x}(N),\hat{\beta})\leq c_f\), and by
\Cref{assum:stabilizability,assum:quad},
\begin{align*}
  V_f(\overline{x}^+(N),\hat{\beta})
  &\leq V_f(\overline{x}(N),\hat{\beta}) -
    \ell(\overline{x}(N),\kappa_f(\overline{x}(N),\hat{\beta}),\hat{\beta}) \\
  &\leq V_f(\overline{x}(N),\hat{\beta}) -
    \underline{\sigma}(Q)|\overline{x}(N)-x_s(\hat{\beta})|^2.
\end{align*}
Since \(\hat{\calB}_c\) is compact and \(\overline{\sigma},P_f\) are
continuous functions, the maximum %
\ifthenelse{\boolean{LongVersion}}{%
  \[
    a_{f,2} \defas \max_{\hat{\beta}\in\hat{\calB}_c}
    \overline{\sigma}(P_f(\hat{\beta}))
  \]
}{%
  \(a_{f,2}\defas\max_{\hat{\beta}\in\hat{\calB}_c}
  \overline{\sigma}(P_f(\hat{\beta}))\) %
}%
exists and is finite, so %
\ifthenelse{\boolean{LongVersion}}{%
  \[
    \frac{c_f}{2} \leq V_f(\overline{x}(N),\hat{\beta}) \leq
    a_{f,2}|\overline{x}(N)-x_s(\hat{\beta})|^2.
  \]
  Then \(|\overline{x}(N)-x_s(\hat{\beta})|\geq \sqrt{\frac{c_f}{2a_{f,2}}}\)
  and
  \begin{equation}\label{eq:mpc:robust:feas:3}
    V_f(\overline{x}^+(N),\hat{\beta}) \leq c_f -
    \frac{c_f\underline{\sigma}(Q)}{2a_{f,2}}.
  \end{equation}
  On the other hand, if \(V_f(\overline{x}(N),\hat{\beta})\leq
  \frac{c_f}{2}\), then we have
  \begin{equation}\label{eq:mpc:robust:feas:4}
    V_f(\overline{x}^+(N),\hat{\beta}) \leq \frac{c_f}{2}.
  \end{equation}
  Finally, combining
  \cref{eq:mpc:robust:feas:2,eq:mpc:robust:feas:3,eq:mpc:robust:feas:4}, we
  have %
}{%
  \(V_f(\overline{x}(N),\hat{\beta}) \leq
  a_{f,2}|\overline{x}(N)-x_s(\hat{\beta})|^2\). Then if
  \(V_f(\overline{x}(N),\hat{\beta})\geq \frac{c_f}{2}\), we have
  \(|\overline{x}(N)-x_s(\hat{\beta})|\geq \sqrt{\frac{c_f}{2a_{f,2}}}\) and
  \(V_f(\overline{x}^+(N),\hat{\beta}) \leq c_f -
  \frac{c_f\underline{\sigma}(Q)}{2a_{f,2}}\). On the other hand, if
  \(V_f(\overline{x}(N),\hat{\beta})\leq \frac{c_f}{2}\), then we have
  \(V_f(\overline{x}^+(N),\hat{\beta}) \leq \frac{c_f}{2}\). In summary, }
\[
  V_f(\hat{x}^+(N),\hat{\beta}^+) \leq c_f - \gamma_f +
  \tilde\sigma_{V_f}(|\tilde{d}|)
\]
where \(\gamma_f\defas\min\set{\frac{c_f}{2},
  \frac{c_f\underline{\sigma}(Q)}{2a_{f,2}}}\) was defined independently of
\((\hat{\beta},\tilde{d})\). Finally, taking
\(\delta_1\defas\min\set{\delta_0,\tilde\sigma_{V_f}^{-1}(\gamma_f)}\), we have
\(V_f(\hat{x}^+(N),\hat{\beta}^+) \leq c_f\) and \(\tilde{\bfu} \in
\calU_N(\hat{x}^+,\hat{\beta}^+)\).

\paragraph*{Part (b)}
By \cref{eq:mpc:descent}, we have
\ifthenelse{\boolean{LongVersion}}{%
  \begin{equation}\label{eq:mpc:robust:cost:decr:2}
    V_N(\overline{x}^+,\tilde{\bfu},\hat{\beta})
    \leq V_N^0(\hat{x},\hat{\beta}) -
    \ell(\hat{x},\kappa_N(\hat{x},\hat{\beta}),\hat{\beta})
    \leq V_N^0(\hat{x},\hat{\beta}) -
    \underline{\sigma}(Q)|\overline{x}(N)-x_s(\hat{\beta})|^2.
  \end{equation}
}{%
  \begin{align}
    V_N(\overline{x}^+,\tilde{\bfu},\hat{\beta})
    &\leq V_N^0(\hat{x},\hat{\beta}) -
      \ell(\hat{x},\kappa_N(\hat{x},\hat{\beta}),\hat{\beta}) \nonumber \\
    &\leq V_N^0(\hat{x},\hat{\beta}) -
      \underline{\sigma}(Q)|\overline{x}(N)-x_s(\hat{\beta})|^2.
      \label{eq:mpc:robust:cost:decr:2}
  \end{align}
}
Combining \cref{eq:mpc:robust:cost:decr:1,eq:mpc:robust:cost:decr:2} gives
\cref{eq:mpc:robust:cost:decr} with \(a_3\defas\underline{\sigma}(Q)\),
which is positive since \(Q\) is positive definite.

\paragraph*{Part (c)}
\ifthenelse{\boolean{LongVersion}}{%
  The proof of this part follows similarly that of part (a). %
}{}%
Since \(\hat{x}\in\calX_N^\rho(\hat{\beta})\), we have
\(V_N^0(\hat{x},\hat{\beta})\leq\rho\). If
\(V_N^0(\hat{x},\hat{\beta})\geq\frac{\rho}{2}\), then, by
\Cref{thm:mpc:nominal}(a), we have %
\ifthenelse{\boolean{LongVersion}}{%
  \[
    \frac{\rho}{2} \leq V_N^0(\hat{x},\hat{\beta}) \leq a_2|\hat{x} -
    x_s(\hat{\beta})|^2
  \]
  for some \(a_2>0\). Therefore \(|\hat{x} -
  x_s(\hat{\beta})|\leq\sqrt{\frac{\rho}{2a_2}}\) and
  \begin{equation}\label{eq:mpc:robust:rpi:2}
    V_N(\overline{x}^+,\tilde{\bfu},\hat{\beta}) \leq
    \rho - \frac{\rho\underline{\sigma}(Q)}{2a_2}.
  \end{equation}
  On the other hand, if \(V_N^0(\hat{x},\hat{\beta})\leq\frac{\rho}{2}\), then
  \begin{equation}\label{eq:mpc:robust:rpi:3}
    V_N(\overline{x}^+,\tilde{\bfu},\hat{\beta}) \leq
    \frac{\rho}{2}.
  \end{equation}
  Combining
  \cref{eq:mpc:robust:cost:decr,eq:mpc:robust:rpi:2,eq:mpc:robust:rpi:3}
  gives %
}{%
  \(V_N^0(\hat{x},\hat{\beta}) \leq a_2|\hat{x} - x_s(\hat{\beta})|^2\)
  for some \(a_2>0\). Therefore \(|\hat{x} -
  x_s(\hat{\beta})|\leq\sqrt{\frac{\rho}{2a_2}}\) and
  \(V_N(\overline{x}^+,\tilde{\bfu},\hat{\beta}) \leq
  \rho - \frac{\rho\underline{\sigma}(Q)}{2a_2}\). On the other hand, if
  \(V_N^0(\hat{x},\hat{\beta})\leq\frac{\rho}{2}\), then
  \(V_N(\overline{x}^+,\tilde{\bfu},\hat{\beta}) \leq
  \frac{\rho}{2}\). In summary, %
}%
\[
  V_N(\hat{x}^+,\tilde{\bfu},\hat{\beta}) \leq \rho - \gamma +
  \tilde{\sigma}_{V_N}(|\tilde{d}|)
\]
where \(\gamma\defas\min\set{\frac{\rho}{2},
  \frac{\rho\underline{\sigma}(Q)}{2a_2}}\). But \(\tilde{\bfu}\) is feasible by
part (a), so by optimality, we have
\[
  V_N^0(\hat{x}^+,\hat{\beta}^+) \leq V_N(\hat{x}^+,\tilde{\bfu},\hat{\beta})
  \leq \rho - \gamma + \tilde{\sigma}_{V_N}(|\tilde{d}|).
\]
Thus, as long as \(|\tilde{d}| \leq \delta \defas
\min\set{\delta_1,\tilde{\sigma}_{V_N}^{-1}(\gamma)}\), we have
\(V_N^0(\hat{x}^+,\hat{\beta}^+) \leq \rho\) and
\(\hat{x}^+\in\calX_N^\rho(\hat{\beta}^+)\). %
% \hspace*{\fill}~\QED
\end{proof}

\ifthenelse{\boolean{LongVersion}}{%
\subsubsection{Reference error bounds}
}{}
\begin{proposition}\label{prop:ref:robust}
  Let \Cref{assum:cont,assum:cons,assum:sstp:exist,assum:stabilizability,%
    assum:quad} hold, \(\rho,\delta>0\), and \(\calB_c\subseteq\calB\) be
  compact. There exist \(\sigma_r,\sigma_g\in\calKinf\) such that
  \begin{subequations}\label{eq:ref:robust}
    \begin{align}
      |g_c(\hat{x},\hat{\beta}) - \rsp|
      &\leq \sigma_r(|\hat{x}-x_s(\hat{\beta})|) \label{eq:ref:robust:a} \\
      |\hat{g}_c(\hat{x},\hat{\beta},\tilde{d}) - \rsp|
      &\leq |g_c(\hat{x},\hat{\beta}) - \rsp| + \sigma_g(|\tilde{d}|)
        \label{eq:ref:robust:b}
    \end{align}
  \end{subequations}
  for all \(\hat{x}\in\calX_N^\rho(\beta)\), \(\hat{\beta} = (\rsp,\zsp,d) \in
  \calB_c\), and \(\tilde{d} \in \tilde{\bbD}_c(\hat{x},\hat{\beta}) \cap
  \delta\bbB^{n_{\tilde{d}}}\). If \(g\) and \(h\) are Lipschitz on
  bounded sets, then we can take \(\sigma_r \defas c_r\id\) and
  \(\sigma_g \defas c_g\id\) for some \(c_r,c_g>0\).
\end{proposition}

\begin{proof}
We already have \cref{eq:ref:robust:a} from \Cref{prop:bound:ref}.
\Cref{prop:cont} gives \(\sigma_g\in\calKinf\) such that
% \begin{multline}\label{eq:ref:robust:0}
%   |g(u_1,h(z_1,d_1)+v_1) - g(u_2,h(z_2,d_2)+v_2)| \\
%   \leq \sigma_g(|(z_1,d_1,v_1)-(z_2,d_2,v_2)|)
% \end{multline}
\begin{align}
  |g(u_1,h(z_1,d_1) + v_1
  &) - g(u_2,h(z_2,d_2)+v_2)| \nonumber \\
  &  \leq \sigma_g(|(z_1,d_1,v_1)-(z_2,d_2,v_2)|) \label{eq:ref:robust:0}
\end{align}
for all \(z_1=(x_1,u_1),z_2=(x_2,u_2)\in\calX_N^\rho(\beta)\times\bbU\),
\(d_1,d_2\in\bbD_c\), and \(v_1\in\bbV_c(z_1,d_1)\), and
\(v_2\in\bbV_c(z_2,d_2)\), where %
\ifthenelse{\boolean{LongVersion}}{%
  \begin{align*}
    \bbD_c &\defas \set{ d\in\bbD | (\ssp,d) \in \calB_c} \\
    \bbV_c(z,d) &\defas \set{ v\in\delta\bbB^{n_y} | h(z,d) + v \in \bbY }
  \end{align*}
}{%
  \(\bbD_c \defas \set{ d\in\bbD | (\ssp,d) \in \calB_c}\) and \(\bbV_c(z,d)
  \defas \set{ v\in\delta\bbB^{n_y} | h(z,d) + v \in \bbY }\). %
}%
Fix \(\hat{x}\in\calX_N^\rho(\hat{\beta})\),
\(\hat{\beta}=(\ssp,\hat{d})\in\calB_c\), and
\(\tilde{d}=(e,e^+,\Delta\ssp,\tilde{w}) \in
\tilde{\bbD}_c(\hat{x},\hat{\beta}) \cap \delta\bbB^{n_{\tilde{d}}}\), where
\(e=(e_x,e_d)\) and \(\tilde{w}=(w,w_d,v)\). %
\ifthenelse{\boolean{LongVersion}}{%
  Substituting \(x_1=\hat{x}+e_x\), \(x_2=\hat{x}\),
  \(u_1=u_2=\kappa_N(\hat{x},\hat{\beta})\), \(d_1=\hat{d}+e_d\),
  \(d_2=\hat{d}\), \(v_1=v\), and \(v_2=0\) into \cref{eq:ref:robust:0} gives,
  independently of \((\hat{x},\hat{\beta},\tilde{d})\),
  \[
    |\hat{g}_c(\hat{x},\hat{\beta},\tilde{d}) - g_c(\hat{x},\hat{\beta})| \leq
    \sigma_g(|(e_x,e_d,v)|) \leq \sigma_g(|\tilde{d}|)
  \]
}{%
  By \cref{eq:bound:ref,eq:ref:robust:0},
  \(|\hat{g}_c(\hat{x},\hat{\beta},\tilde{d}) - g_c(\hat{x},\hat{\beta})| \leq
  \sigma_g(|(e_x,e_d,v)|) \leq \sigma_g(|\tilde{d}|)\), %
}%
and \cref{eq:ref:robust:b} holds by the triangle inequality. If \(g\) and
\(h\) are Lipschitz continuous on bounded sets, we can take
\(\sigma_g\defas c_g\id\) where \(c_g>0\) is the Lipschitz constant
for \(g(u,h(x,u,d)+v)\). %
% \hspace*{\fill}~\QED
\end{proof}

\ifthenelse{\boolean{LongVersion}}{%
\subsubsection{Nominal MPC stability}
}{}
Finally, we use \Cref{prop:mpc:robust:feas,prop:ref:robust} %
\ifthenelse{\boolean{LongVersion}}{%
  and \Cref{thm:mpc:nominal} %
}{}%
to show \Cref{thm:mpc:robust}. %
\ifthenelse{\boolean{LongVersion}}{%
  %% TODO Some intro text
}{}%

\paragraph*{Part (a)}
If \((\hat{x},\hat{\beta})\in\hat{\calS}_N\) and
\(\tilde{d}\in\tilde{\bbD}_c(\hat{x},\hat{\beta})\), then \(\hat{\beta}^+ \defas
\hat{f}_{\beta,c}(\hat{\beta},\tilde{d}) \in \hat{\calB}_c\) by construction of
\(\tilde{\bbD}_c(\hat{x},\hat{\beta})\), and by \Cref{prop:mpc:robust:feas}(c),
there exists \(\delta>0\) such that \(\hat{x}^+ \defas
\hat{f}_c(\hat{x},\hat{\beta},\tilde{d}) \in\calX_N^\rho(\hat{\beta}^+)\) so
long as \(|\tilde{d}|\leq\delta\).

\paragraph*{Part (b)}
\Cref{thm:mpc:nominal} gives~\cref{eq:mpc:robust:lyap:a}, and
\Cref{prop:mpc:robust:feas}(a,b) and the principle of optimality
give~\cref{eq:mpc:robust:lyap:b}.

\paragraph*{Part (c)}
This follows from part (b) due to~\Cref{thm:lyap}.

\paragraph*{Part (d)}
Let \((\hat{\bfx},\hat{\bfbeta},\tilde{\bfd},\bfr)\) satisfy
\cref{eq:model:est:cl}, \((\hat{x}(0),\hat{\beta}(0))\in\hat{\calS}_N^\rho\),
\(\tilde{d} \in \tilde{\bbD}_c(\hat{x},\hat{\beta}) \cap
\delta\bbB^{n_{\tilde{d}}}\), and \(r =
\hat{g}_c(\hat{x},\hat{\beta},\tilde{d})\). Define \(\delta r\defas r-\rsp\) and
\(\delta\hat{r} = g_c(\hat{x},\hat{\beta}) - \rsp\) where
\(\hat{\beta}=(\rsp,\zsp,\hat{d})\). \Cref{prop:ref:robust} and part (b) give
\cref{eq:ref:robust} and %
\ifthenelse{\boolean{LongVersion}}{%
  \[
    \alpha_1(|\delta\hat{r}|) \defas a_1[\sigma_r^{-1}(|\delta\hat{r}|)]^2
    \leq a_1|\delta\hat{x}|^2 \leq V_N^0(\hat{x},\hat{\beta})
  \]
}{%
  \(\alpha_1(|\delta\hat{r}|) \defas a_1[\sigma_r^{-1}(|\delta\hat{r}|)]^2 \leq
  a_1|\delta\hat{x}|^2 \leq V_N^0(\hat{x},\hat{\beta})\) %
}%
for some \(a_1>0\) and \(\sigma_r,\sigma_g\in\calKinf\).
\ifthenelse{\boolean{LongVersion}}{%
  Moreover, \(V_N^0\) is an ISS Lyapunov function on \(\hat{\calS}_N^\rho\) with
  respect to \((\delta\hat{r},\delta\hat{x})\), and RAS on
  \(\hat{\calS}_N^\rho\) with respect to \((\delta\hat{r},\delta\hat{x})\)
  follows by \Cref{thm:lyap}. %
}{%
  \Cref{thm:lyap} implies RAS on \(\hat{\calS}_N^\rho\)
  w.r.t.~\((\delta\hat{r},\delta\hat{x})\). %
}%
Then RAS w.r.t.~\((\delta\hat{r},\delta\hat{x})\), \Cref{prop:ref:robust}, and
\ifthenelse{\boolean{LongVersion}}{%
  Equation~(1) of \cite{rawlings:ji:2012} %
}{%
  \cite[Eq.~(1)]{rawlings:ji:2012} %
}%
give
\begin{align}
  |\delta r(k)|
  &\leq \sigma_r(|\delta\hat{r}(k)|) + \sigma_g(|\tilde{d}(k)|) \nonumber \\
  &\leq \sigma_r(c\lambda^k|\delta\hat{x}(0)| +
    \gamma(\|\tilde{\bfd}\|_{0:k-1})) + \sigma_g(|\tilde{d}(k)|) \nonumber \\
  \ifthenelse{\boolean{LongVersion}}{%
  &\leq \sigma_r(2c\lambda^k|\delta\hat{x}(0)|) +
    \sigma_r(2\gamma(\|\tilde{\bfd}\|_{0:k-1})) + \sigma_g(|\tilde{d}(k)|) \nonumber \\
  }{}
  &\leq \sigma_r(2c\lambda^k|\delta\hat{x}(0)|) +
    (\sigma_r\circ 2\gamma + \sigma_g)(\|\tilde{\bfd}\|_{0:k}) \nonumber \\
  &\asdef \beta_r(|\delta\hat{x}(0)|,k) +
    \gamma_r(\|\tilde{\bfd}\|_{0:k}) \label{eq:mpc:robust:ref:1}
\end{align}
for all \(k\in\nnegint\) and some \(c>0\), \(\lambda\in(0,1)\), and
\(\gamma\in\calK\).

\paragraph*{Part (e)}
If \(g\) and \(h\) are Lipschitz continuous on bounded sets, then by
\Cref{prop:ref:robust}, we can repeat part (d) with \(\sigma_r \defas
c_r\id\) and some \(c_r>0\). %
\hspace*{\fill}~\QED

\subsection{Proof of \Cref{thm:mpc:mismatch}}\label[appendix]{app:mpc:mismatch}
To prove \Cref{thm:mpc:mismatch}, we require two preliminary results. First,
\Cref{prop:bound:noise} establishes a convenient upper bound on \(|\tilde{w}|\).
Second, \Cref{prop:bound:lyap} establishes cost decrease bounds for the
estimator and regulator Lyapunov functions of \cref{eq:model:mismatch:cl}.

%% TODO change the citation if/when the paper is accepted.
\begin{remark}
  \Cref{prop:bound:noise} is similar to the error bound results %
  \ifthenelse{\boolean{LongVersion}}{%
    Section~5.2 of \cite{kuntz:rawlings:2024d}. %
  }{%
    of~\cite[Sec.~5.2]{kuntz:rawlings:2024d}.
    % of~\cite[Section V.B]{kuntz:rawlings:2025a}.
  }%
  The main extension is error on the measurement equation \(v\) and model
  disturbance \(w_d\). Likewise, the bounds
  \cref{eq:bound:lyap:reg,eq:bound:lyap:est} of \Cref{prop:bound:lyap} are
  similar to bounds in~\cite[Section 5.1]{kuntz:rawlings:2024d}.
  % of~\cite[Section V.A]{kuntz:rawlings:2025a}.
  Here, we consider a Lyapunov function of the estimator as well as the
  regulator.
\end{remark}

\ifthenelse{\boolean{LongVersion}}{%
\subsubsection{Estimator noise bound}
}{}
\begin{proposition}\label{prop:bound:noise}
  Suppose \Cref{assum:cont,assum:cons,assum:sstp:exist,assum:sstp,%
    assum:sstp:mismatch,assum:diff} hold. For any compact \(X\subseteq\bbX\),
  there exist \(\sigma_w,\sigma_\alpha\in\calKinf\) for which
  \begin{equation}\label{eq:bound:noise}
    |\tilde{w}| \leq \sigma_w(|\wP|)|z-z_s(\beta)| +
    \sigma_\alpha(|\Delta\alpha|)
  \end{equation}
  for all \(z=(x,u)\in X\times\bbU\) and
  \(\alpha=(\ssp,\wP),\alpha^+\in\calA_c\), where \(\tilde{w}\defas(w,w_d,v)\),
  \(\Delta\alpha\defas\alpha^+-\alpha\), and \cref{eq:noise}.
\end{proposition}
\begin{proof}
  Fix \(z=(x,u)\in X\times\bbU\) and \(\alpha=(\ssp,\wP)\in\calA_c\), and let
  \(\beta\defas(\ssp,d_s(\alpha))\), \(\tilde{w}\defas(w,w_d,v)\), and
  \ifthenelse{\boolean{OneColumn}}{%
    \[
      \Delta\tilde{w}(x,u,\alpha) \defas
      \begin{bmatrix} \fP(x + \Delta
        x_s(\alpha),u,\wP) - f(x,u,\hat{d}_s(\alpha)) - \Delta x_s(\alpha) \\
        \hP(x + \Delta x_s(\alpha),u,\wP) - h(x,u,\hat{d}_s(\alpha))
      \end{bmatrix}
    \]
  }{%
    %% NOTE For some reason multline is producing a lot of space above the
    %% equation, so I simulated the behavior with an align environment. If the
    %% equation is changed, just be careful to adjust the alignment.
    % \begin{multline*}
    %   \Delta\tilde{w}(x,u,\alpha) \\
    %   \defas
    %   \begin{bmatrix} \fP(x + \Delta
    %     x_s(\alpha),u,\wP) - f(x,u,\hat{d}_s(\alpha)) - \Delta x_s(\alpha) \\
    %     \hP(x + \Delta x_s(\alpha),u,\wP) - h(x,u,\hat{d}_s(\alpha)) \end{bmatrix}
    % \end{multline*}
    \begin{align*}
      &\Delta\tilde{w}(x,u,\alpha) \\
      &\defas
        \begin{bmatrix} \fP(x + \Delta
          x_s(\alpha),u,\wP) - f(x,u,\hat{d}_s(\alpha)) - \Delta x_s(\alpha) \\
          \hP(x + \Delta x_s(\alpha),u,\wP) - h(x,u,\hat{d}_s(\alpha)) \end{bmatrix}
    \end{align*}
  }%
  throughout. We also note that, by definition of the SSTP~\cref{eq:sstp} and
  the nominal model assumption~\cref{eq:nominal}, we have
  \begin{align}\label{eq:bound:noise:0}
    \Delta\tilde{w}(z_s(\beta),\alpha) &= 0,
    & \partial_z\Delta\tilde{w}(z,\ssp,0) &= 0.
  \end{align}
  Assume all functions continuously differentiable on \(\bbX\times\bbU\) have
  been extended to continuously differentiable functions on all of
  \(\real^{n+n_u}\) using appropriately defined partitions of unity %
  \ifthenelse{\boolean{LongVersion}}{%
    (cf.~Lemma~2.26 of \cite{lee:2012}). %
  }{%
    (cf.~\cite[Lem.~2.26]{lee:2012}). %
  }%
  Let \(Z_c\) denote the convex hull of \(X\times\bbU\).

  For each \(i\in\intinterval{1}{n+n_y}\), \(\partial_z\Delta\tilde{w}_i\) is
  continuous, and by \Cref{prop:cont}, there exists \(\sigma_i\in\calKinf\) such
  that
  \begin{equation*}
    | \partial_z \Delta\tilde{w}_i(z_1,\alpha_1) -
    \partial_z\Delta\tilde{w}_i(z_2,\alpha_2) | \leq
    \sigma_i(|(z_1,\alpha_1)-(z_2,\alpha_2)|)
  \end{equation*}
  for all \(z_1,z_2\in Z_c\) and \(\alpha_1,\alpha_2\in\calA_c\). Substituting
  \(z_1=z_2=z\), \(\alpha_1=\alpha\), and \(\alpha_2=(\ssp,0)\) into the above
  inequality, we have %
  \begin{equation}\label{eq:bound:noise:1}
    |\partial_z\Delta\tilde{w}_i(z,\alpha)| =
    |\partial_z\Delta\tilde{w}_i(z,\alpha) -
    \partial_z\Delta\tilde{w}(z,\ssp,0)| \leq \sigma_i(|\wP|)
  \end{equation}
  where the equality follows by~\cref{eq:bound:noise:0}. By Taylor's
  theorem~\cite[Thm.~12.14]{apostol:1974}, for each
  \(i\in\intinterval{1}{n+n_y}\), there exist \(z_i(z,\alpha)\in Z_c\) and
  \(t_i(z,\alpha)\in(0,1)\) such that
  \begin{equation}\label{eq:bound:noise:2}
    \Delta\tilde{w}_i(z,\alpha) =
    \partial_z\Delta\tilde{w}_i(\tilde{z}_i(z,\alpha),\alpha)(z-z_s(\beta))
  \end{equation}
  where \(\tilde{z}_i(z,\alpha) \defas t_i(z,\alpha)z_s(\beta) +
  (1-t_i(z,\alpha))z_i(z,\alpha)\in Z_c\) by convexity of \(Z_c\), and the
  zero-order term drops by~\cref{eq:bound:noise:0}.
  Combining~\cref{eq:bound:noise:1,eq:bound:noise:2} gives
  \begin{equation}\label{eq:bound:noise:3}
    |\Delta\tilde{w}(z,\alpha)| \leq \sum_{i=1}^{n+n_y}
    |\Delta\tilde{w}_i(z,\alpha)|
    \ifthenelse{\boolean{OneColumn}}{%
      \leq \sum_{i=1}^{n+n_y} \sigma_i(|\wP|)|z-z_s(\beta)|
    }{}%
    = \sigma_w(|\wP|)|z-z_s(\beta)|
  \end{equation}
  with \(\sigma_w\defas\sum_{i=1}^{n+n_y} \sigma_i\). By \Cref{prop:cont}, since
  \(x_{\mathrm{P},s},x_s,d_s\) are continuous, there exist
  \(\sigma_x,\sigma_d\in\calKinf\) such that
  \begin{subequations}\label{eq:bound:noise:4}
    \begin{align}
      |\Delta x_s(\alpha_1) -
      \Delta x_s(\alpha_2)| &\leq \sigma_x(|\alpha_1-\alpha_2|) \\
      |d_s(\alpha_1) - d_s(\alpha_2)| &\leq \sigma_d(|\alpha_1-\alpha_2|)
    \end{align}
  \end{subequations}
  for all \(\alpha_1,\alpha_2\in\calA_c\). Finally, using
  \cref{eq:bound:noise:3,eq:bound:noise:4} with \(\alpha_1=\alpha\) and
  \(\alpha_2=\alpha^+\) gives %
  \ifthenelse{\boolean{LongVersion}}{%
    \begin{align*}
      |\tilde{w}|
      &\leq |\Delta\tilde{w}(z,\alpha)| + |\Delta x_s(\alpha^+) - \Delta x_s(\alpha)|
        \ifthenelse{\boolean{OneColumn}}{}{ \\
      &\qquad} + |d_s(\alpha^+) - d_s(\alpha)| \\
      &\leq \sigma_w(|\wP|)|z-z_s(\beta)| + \sigma_\alpha(|\Delta\alpha|)
    \end{align*}
  }{%
    \cref{eq:bound:noise} %
  }%
  with \(\sigma_\alpha\defas\sigma_x+\sigma_d\in\calKinf\). %
  % \hspace*{\fill}~\QED
\end{proof}

\ifthenelse{\boolean{LongVersion}}{%
\subsubsection{Lyapunov cost decrease bounds}
}{}
% In \Cref{prop:bound:lyap}, we establish cost decrease bounds for the estimator
% and regulator Lyapunov functions of \cref{eq:model:mismatch:cl}.
% (see \Cref{app:bound:lyap} for proof).
\begin{proposition}\label{prop:bound:lyap}
  Suppose \Cref{assum:cont,assum:cons,assum:sstp:exist,assum:stabilizability,%
    assum:quad,assum:sstp,assum:sstp:mismatch,assum:diff} hold and let
  \(\rho>0\). There exist \(\tilde{c}_e, \tilde{a}_3, \tilde{a}_4, \hat{c}_3,
  \delta, \delta_w > 0\) and \(\tilde{\sigma}_w, \tilde{\sigma}_\alpha,
  \sigma_\alpha, \hat\sigma_w, \hat\sigma_\alpha \in \calKinf\) such that
  \begin{align}
    |\tilde{d}|^2
    &\leq \tilde{c}_e|(e,e^+)|^2 + \tilde{\sigma}_w(|\wP|)|\delta\hat{x}|^2 +
      \tilde{\sigma}_\alpha(|\Delta\alpha|) \label{eq:bound:d} \\
    (V_N^0)^+
    &\leq V_N^0 - \tilde a_3|\delta\hat{x}|^2 + \tilde{a}_4|(e,e^+)|^2 +
      \sigma_\alpha(|\Delta\alpha|) \label{eq:bound:lyap:reg} \\
    V_e^+
    &\leq V_e - \hat{c}_3|e|^2 + \hat\sigma_w(|\wP|)|\delta\hat{x}|^2 +
      \hat\sigma_\alpha(|\Delta\alpha|) \label{eq:bound:lyap:est}
  \end{align}
  so long as \((\hat{x},\hat{\beta}) \in \hat{\calS}_N^\rho\), \(x\in\bbX\),
  \(\alpha=(\ssp,\wP)\in\calA_c(\delta_w)\), \(\Delta\alpha = (\Delta\ssp,
  \Delta\wP) \in \bbA_c(\alpha,\delta_w)\), and \(|\tilde{d}|\leq\delta\), where
  \(\tilde{d} \defas (e,e^+,\Delta\ssp,\tilde{w})\), \(V_N^0 \defas
  V_N^0(\hat{x},\hat{\beta})\), \((V_N^0)^+ \defas
  V_N^0(\hat{x}^+,\hat{\beta}^+)\), \(V_e \defas
  V_e(x,d_s(\alpha),\hat{x},\hat{d})\), \(V_e^+ \defas
  V_e(x^+,d_s(\alpha^+),\hat{x}^+,\hat{d}^+)\),
  \cref{eq:est:err,eq:noise,eq:model:mismatch:cl}.
\end{proposition}
\begin{proof}
  % \subsection{Proof of \Cref{prop:bound:lyap}}\label[appendix]{app:bound:lyap}
  \ifthenelse{\boolean{LongVersion}}{%
    Throughout the proof, fix %
  }{%
    Fix %
  }%
  \((\hat{x},\hat{\beta}) = (\hat{x},\ssp,\hat{d}) \in \hat{\calS}_N^\rho\),
  \(x\in\bbX\), \(\alpha=(\ssp,\wP)\in\calA_c(\delta_w)\), and \(\Delta\alpha =
  (\Delta\ssp, \Delta\wP) \in \bbA_c(\alpha,\delta_w)\). Assume
  \(|\tilde{d}|\leq\delta\). %
  \ifthenelse{\boolean{LongVersion}}{%
    Unless otherwise specified, assume the following constructions are
    independent of \((x,\alpha,\hat{x},\hat{\beta})\) %
  }{} %
  Let \(L_s\) and \(L_f\) denote the Lipschitz constants for \(z_s\) on
  \(\hat{\calB}_c\) and \(f\) on \(\hat{\calS}_N^\rho\), respectively.

\paragraph{Bound~\cref{eq:bound:d}}
%% NOTE \cite[Eq.~(1)]{rawlings:ji:2012} is a looser bound than the identity
%% \((a+b+c)^2\leq 3a^2+3b^2+3c^2\).
By~\Cref{prop:bound:xu,prop:bound:noise} %
\ifthenelse{\boolean{LongVersion}}{%
  and Equation~(1) of \cite{rawlings:ji:2012}, %
}{%
  and~\cite[Eq.~(1)]{rawlings:ji:2012}, %
}%
\begin{align*}
  |\tilde{w}|^2
  &\leq [\sigma_w(|\wP|)|z-z_s(\beta)| + \sigma_\alpha(|\Delta\alpha|)]^2 \\
  \ifthenelse{\boolean{LongVersion}}{%
  &\leq [\sigma_w(|\wP|)|z-z_s(\hat\beta)| + L_s\sigma_w(|\wP|)|e| +
    \sigma_\alpha(|\Delta\alpha|)]^2 \\
  &\leq [\sigma_w(|\wP|)|x-x_s(\hat\beta)| + \sigma_w(|\wP|)|u-u_s(\hat\beta)|
    \ifthenelse{\boolean{OneColumn}}{}{\\
  &\qquad}  + L_s\sigma_w(|\wP|)|e| + \sigma_\alpha(|\Delta\alpha|)]^2 \\
  &\leq [(1+c_u)\sigma_w(|\wP|)|\hat{x}-x_s(\hat{\beta})| +
    (L_s + 1)\sigma_w(|\wP|)|e|
    \ifthenelse{\boolean{OneColumn}}{}{\\
  &\qquad} + \sigma_\alpha(|\Delta\alpha|)]^2 \\
  }{}
  &\leq 9(1+c_u)^2[\sigma_w(|\wP|)]^2|\hat{x}-x_s(\beta)|^2
    \ifthenelse{\boolean{OneColumn}}{}{\\
  &\qquad} + 9(L_s + 1)^2[\sigma_w(|\wP|)]^2|e|^2 +
    9[\sigma_\alpha(|\Delta\alpha|)]^2
\end{align*}
where \(c_u>0\) and \(\sigma_w,\sigma_\alpha\in\calKinf\) satisfy
\cref{eq:bound:u,eq:bound:noise}. Therefore~%
\ifthenelse{\boolean{LongVersion}}{%
  \begin{align*}
    |\tilde{d}|^2
    &= |(e,e^+)|^2 + |\Delta\ssp|^2 + |\tilde{w}|^2 \\
    &\leq 9(1+c_u)^2(\sigma_w(|\wP|))^2|\hat{x}-x_s(\beta)|^2 \\
    &\qquad + (1 + 9(L_s + 1)^2(\sigma_w(\delta_w))^2)|(e,e^+)|^2
      \ifthenelse{\boolean{OneColumn}}{}{\\
    &\qquad} + |\Delta\alpha|^2 + 9\sigma_\alpha(|\Delta\alpha|))^2
  \end{align*}
  so~%
}{}%
\cref{eq:bound:d} holds with \(\tilde c_e \defas 1 + 9(L_s + 1)^2
[\sigma_w(\delta_w)]^2 > 0\), \(\tilde\sigma_w \defas 9(1+c_u)^2\sigma_w^2 \in
\calKinf\), and \(\tilde{\sigma}_\alpha \defas \id^2 +
9\sigma_\alpha\in\calKinf\).

\paragraph{Intermediate result}
To show \cref{eq:bound:lyap:reg}, it is first necessary to derive the following
inequality: %
\ifthenelse{\boolean{OneColumn}}{%
  \begin{equation}\label{eq:bound:VN}
    |V_N(\hat{x}^+,\tilde{\bfu}(\hat{x},\hat{\beta}),\hat{\beta}^+) -
    V_N(\overline{x}^+,\tilde{\bfu}(\hat{x},\hat{\beta}),\hat{\beta})|
    \leq a_{V_N,1}|\hat{x} - x_s(\hat{\beta})|^2 + a_{V_N,2}|\tilde{d}|^2
  \end{equation}
}{%
  %% NOTE For some reason multline is producing a lot of space above the
  %% equation, so I simulated the behavior with an align environment. If the
  %% equation is changed, just be careful to adjust the alignment.
  % \begin{multline}\label{eq:bound:VN}
  %   |V_N(\hat{x}^+,\tilde{\bfu}(\hat{x},\hat{\beta}),\hat{\beta}^+) -
  %   V_N(\overline{x}^+,\tilde{\bfu}(\hat{x},\hat{\beta}),\hat{\beta})| \\
  %   \leq a_{V_N,1}|\hat{x} - x_s(\hat{\beta})|^2 + a_{V_N,2}|\tilde{d}|^2
  % \end{multline}
  \begin{align}
    |V_N(\hat{x}^+,\tilde{\bfu}(\hat{x},\hat{\beta}),\hat{\beta}^+)
    &- V_N(\overline{x}^+,\tilde{\bfu}(\hat{x},\hat{\beta}),\hat{\beta})| \nonumber \\
    \leq \,& a_{V_N,1}|\hat{x} - x_s(\hat{\beta})|^2 + a_{V_N,2}|\tilde{d}|^2 \label{eq:bound:VN}
  \end{align}
}%
for some \(a_{V_N,1}\in(0,\underline{\sigma}(Q))\), \(a_{V_N,2}>0\), and
\(\sigma_{V_N}\in\calKinf\), where \(\overline{x}^+\defas
f_c(\hat{x},\hat{\beta})\) and \cref{eq:model:est:cl}.

By \Cref{prop:cont}, we have \(\sigma_{P_f}\in\calKinf\) such that
\begin{equation}\label{eq:bound:Pf}
  \overline{\sigma}(P_f(\beta_1) - P_f(\beta_2)) \leq
  \sigma_{P_f}(|\beta_1-\beta_2|)
\end{equation}
for all \(\beta_1,\beta_2\in\hat{\calB}_c\). Moreover, since \(\hat{\calB}_c\)
is compact and \(P_f(\cdot)\) is continuous and positive definite,
\(\gamma\defas\max_{\hat{\beta}\in\hat{\calB}_c}
\overline{\sigma}(P_f(\hat{\beta}))\) and
\(\gamma_0\defas\max_{\hat{\beta}\in\hat{\calB}_c}
\underline{\sigma}(P_f(\hat{\beta}))\) exist and are positive and finite. For
ease of notation, let \(\delta{\hat{x}} \defas \hat{x} - x_s(\hat{\beta})\),
\(\tilde{\bfu} \defas \tilde{\bfu}(\hat{x},\hat{\beta})\), \(\overline{x}^+(k)
\defas \phi(k;\overline{x}^+,\tilde{\bfu},\hat{\beta})\), and \(\hat{x}^+(k)
\defas \phi(k;\hat{x}^+,\tilde{\bfu},\hat{\beta}^+)\).

%% Part 0. Misc. facts
By \Cref{assum:diff}, we have
\begin{equation}\label{eq:bound:VN:f}
  |\overline{x}^+ - \hat{x}^+| \leq L_f|e| + |w| + |e_x^+| \leq
  L_f'|\tilde{d}|
\end{equation}
where \(L_f'\defas L_f+2\). By \Cref{assum:sstp:mismatch}(b), we have
\begin{equation}\label{eq:bound:VN:zs}
  |z_s(\hat{\beta}^+)-z_s(\hat{\beta})| \leq
  \ifthenelse{\boolean{OneColumn}}{L_s|\hat{\beta}^+-\hat{\beta}| \leq}{}
  L_s(|\Delta\beta|+|e_d|+|e_d^+|) \leq 3L_s|\tilde{d}|
\end{equation}
and by \Cref{prop:bound:xu}, we have \(c_x,c_u>0\) such that
\begin{align}
  |\overline{x}^+(j) - x_s(\hat{\beta})|
  &\leq c_x|\delta\hat{x}| \label{eq:bound:VN:x} \\
  |\tilde{u}(k) - u_s(\hat{\beta})|
  &\leq c_u|\delta\hat{x}| \label{eq:bound:VN:u}
\end{align}
for each \(j\in\intinterval{0}{N-1}\) and \(k\in\intinterval{0}{N-2}\).

%% Part 3: Terminal state/input upper bound
\ifthenelse{\boolean{LongVersion}}{%
  By \Cref{assum:stabilizability,assum:quad}, we have
  \begin{align*}
    \gamma_0 \ifthenelse{\boolean{OneColumn}}{}{
    &}|\overline{x}^+(N) - x_s(\hat{\beta})|^2 \ifthenelse{\boolean{OneColumn}}{}{\\}
    &\leq V_f(\overline{x}^+(N-1),\hat{\beta}) \\
    &\leq V_f(\overline{x}^+(N-1),\hat{\beta})
      - \underline{\sigma}(Q)|\overline{x}^+(N-1) - x_s(\hat{\beta})|^2 \\\
    &\leq [\gamma - \underline{\sigma}(Q)]|\overline{x}^+(N-1) - x_s(\hat{\beta})|^2 \\
    &\overset{\cref{eq:bound:VN:x}}{\leq}
      [\gamma - \underline{\sigma}(Q)]c_x^2|\delta\hat{x}|^2.
  \end{align*}
  Therefore
  \begin{subequations}\label{eq:bound:VN:5}
    \begin{equation}\label{eq:bound:VN:5:x}
      |\overline{x}^+(N) - x_s(\hat{\beta})| \leq c_{x,f}|\delta\hat{x}|
    \end{equation}
    where \(c_{x,f}\defas c_x\sqrt{\frac{\gamma -
        \underline{\sigma}(Q)}{\gamma_0}}\). Similarly, using the fact that
    \(V_f(\overline{x}^+(N),\hat{\beta})\geq 0\), we have
    \begin{align*}
      \underline{\sigma}(R) \ifthenelse{\boolean{OneColumn}}{}{
      &}|\tilde{u}(N-1) - u_s(\hat{\beta})|^2 \ifthenelse{\boolean{OneColumn}}{}{\\}
      &\leq V_f(\overline{x}^+(N-1),\hat{\beta})
        - \underline{\sigma}(Q)|\overline{x}^+(N-1) - x_s(\hat{\beta})|^2 \\
      &\leq [\gamma - \underline{\sigma}(Q)]
        |\overline{x}^+(N-1) - x_s(\hat{\beta})|^2 \\
      &\overset{\cref{eq:bound:VN:x}}{\leq}
        [\gamma - \underline{\sigma}(Q)]c_x^2|\delta\hat{x}|^2
    \end{align*}
    and therefore
    \begin{equation}\label{eq:bound:VN:5:u}
      |\tilde{u}(N-1) - u_s(\hat{\beta})| \leq c_{u,f}|\delta\hat{x}|
    \end{equation}
  \end{subequations}
  with \(c_{u,f} \defas c_x
  \sqrt{\frac{\gamma-\underline{\sigma}(Q)}{\underline{\sigma}(R)}}\). %
}{%
  By \Cref{assum:stabilizability,assum:quad} and \cref{eq:bound:VN:x}, we have
  \(\gamma_0|\overline{x}^+(N) - x_s(\hat{\beta})|^2 \leq [\gamma -
  \underline{\sigma}(Q)]c_x^2|\delta\hat{x}|^2\), \(\underline{\sigma}(R)
  |\tilde{u}(N-1) - u_s(\hat{\beta})|^2 \leq [\gamma -
  \underline{\sigma}(Q)]c_x^2|\delta\hat{x}|^2\), and
  \begin{subequations}\label{eq:bound:VN:5}
    \begin{align}
      |\overline{x}^+(N) - x_s(\hat{\beta})|
      &\leq c_{x,f}|\delta\hat{x}| \label{eq:bound:VN:5:x} \\
      |\tilde{u}(N-1) - u_s(\hat{\beta})|
      &\leq c_{u,f}|\delta\hat{x}| \label{eq:bound:VN:5:u}
    \end{align}
  \end{subequations}
  where \(c_{x,f}\defas c_x \sqrt{\frac{\gamma -
      \underline{\sigma}(Q)}{\gamma_0}}\) and \(c_{u,f} \defas c_x
  \sqrt{\frac{\gamma-\underline{\sigma}(Q)}{\underline{\sigma}(R)}}\). %
}%

%% Part 4: Trajectory deviation upper bound(s)
Next, Lipschitz continuity of \(f\) on \(\hat{\calS}_N^\rho\) gives %
\ifthenelse{\boolean{LongVersion}}{%
  \begin{align*}
    |\hat{x}^+(k) \ifthenelse{\boolean{OneColumn}}{}{
    &}- \overline{x}^+(k)|
      \ifthenelse{\boolean{OneColumn}}{}{\\}
    &= |f(\hat{x}^+(k-1),\tilde{u}(k),\hat{d}^+) -
      f(\overline{x}^+(k-1),\tilde{u}(k),\hat{d})| \\
    &\leq L_f|\hat{x}^+(k-1) - \overline{x}^+(k-1)| +
      L_f|\hat{d}^+ - \hat{d}|
  \end{align*}
  Applying this inequality recursively, we have %
}{%
  \(|\hat{x}^+(k) - \overline{x}^+(k)|\leq L_f|\hat{d}^+ - \hat{d}|\), and
  therefore %
}%
\begin{equation}
  |\hat{x}^+(k) - \overline{x}^+(k)|
  \leq L_f^k|\hat{x}^+ - \overline{x}^+| + L_k|\hat{d}^+-\hat{d}|
  \leq L_k'|\tilde{d}| \label{eq:bound:VN:6}
\end{equation}
for all \(k\in\intinterval{0}{N}\), where \(L_k \defas \sum_{i=1}^k L_f^i\)
and \(L_k' \defas L_f^kL_f' + 3L_k\), and we have used
\cref{eq:bound:VN:f} and the fact that
\(|\hat{d}^+-\hat{d}|\leq|w_d|+|e_d|+|e_d^+|\leq 3|\tilde{d}|\).
%% Part 5: Another error bound
Moreover,
\begin{align}
  |\hat{x}^+(k) - x_s(\hat{\beta})|
  &\overset{\cref{eq:bound:VN:x},\cref{eq:bound:VN:6}}{\leq}
    c_x|\delta\hat{x}| + L_k'|\tilde{d}| \label{eq:bound:VN:7} \\
  |\hat{x}^+(N) - x_s(\hat{\beta})|
  &\overset{\cref{eq:bound:VN:5},\cref{eq:bound:VN:6}}{\leq}
    c_{x,f}|\delta\hat{x}| + L_N'|\tilde{d}|. \label{eq:bound:VN:8}
\end{align}

%% Part 1 (pre-rearrangement). Get rid of \(\hat\beta^+\) in the terminal
%% weight
Using the inequalities, \(||\xi|_{M_1}^2 - |\xi|_{M_2}^2| \leq
\overline{\sigma}(M_1-M_2) |\xi|^2\), \(|\xi_1+\xi_2|^2 \leq 2|\xi_1|^2 +
2|\xi_2|^2\), \cref{eq:bound:Pf}, \cref{eq:bound:VN:8}, and
\(|\hat{\beta}^+-\hat{\beta}| \leq |\Delta\beta|+|e_d|+|e_d^+| \leq
3|\tilde{d}|\), we have %
\ifthenelse{\boolean{LongVersion}}{%
  \begin{align}
    V_f(\hat{x}^+(N),\hat{\beta}^+)
    &\overset{\cref{eq:bound:Pf}}{\leq}
      |\hat{x}^+(N) - x_s(\hat{\beta}^+)|_{P_f(\hat{\beta})}^2
      \ifthenelse{\boolean{OneColumn}}{}{ \nonumber \\
    &\qquad } + \sigma_{P_f}(3|\tilde{d}|)
      |\hat{x}^+(N) - x_s(\hat{\beta}^+)|^2 \nonumber \\
    &\overset{\cref{eq:bound:VN:8}}{\leq}
      |\hat{x}^+(N) - x_s(\hat{\beta}^+)|_{P_f(\hat{\beta})}^2
      \ifthenelse{\boolean{OneColumn}}{}{ \nonumber \\
    &\qquad } + \sigma_{P_f}(3|\tilde{d}|)
      [c_{x,f}|\delta\hat{x}| + L_N'|\tilde{d}|]^2 \nonumber \\
    &\leq |\hat{x}^+(N) - x_s(\hat{\beta}^+)|_{P_f(\hat{\beta})}^2
      \ifthenelse{\boolean{OneColumn}}{}{ \nonumber \\
    &\qquad } + \sigma_{P_f,x}(|\tilde{d}|)|\delta\hat{x}|^2 +
      \sigma_{P_f,d}(|\tilde{d}|)|\tilde{d}|^2 \label{eq:bound:VN:1}
  \end{align}
  where \(\sigma_{P_f,x} \defas 2c_{x,f}^2\sigma_{P_f}\circ 3\id \in \calKinf\)
  and \(\sigma_{P_f,d} \defas 2(L_N')^2\sigma_{P_f}\circ 3\id \in \calKinf\).
}{%
  \ifthenelse{\boolean{OneColumn}}{%
    \begin{equation}\label{eq:bound:VN:1}
      V_f(\hat{x}^+(N),\hat{\beta}^+)
      \leq |\hat{x}^+(N) - x_s(\hat{\beta}^+)|_{P_f(\hat{\beta})}^2
      + \sigma_{P_f,x}(|\tilde{d}|)|\delta\hat{x}|^2 +
      \sigma_{P_f,d}(|\tilde{d}|)|\tilde{d}|^2
    \end{equation}
  }{%
    %% NOTE For some reason multline is producing a lot of space above the
    %% equation, so I simulated the behavior with an align environment. If the
    %% equation is changed, just be careful to adjust the alignment.
    % \begin{multline}\label{eq:bound:VN:1}
    %   V_f(\hat{x}^+(N),\hat{\beta}^+)
    %   \leq |\hat{x}^+(N) - x_s(\hat{\beta}^+)|_{P_f(\hat{\beta})}^2 \\
    %   + \sigma_{P_f,x}(|\tilde{d}|)|\delta\hat{x}|^2 +
    %   \sigma_{P_f,d}(|\tilde{d}|)|\tilde{d}|^2
    % \end{multline}
    \begin{align}
      V_f(\hat{x}^+(N),\hat{\beta}^+)
      &\leq |\hat{x}^+(N) - x_s(\hat{\beta}^+)|_{P_f(\hat{\beta})}^2 \nonumber \\
      &\quad + \sigma_{P_f,x}(|\tilde{d}|)|\delta\hat{x}|^2 +
      \sigma_{P_f,d}(|\tilde{d}|)|\tilde{d}|^2 \label{eq:bound:VN:1}
    \end{align}
  }%
  where \(\sigma_{P_f,x} \defas 2c_{x,f}^2\sigma_{P_f}\circ 3\id\) and
  \(\sigma_{P_f,d} \defas 2(L_N')^2\sigma_{P_f}\circ 3\id\). %
}%

%% Part 2 (pre-rearrangement): Get rid of the \(\hat{\beta}^+\) in ALL the
%% targets
For the remainder of this part, we let \(\lambda>0\) (to be defined) and use the
identity \(2ab\leq\lambda a^2 + \lambda^{-1}b^2\). Expanding quadratics and
using \cref{eq:bound:VN:zs,eq:bound:VN:8}, we have %
\ifthenelse{\boolean{LongVersion}}{%
  \begin{align}
    \Big| |\hat{x}^+(N)
    &- x_s(\hat{\beta}^+)|_{P_f(\hat{\beta})}^2
      - |\hat{x}^+(N) - x_s(\hat{\beta})|_{P_f(\hat{\beta})}^2 \Big| \nonumber \\
    &\leq 2\gamma|\hat{x}^+(N) - x_s(\hat{\beta})||x_s(\hat{\beta}^+) -
      x_s(\hat{\beta})| + \gamma|x_s(\hat{\beta}^+) - x_s(\hat{\beta})|^2 \nonumber \\
    &\overset{\cref{eq:bound:VN:zs}}{\leq}
      6\gamma L_s|\hat{x}^+(N) - x_s(\hat{\beta})||\tilde{d}| +
      9\gamma L_s^2|\tilde{d}|^2 \nonumber \\
    &\overset{\cref{eq:bound:VN:8}}{\leq}
      6\gamma L_sc_{x,f}|\delta\hat{x}||\tilde{d}| +
      (6\gamma L_sL_N'+9\gamma L_s^2)|\tilde{d}|^2 \nonumber \\
    &\leq 3\lambda\gamma L_sc_{x,f}|\delta\hat{x}|^2 +
      (6\gamma L_sL_N' + 9\gamma L_s^2 + 3\lambda^{-1}\gamma L_sc_{x,f})
      |\tilde{d}|^2 \nonumber \\
    &= \lambda\hat{L}_{1,N}|\delta\hat{x}|^2 +
      \hat{L}_{2,N}(\lambda)|\tilde{d}|^2 \label{eq:bound:VN:2}
  \end{align}
}{%
  %% NOTE For some reason multline is producing a lot of space above the
  %% equation, so I simulated the behavior with an align environment. If the
  %% equation is changed, just be careful to adjust the alignment.
  % \begin{multline}\label{eq:bound:VN:2}
  %   \Big| |\hat{x}^+(N) - x_s(\hat{\beta}^+)|_{P_f(\hat{\beta})}^2 -
  %   |\hat{x}^+(N) - x_s(\hat{\beta})|_{P_f(\hat{\beta})}^2 \Big| \\
  %   \leq \lambda\hat{L}_{1,N}|\delta\hat{x}|^2 +
  %   \hat{L}_{2,N}(\lambda)|\tilde{d}|^2
  % \end{multline}
  \begin{align}
    \Big| |\hat{x}^+(N) - x_s(\hat{\beta}^+)|_{P_f(\hat{\beta})}^2
    &- |\hat{x}^+(N) - x_s(\hat{\beta})|_{P_f(\hat{\beta})}^2 \Big| \nonumber \\
    \leq \, &\lambda\hat{L}_{1,N}|\delta\hat{x}|^2 +
    \hat{L}_{2,N}(\lambda)|\tilde{d}|^2 \label{eq:bound:VN:2}
  \end{align}
}%
where \(\hat{L}_{1,N} \defas 3\gamma L_sc_{x,f}\) and \(\hat{L}_{2,N}(\lambda)
\defas 6\gamma L_sL_N'+9\gamma L_s^2 + 3\lambda^{-1}\gamma L_sc_{x,f}\).
Similarly, using \cref{eq:bound:VN:zs,eq:bound:VN:7,eq:bound:VN:u}, we have %
\ifthenelse{\boolean{LongVersion}}{%
  \begin{align}
    \Big| |\hat{x}^+(k)
    &- x_s(\hat{\beta}^+)|_Q^2
      - |\hat{x}^+(k) - x_s(\hat{\beta})|_Q^2 \Big| \nonumber \\
    &\leq 2\underline{\sigma}(Q)|\hat{x}^+(k) - x_s(\hat{\beta})|
      |x_s(\hat{\beta}^+) - x_s(\hat{\beta})| + \underline{\sigma}(Q)
      |x_s(\hat{\beta}^+) - x_s(\hat{\beta})|^2 \nonumber \\
    &\overset{\cref{eq:bound:VN:zs}}{\leq}
      6\underline{\sigma}(Q) L_s|\hat{x}^+(k) - x_s(\hat{\beta})||\tilde{d}| +
      9\underline{\sigma}(Q) L_s^2|\tilde{d}|^2 \nonumber \\
    &\overset{\cref{eq:bound:VN:7}}{\leq}
      6\underline{\sigma}(Q) L_sc_x|\delta\hat{x}||\tilde{d}| +
      (6\underline{\sigma}(Q) L_sL_k' + 9\underline{\sigma}(Q) L_s^2)
      |\tilde{d}|^2 \nonumber \\
    &\leq 3\lambda \underline{\sigma}(Q) L_sc_x|\delta\hat{x}|^2 +
      (6\underline{\sigma}(Q) L_sL_k' + 9\underline{\sigma}(Q) L_s^2 +
      3\lambda^{-1}\gamma L_sc_x) |\tilde{d}|^2 \nonumber \\
    &\leq \lambda\hat{L}_{1,k}|\delta\hat{x}|^2 +
      \hat{L}_{2,k}(\lambda)|\tilde{d}|^2 \label{eq:bound:VN:3}
  \end{align}
  and
  \begin{align}
    \Big| |\tilde{u}(k)
    &- u_s(\hat{\beta}^+)|_R^2
      - |\tilde{u}(k) - u_s(\hat{\beta})|_R^2 \Big| \nonumber \\
    &\leq 2\underline{\sigma}(R)|\tilde{u}(k) - u_s(\hat{\beta})|
      |u_s(\hat{\beta}^+) - u_s(\hat{\beta})| + \underline{\sigma}(R)
      |u_s(\hat{\beta}^+) - u_s(\hat{\beta})|^2 \nonumber \\
    &\overset{\cref{eq:bound:VN:zs}}{\leq}
      6\underline{\sigma}(R) L_s|\tilde{u}(k) - u_s(\hat{\beta})||\tilde{d}|
      + 9\underline{\sigma}(R) L_s^2|\tilde{d}|^2 \nonumber \\
    &\overset{\cref{eq:bound:VN:u}}{\leq}
      6\underline{\sigma}(R) L_sc_{u,k}|\delta\hat{x}||\tilde{d}|
      + 9\underline{\sigma}(R) L_s^2|\tilde{d}|^2 \nonumber \\
    &\leq 3\lambda\underline{\sigma}(R) L_sc_{u,k}|\delta\hat{x}|^2 +
      (9\underline{\sigma}(R) L_s^2 + 3\lambda^{-1}\underline{\sigma}(R)
      L_sc_{u,k})|\tilde{d}|^2 \nonumber \\
    &\leq \lambda\tilde{L}_{1,k}|\delta\hat{x}|^2 +
      \tilde{L}_{2,k}(\lambda)|\tilde{d}|^2 \label{eq:bound:VN:4}
  \end{align}
}{%
  \begin{align}
    \Big| |\hat{x}^+(k) - x_s(\hat{\beta}^+)|_Q^2
    &- |\hat{x}^+(k) - x_s(\hat{\beta})|_Q^2 \Big| \nonumber \\
    &\leq \lambda\hat{L}_{1,k}|\delta\hat{x}|^2 +
      \hat{L}_{2,k}(\lambda)|\tilde{d}|^2 \label{eq:bound:VN:3} \\
    \Big| |\tilde{u}(k) - u_s(\hat{\beta}^+)|_R^2
    &- |\tilde{u}(k) - u_s(\hat{\beta})|_R^2 \Big| \nonumber \\
    &\leq \lambda\tilde{L}_{1,k}|\delta\hat{x}|^2 +
      \tilde{L}_{2,k}(\lambda)|\tilde{d}|^2 \label{eq:bound:VN:4}
  \end{align}
}%
for each \(k\in\intinterval{0}{N-1}\), where \(\hat{L}_{1,k} \defas
3\underline{\sigma}(Q) L_sc_x\), \(\hat{L}_{2,k}(\lambda) \defas
6\underline{\sigma}(Q) L_sL_k' + 9\underline{\sigma}(Q) L_s^2 +
3\lambda^{-1}\gamma L_sc_x\), \(\tilde{L}_{1,k} \defas 3\underline{\sigma}(R) L_sc_{u,k}\),
\(\tilde{L}_{2,k}(\lambda) \defas 9\underline{\sigma}(R) L_s^2 +
3\lambda^{-1}\underline{\sigma}(R) L_sc_{u,k}\), \(c_{u,k} \defas c_u\) if
\(k\in\intinterval{0}{N-2}\), and \(c_{u,N-1} \defas c_{u,f}\).

%% Part 6: Uniform \(\hat{\beta}\) case
\ifthenelse{\boolean{LongVersion}}{%
  For the uniform \(\hat{\beta}\) bound, we have
  \begin{align}
    |V_N(\hat{x}^+,\tilde{\bfu},\hat{\beta})
    &- V_N(\overline{x}^+,\tilde{\bfu},\hat{\beta})| \nonumber \\
    &\leq \sum_{k=0}^{N-1} 2\overline{\sigma}(Q)|\hat{x}^+(k)-\overline{x}^+(k)|
      |\overline{x}^+(k) - x_s(\hat{\beta})|
      \ifthenelse{\boolean{OneColumn}}{}{\nonumber \\
    &\qquad} + \overline{\sigma}(Q)|\hat{x}^+(k)-\overline{x}^+(k)|^2 \nonumber \\
    &\qquad + 2\gamma|\hat{x}^+(N)-\overline{x}^+(N)|
      |\overline{x}^+(N) - x_s(\hat{\beta})|
      \ifthenelse{\boolean{OneColumn}}{}{\nonumber \\
    &\qquad} + \gamma|\hat{x}^+(N)-\overline{x}^+(N)|^2 \nonumber \\
    &\overset{\cref{eq:bound:VN:x}, \cref{eq:bound:VN:5:x},
      \cref{eq:bound:VN:6}}{\leq}
      \sum_{k=0}^{N-1} 2\overline{\sigma}(Q) c_xL_k' |\delta\hat{x}|
      |\tilde{d}| + \overline{\sigma}(Q) (L_k')^2|\tilde{d}|^2 \nonumber \\
    &\qquad + 2\gamma c_{x,f} L_N' |\delta\hat{x}| |\tilde{d}|
      + \gamma (L_N')^2|\tilde{d}|^2 \nonumber \\
    &\leq \sum_{k=0}^{N-1} \lambda\overline{\sigma}(Q) c_xL_k'
      |\delta\hat{x}|^2
      \ifthenelse{\boolean{OneColumn}}{}{\nonumber \\
    &\qquad} + (\overline{\sigma}(Q) (L_k')^2 +
      \lambda^{-1}\overline{\sigma}(Q) c_xL_k')|\tilde{d}|^2 \nonumber \\
    &\qquad + \lambda\gamma c_{x,f} L_N' |\delta\hat{x}|^2
      \ifthenelse{\boolean{OneColumn}}{}{\nonumber \\
    &\qquad} + (\gamma (L_N')^2 + \lambda^{-1}\gamma c_{x,f} L_N')
      |\tilde{d}|^2 \nonumber \\
    &\leq \lambda L_1 |\delta\hat{x}|^2 + L_2(\lambda)|\tilde{d}|^2
      \label{eq:bound:VN:9}
  \end{align}
}{%
  Combining~\cref{eq:bound:VN:x,eq:bound:VN:5:x,eq:bound:VN:6}, we have
  \begin{equation}\label{eq:bound:VN:9}
    |V_N(\hat{x}^+,\tilde{\bfu},\hat{\beta}) -
    V_N(\overline{x}^+,\tilde{\bfu},\hat{\beta})|
    \leq \lambda L_1 |\delta\hat{x}|^2 + L_2(\lambda)|\tilde{d}|^2
  \end{equation}
}%
where \(L_1 \defas \sum_{k=0}^{N-1} \overline{\sigma}(Q) c_xL_k' + \gamma
c_{x,f} L_N'\) and \(L_2(\lambda) \defas \sum_{k=0}^{N-1} \overline{\sigma}(Q)
(L_k')^2 + \lambda^{-1}\overline{\sigma}(Q) c_xL_k' + \gamma (L_N')^2 +
\lambda^{-1}\gamma c_{x,f} L_N'\).

%% Part 7: The whole shebang
\ifthenelse{\boolean{LongVersion}}{%
  Compiling the above results, we have
  \begin{align}
    & \left| |\hat{x}^+(N) - x_s(\hat{\beta}^+)|_{P_f(\hat{\beta}^+)}^2 -
      |\overline{x}^+(N) - x_s(\hat{\beta})|_{P_f(\hat{\beta})}^2  \right| \nonumber \\
    &\overset{\cref{eq:bound:VN:1}}{\leq}
      \left| |\hat{x}^+(N) - x_s(\hat{\beta}^+)|_{P_f(\hat{\beta})}^2 -
      |\overline{x}^+(N) - x_s(\hat{\beta})|_{P_f(\hat{\beta})}^2 \right|
      \ifthenelse{\boolean{OneColumn}}{}{\nonumber \\
    &\qquad} + \sigma_{P_f,x}(|\tilde{d}|)|\delta\hat{x}|^2 +
      \sigma_{P_f,d}(|\tilde{d}|)|\tilde{d}|^2 \nonumber \\
    &\overset{\cref{eq:bound:VN:2}}{\leq}
      \left| |\hat{x}^+(N) - x_s(\hat{\beta}^+)|_{P_f(\hat{\beta})}^2 -
      |\overline{x}^+(N) - x_s(\hat{\beta})|_{P_f(\hat{\beta})}^2 \right| \nonumber \\
    &\qquad + (\sigma_{P_f,x}(|\tilde{d}|) + \lambda\hat{L}_{1,N})
      |\delta\hat{x}|^2
      \ifthenelse{\boolean{OneColumn}}{}{\nonumber \\
    &\qquad} + (\sigma_{P_f,d}(|\tilde{d}|) + \hat{L}_{2,N}(\lambda))|\tilde{d}|^2
      \label{eq:bound:VN:10}
  \end{align}
  and therefore
  \begin{align*}
    |V_N(\hat{x}^+,\tilde{\bfu},\hat{\beta}^+)
    &- V_N(\hat{x}^+,\tilde{\bfu},\hat{\beta})| \\
    &\overset{\cref{eq:bound:VN:3}, \cref{eq:bound:VN:4}}{\leq}
      \sum_{k=0}^{N-1} \lambda(\hat{L}_{1,k}+\tilde{L}_{1,k})
      |\delta\hat{x}|^2
      \ifthenelse{\boolean{OneColumn}}{}{\nonumber \\
    &\qquad} + (\hat{L}_{2,k}(\lambda)+\tilde{L}_{2,k}(\lambda))
      |\tilde{d}|^2 \\
    &\qquad + \left| |\hat{x}^+(N) - x_s(\hat{\beta}^+)|_{P_f(\hat{\beta}^+)}^2 -
      |\overline{x}^+(N) - x_s(\hat{\beta})|_{P_f(\hat{\beta})}^2  \right| \\
    &\overset{\cref{eq:bound:VN:10}}{\leq}
      \sum_{k=0}^{N-1} \lambda(\hat{L}_{1,k}+\tilde{L}_{1,k})
      |\delta\hat{x}|^2
      \ifthenelse{\boolean{OneColumn}}{}{\nonumber \\
    &\qquad} + (\hat{L}_{2,k}(\lambda)+\tilde{L}_{2,k}(\lambda))
      |\tilde{d}|^2 \\
    &\qquad + (\sigma_{P_f,x}(|\tilde{d}|) + \lambda\hat{L}_{1,N})
      |\delta\hat{x}|^2
      \ifthenelse{\boolean{OneColumn}}{}{\nonumber \\
    &\qquad} + (\sigma_{P_f,d}(|\tilde{d}|) +
      \hat{L}_{2,N}(\lambda))|\tilde{d}|^2
  \end{align*}
  Finally~\cref{eq:bound:VN} %
}{%
  Combining \cref{eq:bound:VN:9,eq:bound:VN:3,eq:bound:VN:4,%
    eq:bound:VN:1,eq:bound:VN:2}, we have that~\cref{eq:bound:VN} %
}%
holds so long as \(|\tilde{d}|\leq\delta\), with
\ifthenelse{\boolean{LongVersion}}{%
  \begin{align*}
    a_{V_N,1}
    &\defas \sigma_{P_f,x}(\delta) + \lambda\left( L_1 + \hat{L}_{1,N} +
      \sum_{k=0}^{N-1} \overline L_{1,k} \right) \\
    a_{V_N,2}
    &\defas \sigma_{P_f,d}(\delta) + L_2(\lambda) + \hat{L}_{2,N}(\lambda) +
      \sum_{k=0}^{N-1} \overline L_{2,k}(\lambda)
  \end{align*}
}{%
  \(a_{V_N,1} \defas \sigma_{P_f,x}(\delta) + \lambda\left( L_1 + \hat{L}_{1,N}
    + \sum_{k=0}^{N-1} \overline L_{1,k} \right)\) and \(a_{V_N,2} \defas
  \sigma_{P_f,d}(\delta) + L_2(\lambda) + \hat{L}_{2,N}(\lambda) +
  \sum_{k=0}^{N-1} \overline L_{2,k}(\lambda)\), %
}%
where \(\overline{L}_{1,k} \defas \hat{L}_{1,k} + \tilde{L}_{1,k}\) and
\(\overline{L}_{2,k}(\lambda) \defas \hat{L}_{2,k}(\lambda) +
\tilde{L}_{2,k}(\lambda)\). Finally, to ensure \(a_{V_N,1} <
\underline{\sigma}(Q)\), we can simply choose \(\lambda \in \left(0,
  \frac{\underline{\sigma}(Q) - \sigma_{P_f,x}(\delta)}{L_1 + \hat{L}_{1,N} +
    \sum_{k=0}^{N-1} \overline{L}_{1,k}} \right)\) and \(\delta \in (0,
\sigma_{P_f,x}^{-1}(\underline{\sigma}(Q)))\).

\paragraph{Bound~\cref{eq:bound:lyap:reg}}
Now we have \(a_{V_N,1}\in(0,\underline{\sigma}(Q))\),
\(a_{V_N,2},\tilde{c}_e,\delta,\delta_w>0\), and
\(\tilde\sigma_w,\tilde\sigma_\alpha\in\calKinf\) such that %
\ifthenelse{\boolean{LongVersion}}{%
  \begin{multline*}
    |V_N(\hat{x}^+,\tilde{\bfu}(\hat{x},\hat{\beta}),\hat{\beta}^+)
    - V_N(\overline{x}^+,\tilde{\bfu}(\hat{x},\hat{\beta}),\hat{\beta})| \\
    \leq (a_{V_N,1} + \tilde\sigma_w(|\wP|))|\delta\hat{x}|^2 +
    a_{V_N,2}\tilde{c}_e|(e,e^+)|^2 + a_{V_N,2}\tilde\sigma_\alpha(|\Delta\alpha|)
  \end{multline*}
}{%
  \cref{eq:bound:d,eq:bound:VN} hold %
}%
so long as \(|\tilde{d}|\leq\delta\), \(\alpha\in\calA_c(\delta_w)\), and
\(\Delta\alpha\in\bbA_c(\alpha,\delta_w)\). Without loss of generality, assume
\(\delta_w < \tilde{\sigma}_w^{-1}(\underline{\sigma}(Q) - a_{V_N,1})\). By
\Cref{prop:mpc:robust:feas}, we can choose \(\delta>0\) such that
\(\tilde{\bfu}(\hat{x},\hat{\beta})\in\calU_N(\hat{x}^+,\hat{\beta}^+)\), so
\begin{align*}
  V_N^0(\hat{x}^+,\hat{\beta}^+)
  &\leq V_N(\hat{x}^+,\tilde{\bfu}(\hat{x},\hat{\beta}),\hat{\beta}^+) \\
  &\leq V_N(\overline{x}^+,\tilde{\bfu}(\hat{x},\hat{\beta}),\hat{\beta}) +
    (a_{V_N,1} + \tilde\sigma_w(\delta_w))|\delta\hat{x}|^2 \\
  &\qquad + a_{V_N,2}c_e|(e,e^+)|^2 + a_{V_N,2}\tilde\sigma_\alpha(|\Delta\alpha|) \\
  &\leq V_N^0(\hat{x},\hat{\beta}) - (\underline{\sigma}(Q) -
    a_{V_N,1} - \tilde\sigma_w(\delta_w))|\delta\hat{x}|^2 \\
  &\qquad + a_{V_N,2}c_e|(e,e^+)|^2 + a_{V_N,2}\tilde\sigma_\alpha(|\Delta\alpha|).
\end{align*}
where the first and third inequalities follow by optimality
and~\cref{eq:mpc:descent}. Thus,~\cref{eq:bound:lyap:reg} holds with
\(\tilde{a}_3 \defas \underline{\sigma}(Q) - a_{V_N,1} -
\tilde\sigma_w(\delta_w) > 0\), \(\tilde{a}_4 \defas a_{V_N,2}c_e > 0\), and
\(\sigma_\alpha \defas a_{V,2} \tilde{\sigma}_\alpha \in \calKinf\).

\paragraph{Bound~\cref{eq:bound:lyap:est}}
With \(\delta_w\in(0,\sigma_w^{-1}(\sqrt{\frac{c_3}{4c_4L_s^2}}))\), we
can combine~\cref{eq:est:lyap:b,eq:bound:u,eq:bound:noise} (from
\Cref{assum:est,prop:bound:xu,prop:bound:noise}, respectively)
%% NOTE \cite[Eq.~(1)]{rawlings:ji:2012} is a looser bound than the identity
%% \((a+b)^2\leq 2a^2+2b^2\).
and the identity \((a+b)^2\leq 2a^2+2b^2\) to give%
\ifthenelse{\boolean{LongVersion}}{%
  \begin{align*}
    |\tilde{w}|^2
    &\leq [\sigma_w(|\wP|)|z-z_s(\beta)| + \sigma_\alpha(|\Delta\alpha|)]^2 \\
    &\leq 2[\sigma_w(|\wP|)]^2|z-z_s(\beta)|^2 + 2[\sigma_\alpha(|\Delta\alpha|)]^2 \\
    &\leq 2[\sigma_w(|\wP|)]^2[(1+c_u)|\hat{x}-x_s(\hat{\beta})| + L_s|e|]^2 +
      2[\sigma_\alpha(|\Delta\alpha|)]^2 \\
    &\leq 4[\sigma_w(|\wP|)]^2(1+c_u)^2|\hat{x}-x_s(\hat{\beta})|^2 +
      4[\sigma_w(|\wP|)]^2L_s^2|e|^2 + 2[\sigma_\alpha(|\Delta\alpha|)]^2
  \end{align*}
  and therefore~\cref{eq:bound:lyap:est}, where \(\hat{c}_3 \defas c_3 -
  4c_4[\sigma_w(\delta_w)]^2L_s^2 > 0 \), \(\hat{\sigma}_w(\cdot) \defas
  4c_4[\sigma_w(\cdot)]^2(1+c_u)^2 \in \calKinf\), \(\hat{\sigma}_\alpha(\cdot)
  \defas 2c_4[\sigma_\alpha(\cdot)]^2 \in \calKinf\), and \(L_s>0\) is the
  Lipschitz constant for \(z_s\). %
}{%
  ~\cref{eq:bound:lyap:est}, where \(\hat{c}_3 \defas c_3 -
  4c_4[\sigma_w(\delta_w)]^2L_s^2 > 0 \), \(\hat{\sigma}_w \defas
  4c_4\sigma_w^2(1+c_u)^2, \hat{\sigma}_\alpha \defas 2c_4\sigma_\alpha^2 \in
  \calKinf\), and \(L_s>0\) is the Lipschitz constant for \(z_s\). %
}%
% \hspace*{\fill}~\QED
\end{proof}

\ifthenelse{\boolean{LongVersion}}{%
\subsubsection{Robust stability of offset-free MPC with mismatch}
}{}
%% NOTE This ifthenelse statement just cancels out any latexdiff markup in
%% this section. Otherwise, the spacing gets wonky and text spills onto the
%% next page.
\ifthenelse{\boolean{true}}{%
Finally, we return to the proof of \Cref{thm:mpc:mismatch}.
\paragraph*{Part (a)}
By \Cref{thm:mpc:robust}, we already have
\((\hat{x},\hat{\beta})\in\hat{\calS}_N^\rho\) and
\(\tilde{d}\in\tilde{\bbD}_c(\hat{x},\hat{\beta}) \cap \delta\bbB^{n_d}\)
implies \((\hat{x}^+,\hat{\beta}^+)\in\hat{\calS}_N^\rho\) for some
\(\delta>0\). To ensure \((x,\alpha,\hat{x},\hat{\beta})\) in
\(\calS_N^{\rho,\tau}\) at all times, it suffices to find
\(\tau,\delta_w,\delta_\alpha>0\) such that \(\alpha\in\calA_c(\delta_w)\),
\(\Delta\alpha\in\bbA_c(\alpha,\delta_w)\cap\delta_\alpha\bbB^{n_\alpha}\), and
\(V_e \defas V_e(x,d_s(\alpha),\hat{x},\hat{d})\leq\tau\) implies \(V_e^+ \defas
V_e(x^+,\hat{x}^+)\leq\tau\) and \(|(e,e^+,w)|\leq\delta\).

By \Cref{prop:bound:lyap}, there exist %
\ifthenelse{\boolean{LongVersion}}{%
  constants \(\hat{c}_3,\tilde{c}_e,\delta_w>0\) and functions \(\hat{\sigma}_w,
  \hat{\sigma}_\alpha, \tilde{\sigma}_w, \tilde{\sigma}_\alpha\in\calKinf\) %
}{%
  \(\hat{c}_3,\tilde{c}_e,\delta_w>0\) and \(\hat{\sigma}_w,
  \hat{\sigma}_\alpha, \tilde{\sigma}_w, \tilde{\sigma}_\alpha\in\calKinf\) %
}%
satisfying \cref{eq:bound:d,eq:bound:lyap:est},
% TODO is this ``so long as'' part comprehensive?
so long as \(\alpha=(\ssp,\wP)\in\calA_c(\delta_w)\) and \(\Delta\alpha \in
\bbA_c(\alpha,\delta_w)\). Assume, without loss of generality, that %
\ifthenelse{\boolean{LongVersion}}{%
  \[
    \delta_w < \delta_{w,1} \defas \left(
      \frac{4c_2\tilde{c}_3}{a_1c_1\hat{c}_3}\hat{\sigma}_w + \tilde{\sigma}_w
    \right)^{-1} \left( \frac{a_1\delta^2}{\rho} \right)
  \]
  which implies
  \begin{align*}
    \frac{2c_2\hat{\sigma}_w(\delta_w)\rho}{a_1\hat{c}_3}
    &< \left( \delta^2 - \frac{\tilde{\sigma}_w(\delta_w)\rho}{a_1} \right)
      \frac{c_1}{2\tilde{c}_e}
      \ifthenelse{\boolean{OneColumn}}{, &}{\\}
    \frac{\tilde{\sigma}_w(\delta_w)\rho}{a_1}
    &< \delta^2.
  \end{align*}
}{%
  \(\delta_w < \delta_{w,1} \defas
  \left(\frac{4c_2\tilde{c}_3}{a_1c_1\hat{c}_3}\hat{\sigma}_w +
    \tilde{\sigma}_w\right)^{-1}\left(\frac{a_1\delta^2}{\rho}\right)\), which
  implies \(\frac{2c_2\hat{\sigma}_w(\delta_w)\rho}{a_1\hat{c}_3} < \left(
    \delta^2 - \frac{\tilde{\sigma}_w(\delta_w)\rho}{a_1} \right)
  \frac{c_1}{2\tilde{c}_e}\) and \(\frac{\tilde{\sigma}_w(\delta_w)\rho}{a_1}
  < \delta^2\). %
}%
Then we can take %
\ifthenelse{\boolean{LongVersion}}{%
  \[
    \tau \in \left( \frac{2c_2\hat{\sigma}_w(\delta_w)\rho}{a_1\hat{c}_3},
      \left( \delta^2 - \frac{\tilde{\sigma}_w(\delta_w)\rho}{a_1} \right)
      \frac{c_1}{2\tilde{c}_e} \right)
  \]
  which implies
  \(\frac{\tau\hat{c}_3}{2c_2} > \frac{\hat{\sigma}_w(\delta_w)\rho}{a_1}\) and
  \(\delta^2 > \frac{2\tilde{c}_e\tau}{c_1} +
  \frac{\tilde{\sigma}_w(\delta_w)\rho}{a_1}\).

  From \cref{eq:bound:lyap:est}, we have %
}{%
  \(\tau \in \left( \frac{2c_2\hat{\sigma}_w(\delta_w)\rho}{a_1\hat{c}_3},
    \left( \delta^2 - \frac{\tilde{\sigma}_w(\delta_w)\rho}{a_1} \right)
    \frac{c_1}{2\tilde{c}_e} \right)\), and from \cref{eq:bound:lyap:est}, %
}%
\begin{equation*}
  V_e^+ \leq
  \begin{cases} \frac{\tau}{2} +
    \frac{\hat{\sigma}_w(\delta_w)\rho}{a_1} + \hat{\sigma}_\alpha(|\Delta\alpha|),
    & V_e\leq\frac{\tau}{2} \\
    \tau - \frac{\tau\hat{c}_3}{2c_2} +
    \frac{\hat{\sigma}_w(\delta_w)\rho}{a_1} + \hat{\sigma}_\alpha(|\Delta\alpha|),
    & \frac{\tau}{2} < V_e \leq \tau. \end{cases}
\end{equation*}
But \(\hat{c}_3\leq c_2\) (otherwise we could show \(V_e<0\) with \(\wP=0\),
\(\Delta\alpha=0\), and \(e\neq 0\)) so %
\ifthenelse{\boolean{LongVersion}}{%
  \[
    \frac{\tau}{2} \geq \frac{\tau\hat{c}_3}{2c_2} >
    \frac{\hat{\sigma}_w(\delta_w)\rho}{a_1}
  \]
}{%
  \(\frac{\tau}{2} \geq \frac{\tau\hat{c}_3}{2c_2} >
  \frac{\hat{\sigma}_w(\delta_w)\rho}{a_1}\) %
}%
and we have \(V_e^+ \leq \tau\) so long as %
\ifthenelse{\boolean{LongVersion}}{%
  \[
    |\Delta\alpha| \leq \delta_{\alpha,1} \defas \hat\sigma_\alpha^{-1}\left(
      \frac{\tau\hat{c}_3}{2c_2} - \frac{\hat{\sigma}_w(\delta_w)\rho}{a_1}
    \right)
  \]
}{%
  \(|\Delta\alpha| \leq \delta_{\alpha,1} \defas \hat\sigma_\alpha^{-1}\left(
    \frac{\tau\hat{c}_3}{2c_2} - \frac{\hat{\sigma}_w(\delta_w)\rho}{a_1}
  \right)\), %
}%
which is positive by construction. Moreover, \(V_e,V_e^+\leq\tau\) implies
\(|(e,e^+)|^2 = |e|^2 + |e^+|^2 \leq \frac{2\tau}{c_1}\) and by
\cref{eq:bound:d}, \ifthenelse{\boolean{LongVersion}}{%
  \begin{align*}
    |\tilde{d}|^2
    &\leq \tilde{c}_e|(e,e^+)|^2 + \tilde{\sigma}_w(|\wP|)|\hat{x}-x_s(\hat{\beta})|^2
      + \tilde{\sigma}_\alpha(|\Delta\alpha|) \\
    &\leq \frac{2\tilde{c}_e\tau}{c_1} + \tilde{\sigma}_w(\delta_w)\rho^2 +
      \tilde{\sigma}_\alpha(\delta_\alpha) \\
    &\leq \delta^2
  \end{align*}
}{%
  \(|\tilde{d}|^2 \leq \frac{2\tilde{c}_e\tau}{c_1} +
  \frac{\tilde{\sigma}_w(\delta_w)\rho}{a_1} +
  \tilde{\sigma}_\alpha(|\Delta\alpha|) \leq \delta^2\) %
}%
so long as \(|\Delta\alpha| \leq \delta_{\alpha,2} \defas
\tilde{\sigma}_\alpha^{-1}\left(\delta^2 - \frac{2\tilde{c}_e\tau}{c_1} -
  \frac{\tilde{\sigma}_w(\delta_w)\rho}{a_1}\right)\), which exists and is
positive by construction. Finally, we can take
\(\delta_\alpha\defas\min\set{\delta_{\alpha,1},\delta_{\alpha,2}}\) to achieve
\((x,\alpha,\hat{x},\hat{\beta})\in\calS_N^{\rho,\tau}\) at all times.

\paragraph*{Part (b)}
From part (a), we already have \(\tau,\delta_w,\delta_\alpha>0\) such that
\(\calS_N^{\rho,\tau}\) is RPI.\@ By \Cref{assum:est,thm:mpc:robust} we have
\cref{eq:est:lyap:a,eq:mpc:robust:lyap:a} at all times for some
\(a_1,a_2,c_1,c_2>0\). By \Cref{prop:bound:lyap}, there exist
\(\hat{c}_3,\tilde{a}_3,\tilde{a}_4>0\) and \(\hat{\sigma}_w,
\hat{\sigma}_\alpha, \sigma_\alpha\in\calKinf\) such that
\cref{eq:bound:lyap:reg,eq:bound:lyap:est} at all times. Assume, without loss of
generality, that %
\ifthenelse{\boolean{LongVersion}}{%
  \[
    \delta_w < \delta_{w,2} \defas \hat{\sigma}_w^{-1}\left( \min\left\{
        \frac{c_1\tilde{a}_3}{\tilde{a}_4}, \;
        \frac{a_3\hat{c}_3}{a_4}\frac{c_1}{c_1+c_2} \right\} \right)
  \]
}{%
  \(\delta_w < \delta_{w,2} \defas
  \hat{\sigma}_w^{-1}(\min\set{\frac{c_1\tilde{a}_3}{\tilde{a}_4},
    \frac{a_3\hat{c}_3}{a_4}\frac{c_1}{c_1+c_2}})\) %
}%
in addition to \(\delta_w < \delta_{w,1}\). By \Cref{thm:smallgain}, the system
is RES on \(\calS_N^{\rho,\tau}\) w.r.t.~\(\delta\hat{x}\).

\paragraph*{Part (c)}
By \Cref{prop:ref:robust}, there exist \(c_r,c_g>0\) such that \(|\delta r| \leq
c_r|\delta\hat{x}| + c_g|\tilde{d}|\) where
\(\tilde{d}\defas(e,e^+,\Delta\ssp,\tilde{w})\). Combining this inequality with
\cref{eq:est:lyap:a,eq:bound:lyap:est,eq:bound:d} gives
\[
  |\delta r| \leq c_{r,x}|\delta\hat{x}| + c_{r,e}|e| +
  \tilde{\gamma}_r(|\Delta\alpha|)
\]
where \(c_{r,x} \defas c_r + c_g(\sqrt{\tilde{\sigma}_\alpha(\delta_w)} +
\sqrt{\tilde{c}_e\hat{\sigma}_\alpha(\delta_w)})\), \(c_{r,e} \defas
c_g\sqrt{\tilde{c}_e}(1 + \sqrt{c_2-\hat{c}_3})\), and \(\tilde{\gamma}_r \defas
c_g(\sqrt{\tilde{\sigma}_\alpha} + \sqrt{\tilde{c}_e\hat{\sigma}_\alpha})\).
Then
\[
  |(\delta r,e)| \leq \tilde{c}_r|(\delta\hat{x},e)| +
  \tilde{\gamma}_r(|\Delta\alpha|)
\]
where \(\tilde{c}_r \defas c_{r,x} + c_{r,e} + 1\). Finally, RES
w.r.t.~\(\delta\hat{x}\) gives
\[
  |(\delta\hat{x}(k),e(k))| \leq
  \tilde{c}\lambda^k|(\delta\hat{x}(0),\overline{e})| + \sum_{j=0}^k \lambda^j
  \tilde\gamma(|\Delta\alpha(k-j)|)
\]
for some \(\tilde{c}>0\), \(\lambda\in(0,1)\), and \(\tilde\gamma\in\calK\), and
therefore
\begin{align*}
  |(\delta r(k),e(k))|
  &\leq \tilde{c}_r|(\delta\hat{x}(k),e(k))| + \tilde\gamma_r(|\Delta\alpha(k)|) \\
  &\leq c\lambda^k|(\delta\hat{x}(0),\overline{e})| +
    \sum_{j=0}^k \lambda^j \gamma(|\Delta\alpha(k-j)|)
\end{align*}
where \(c\defas\tilde{c}_r\tilde{c} > 0\) and \(\gamma\defas
\tilde{c}_r\tilde\gamma + \tilde{\gamma}_r\in\calKinf\). %
\hspace*{\fill}~\QED
}{}

\ifthenelse{\boolean{LongVersion}}{%
\section{Establishing steady-state target problem assumptions}\label[appendix]{app:ssop}
}{}%
\subsection{Proof of \Cref{lem:sstp}}\label[appendix]{app:sstp}
To show \Cref{lem:sstp}, we require the following result on sensitivity of
optimization problems.

\begin{proposition}\label{prop:opt}
  Suppose \(F:\real^{n_\xi}\times\real^{n_\omega}\rightarrow\nnegreal\),
  \ifthenelse{\boolean{true}}{%
    \(G:\real^{n_\xi}\times\real^{n_\omega}\rightarrow\real^{n_G}\), %
  }{}%
  and \(H:\real^{n_\xi}\times\real^{n_\omega}\rightarrow\real^{n_H}\) are
  continuously differentiable. Consider the optimization problem
  \begin{equation}\label{eq:opt}
    \min_{\xi\in\Xi(\omega)} F(\xi,\omega)
  \end{equation}
  where \(\Xi(\omega) \defas \set{ \xi \in \real^{n_\xi} | G(\xi,\omega) = 0, \;
    H(\xi,\omega) \leq 0 }\).
  \ifthenelse{\boolean{LongVersion}}{%
    Suppose the following conditions hold.
    \begin{enumerate}[(i)]
    \item \emph{Local uniqueness}: \(\xi_0\) uniquely solves \cref{eq:opt} at
      \(\omega_0\).
    \item \emph{Inf-compactness}: There exist \(\alpha,\delta>0\) and a compact
      set \(C\subseteq\real^{n_\xi}\) such that, for each
      \(|\omega|\leq\delta\), the level set
      \[
        L_\alpha(\omega) \defas \set{ \xi \in \Xi(\omega) | F(\xi,\omega) \leq
          \alpha }
      \]
      is nonempty and contained in \(C\).
    \item \emph{Regularity}: \(\partial_\xi G(\xi_0,\omega_0)\) is full row rank.
    \item \emph{Locally inactive constraints}: \(H(\xi_0,\omega_0) < 0\).
    \end{enumerate}
    Then %
  }{%
    If (i) \(\xi_0\) uniquely solves \cref{eq:opt} at \(\omega_0\); (ii) there
    exist \(\alpha,\delta>0\) and a compact set \(C\subseteq\real^{n_\xi}\) such
    that, for each \(|\omega|\leq\delta\), the set \(\Xi(\omega) \cap
    \lev_\alpha F(\cdot,\omega)\) is nonempty and contained in \(C\); (iii)
    \(\partial_\xi G(\xi_0,\omega_0)\) is full row rank; and (iv)
    \(H(\xi_0,\omega_0) < 0\); then %
  }%
  there exists a continuous function
  \(\xi^0:\real^{n_\omega}\rightarrow\real^{n_\xi}\) that uniquely solves
  \cref{eq:opt} in a neighborhood of \(\omega=\omega_0\).
\end{proposition}
\begin{proof}
  %% Prop.~4.4(i)   <= Assumption 1
  %% Prop.~4.4(ii)  <= Continuity of (G,H)
  %% Prop.~4.4(iii) <= (ii) above
  %% Prop.~4.4(iv)  <= Continuous differentiability of (G,H), (iii) and (iv)
  %%                   above, p. 71 of B&S (2000)
  It follows immediately from %
  \ifthenelse{\boolean{LongVersion}}{%
    Proposition~4.4~ of \cite{bonnans:shapiro:2000} %
  }{%
    \cite[Prop.~4.4]{bonnans:shapiro:2000} %
  }%
  and the discussions in \cite[pp.~71,~264]{bonnans:shapiro:2000} that
  \(\calS(\omega) \defas \argmin_{\xi\in\Xi(\omega)} F(\xi,\omega)\) is outer
  semicontinuous\footnote{A function \(\mathcal{F} : \real^m \rightarrow
    \mathcal{P}(\real^n)\) is \emph{outer semicontinuous at \(x=x_0\)} if
    \(\limsup_{x\rightarrow x_0} \mathcal{F}(x) \subseteq \mathcal{F}(x_0)\).}
  at \(\omega=\omega_0\). But \(\calS(\omega_0)=\set{\xi_0}\) is a singleton,
  so, for it to be outer semicontinuous at \(\omega=\omega_0\), it must be a
  singleton in a neighborhood of \(\omega=\omega_0\). In other words, there
  exists a continuous function \(\xi^0 : \real^{n_\omega} \rightarrow
  \real^{n_\xi}\) such that \(\calS(\omega) = \set{ \xi^0(\omega) }\) in a
  neighborhood of \(\omega=\omega_0\).
\end{proof}

Returning to the proof of \Cref{lem:sstp}, we have the following relationships
between the conditions of \Cref{lem:sstp,prop:opt}: (e,f) \(\Rightarrow\) (i),
\Cref{assum:sstp:exist} \(\Rightarrow\) (ii), (b) \(\Rightarrow\) (iii), and
(a,c,d) \(\Rightarrow\) (iv).
% \Cref{lem:sstp}(e,f) implies
% \Cref{prop:opt}(i), \Cref{assum:sstp:exist} implies \Cref{prop:opt}(ii), and
% \Cref{lem:sstp}(b) implies \Cref{prop:opt}(iii), and \Cref{lem:sstp}(a,c,d)
% implies \Cref{prop:opt}(iv).
Thus, there exists \(\delta_1>0\) and a continuous function
\(z_s:\calB\rightarrow\bbX\times\bbU\) such that \(z_s(\beta)\) uniquely solves
\cref{eq:sstp} for all \(|\beta|\leq\delta_1\). Let \(0 < \delta < \delta_1\),
\(\delta_0 \defas \delta - \delta_1\), \(\calB_c \defas \delta\bbB^{n_\beta}\),
and \(\overline{\calB}_c \defas \delta_1\bbB^{n_\beta}\). Defining
\(\hat{\calB}_c\) as in \Cref{assum:sstp}(a), we have
\(|\hat\beta|\leq|\beta|+|e_d|\leq\delta+\delta_0=\delta_1\) for each
\(\hat{\beta}=(\ssp,\hat{d})\in\hat{\calB}_c\), and therefore \(\calB_c
\subseteq \hat{\calB}_c \subseteq \overline{\calB}_c \subseteq \calB\), which
completes the proof. %
\hspace*{\fill}~\QED

\subsection{Proof of
  \Cref{lem:sstp:mismatch}}\label[appendix]{app:sstp:mismatch}
%% TODO
% Assume all twice continuously differentiable functions on subsets of Euclidean
% spaces have been extended to twice continuously differentiable functions on
% the whole Euclidean space using appropriately defined partitions of unity (cf.
% \cite[Lem.~2.26]{lee:2012}).
%
From \Cref{lem:sstp}, there exists a neighborhood of the origin
\(\calB_c\subseteq\calB\) and a continuous function \(z_s \defas (x_s,u_s) :
\calB \rightarrow \bbX \times \bbU\) satisfying \Cref{assum:sstp} and uniquely
solving \cref{eq:sstp} on \(\hat{\calB}_c\). For convenience, we define
\(z\defas (x,u)\), \(\zP \defas (\xP,d)\), \(\alpha\defas(\ssp,\wP)\),
\(\beta\defas(\ssp,d)\), and
\begin{align*}
  G_1(z,\beta)
  &\defas \begin{bmatrix} f(x,u,d) - x \\ g(u,h(x,u,d)) - \rsp \end{bmatrix}, \\
  G_2(z,\zP,\alpha)
    &\defas \begin{bmatrix} \fP(\xP,u,\wP) - \xP \\ \hP(\xP,u,\wP) - h(x,u,d)
            \end{bmatrix}, \\
  \mathcal{L}(z,\beta,\lambda) &\defas V_s(z,\beta) + \lambda^\top G_1(z,\beta).
\end{align*}
The system of equations
\begin{equation}\label{eq:sstp:mismatch:a}
  \mathcal{F}(z,\zP,\lambda,\alpha) \defas
  \begin{bmatrix} \partial_z \mathcal{L}(z,\beta,\lambda) \\
    G_1(z,\beta) \\ G_2(z,\zP,\alpha) \end{bmatrix} = 0
\end{equation}
is the combination of the stationary point condition for the Lagrangian of
\cref{eq:sstp} with the steady-state disturbance problem \cref{eq:ssop}. We seek
to use the implicit function theorem on \cref{eq:sstp:mismatch:a} to solve these
problems simultaneously.

We already have \(\mathcal{F}(0,0,0,0) = 0\) by assumption. Next, we need to
show \(M_0 \defas \partial_{(z,\zP,\lambda)} \mathcal{F}(0,0,0,0)\) is
invertible. Evaluating derivatives, we have
\begin{equation*}
  M_0 = \begin{bmatrix} M_3^\top\partial_z^2 \ell_s(0,0)M_4 & M_1^\top \\
          \partial_{(\zP,\lambda)} G(0,0,0) & 0 \end{bmatrix}
\end{equation*}
where \(G \defas \begin{bmatrix} G_1^\top & G_2^\top \end{bmatrix}^\top\), \(M_3
\defas \begin{bsmallmatrix} 0 & I \\ C & D \end{bsmallmatrix}\), and \(M_4
\defas \begin{bsmallmatrix} 0 & I & 0 & 0 \\ C & D & 0 &
  C_d \end{bsmallmatrix}\). Defining the invertible matrices
\begin{align*}
  T_1 &\defas \begin{bsmallmatrix} I_n & 0 & 0 & 0 \\ 0 & I_{n_r} & 0 & 0 \\
                I_n & 0 & -I_n & 0 \\ 0 & 0 & 0 & I_{n_y} \end{bsmallmatrix},
  & T_2 &\defas \begin{bsmallmatrix} I_n & 0 & 0 & 0 \\ 0 & I_{n_u} & 0 & 0 \\
                  I_n & 0 & -I_n & 0 \\ 0 & 0 & 0 & I_{n_d} \end{bsmallmatrix},
\end{align*}
we have %
\ifthenelse{\boolean{LongVersion}}{%
  \[
    T_1\partial_{(z,\zP)} G(0,0,0) T_2 =
    \begin{bmatrix} M_1 & * \\ 0 & M_2 \end{bmatrix}.
  \]
}{%
  \(T_1\partial_{(z,\zP)} G(0,0,0) T_2 =
  \begin{bsmallmatrix} M_1 & * \\ 0 & M_2 \end{bsmallmatrix}\). %
}%
Note that \(M_4T_2 = M_4\) and \(M_4=\begin{bmatrix} M_3 & * \end{bmatrix}\).
Define the invertible matrices %
\ifthenelse{\boolean{LongVersion}}{%
  \begin{align*}
    T_3 &\defas\begin{bsmallmatrix} I_{n+n_u} \\ & T_1 \end{bsmallmatrix},
    & T_4 &\defas\begin{bsmallmatrix} T_2 \\ & I_{n+n_d} \end{bsmallmatrix},
            \ifthenelse{\boolean{OneColumn}}{&}{\\}
    P &\defas\begin{bsmallmatrix} I_{n+n_u} & 0 & 0 \\ 0 & 0 & I_{n+n_d} \\
               0 & I_{n+n_r} & 0 \end{bsmallmatrix}.
  \end{align*}
}{%
  \(T_3 \defas \begin{bsmallmatrix} I_{n+n_u} \\ & T_1 \end{bsmallmatrix}\),
  \(T_4 \defas \begin{bsmallmatrix} T_2 \\ & I_{n+n_d} \end{bsmallmatrix}\), \(P
  \defas
  \begin{bsmallmatrix} I_{n+n_u} & 0 & 0 \\ 0 & 0 & I_{n+n_d} \\
    0 & I_{n+n_r} & 0 \end{bsmallmatrix}\). %
}%
Using these invertible matrices, we have
\begin{equation*}
  T_3M_0T_4P = \begin{bmatrix} M_5 & * \\ 0 & M_2 \end{bmatrix}
\end{equation*}
where \(M_5 \defas \begin{bsmallmatrix} M_3^\top\partial_{(u,y)}^2\ell_s(0,0)M_3
  & M_1^\top \\ M_1 & 0 \end{bsmallmatrix}\), and therefore \(M_0\) is
invertible if and only if both \(M_2\) and \(M_5\) are as well. We already have
\(M_2\) invertible by assumption. Next,
\[
  \begin{bmatrix} M_3 \\ M_1 \end{bmatrix} =
  \begin{bsmallmatrix} 0 & I \\ C & D \\ A-I & B \\ H_yC & H_yD+H_u
  \end{bsmallmatrix}
\]
is full column rank, which implies \(M_5 =
\begin{bsmallmatrix} M_3 \\ M_1 \end{bsmallmatrix}^\top
\begin{bsmallmatrix} \partial_{(u,y)}^2\ell_s(0,0) \\ & I \end{bsmallmatrix}
\begin{bsmallmatrix} M_3 \\ M_1 \end{bsmallmatrix}\) is invertible since
\(\partial_{(u,y)}^2\ell_s(0,0)\) is invertible. Finally, \(M_0\) is invertible.

By the implicit function theorem~\cite[Thm.~13.7]{apostol:1974}, there exist
\(\delta_1>0\) and continuously differentiable functions
\((z_s^*,\zPs,\lambda^*) : \real^{n_\alpha} \rightarrow \real^{n+n_u} \times
\real^{n+n_d} \times \real^{n+n_r}\) such that \((z,\zP,\lambda) =
(z_s^*(\alpha),\zPs(\alpha),\lambda^*(\alpha))\) solve \cref{eq:sstp:mismatch:a}
for all \(|\alpha|\leq\delta_1\). Since \(\calB_c\) contains a neighborhood of
the origin, there exists \(0<\delta\leq\delta_1\) such that
\(\beta=(\ssp,d_s(\alpha))\in\calB_c\) for all \(|(\ssp,\wP)|\leq\delta\). But
\(z_s(\beta)\) uniquely solves \cref{eq:sstp} for all \(\beta\in\calB_c\), and
(since \(M_1\) is full row) we have the necessary condition
\(\partial_{(z,\lambda)} \mathcal{L}(z_s(\beta),\beta,\lambda) = 0\) for some
\(\lambda\) and each \(\beta\in\calB_c\). Therefore \(z_s(\ssp,d_s(\alpha)) =
z_s^*(\alpha)\) for all \(\alpha=(\ssp,\wP) \in \calA_c \defas
\delta\bbB^{n_\alpha}\). %
\ifthenelse{\boolean{LongVersion}}{%
  Finally, \Cref{assum:sstp:mismatch}(e) follows automatically from the fact
  that the set \(\calA_c\) is a ball centered at the origin. %
}{}%
\hspace*{\fill}~\QED

\ifthenelse{\boolean{LongVersion}}{%
\section{Construction of terminal ingredients}\label[appendix]{app:terminal}
}{%
\subsection{Construction of terminal ingredients}\label[appendix]{app:terminal}
}%
Let \(Q\in\realm{n}{n}\) and \(R\in\realm{n_u}{n_u}\) be positive definite.
Suppose \Cref{assum:cont,assum:cons,assum:sstp:exist,assum:sstp} hold with
\(\calB=\hat{\calB}_c\)
%% TODO can soft constraints be reintroduced?
and \(n_c=0\), %
\(\partial_{(x,u)}^2 f_i,i\in\intinterval{1}{n}\) exist and are locally bounded,
and
\[
  (A(\beta),B(\beta)) \defas (\partial_x f(z_s(\beta),d), \partial_u f(z_s(\beta),d))
\]
is stabilizable for each \(\beta=(\ssp,d)\in\calB\).

Fix \(\beta=(\ssp,d)\in\calB\). Since \((A,B)\) is stabilizable, there exists a
positive definite \(P = P(A,B)\) that uniquely solves the following discrete
algebraic Riccati equation,
\[
  P = A^\top PA + Q - A^\top PB(B^\top P B + R)^{-1} B^\top PA
\]
where dependence on \(\beta\) has been suppressed for brevity. The solution
\(P\) is continuous at each \((A,B)\) such that \((A,B)\) is stabilizable and
\((Q,R)\) are positive definite~\citep{sun:1998}.%
\ifthenelse{\boolean{LongVersion}}{%
  \footnote{In fact, \cite{sun:1998} needed only \((A,Q^{1/2})\) detectable to
    derive perturbation bounds. However, \Cref{assum:quad} guarantees positive
    definiteness of \(Q\), so we get this automatically.} %
}{} %
Moreover, since \(f\) is twice differentiable and \((x_s,u_s)\) are continuous
on \(\calB\), so \((A(\beta),B(\beta))\) and \(P(\beta) \defas
P(A(\beta),B(\beta))\) must be continuous on \(\calB\), \Cref{assum:quad} holds
for \(P_f(\beta) \defas 2P(\beta)\).

Next, with \(K \defas PB(B^\top PB + R)^{-1}\), \(A_K \defas A - BK\), and \(Q_K
\defas Q + K^\top RK\), we have \(A_K^\top P_fA_K - P_f = -2Q_K\), where
dependence on \(\beta\) has been suppressed for brevity. Then
\begin{equation}\label{eq:terminal:1}
  V_f(\overline{x}^+,\beta) - V_f(x,\beta) \leq -2|\delta x|_{Q_K(\beta)}^2
\end{equation}
where \(\overline{x}^+ \defas A_K(\beta)\delta x + x_s(\beta)\) and \(\delta
x\defas x-x_s(\beta)\). Since the second derivatives \(H_i(x,\beta) \defas
\partial_{(x,u)}^2 f_i(x,\kappa_N(x,\beta),d)\) are locally bounded, %
\ifthenelse{\boolean{LongVersion}}{%
  the maximum
  \[
    c_x \defas \max_{(x,\beta) \in \hat{\calS}_N^\rho} \sum_{i=1}^n
    \overline{\sigma}(\partial_{(x,u)}^2 H_i(x,\beta))
  \]
}{%
  \(c_x \defas \max_{(x,\beta) \in \hat{\calS}_N^\rho} \sum_{i=1}^n
  \overline{\sigma}(\partial_{(x,u)}^2 H_i(x,\beta))\) %
}%
exists (independently of \(\beta\)). By Taylor's
theorem~\cite[Thm.~12.14]{apostol:1974}, \(|x^+ - \overline{x}^+| \leq
c_x|\delta x|^2\) where \(x^+ \defas f(x,\kappa_f(x,\beta),d)\) and
\(\kappa_f(x,\beta) \defas -K(\beta)\delta x + u_s(\beta)\). With
\(a(\beta)\defas 2c_x\overline{\sigma}([A_K(\beta)]^\top P_f(\beta))\) and
\(b(\beta) \defas c_x^2\overline{\sigma}(P_f(\beta))\),
\begin{equation}\label{eq:terminal:2}
  |V_f(x^+,\beta) - V_f(\overline{x}^+,\beta)| \leq a(\beta)|\delta x|^3 +
  b(\beta)|\delta x|^4
\end{equation}
and combining \cref{eq:terminal:1} with \cref{eq:terminal:2}, we have
\begin{align}
  V_f\ifthenelse{\boolean{OneColumn}}{}{
  &} (x^+,\beta) - V_f(x,\beta) + \ell(x,\kappa_f(x,\beta),\beta)
    \ifthenelse{\boolean{OneColumn}}{}{\nonumber \\}
  &\leq -|\delta x|_{Q_K(\beta)}^2 + V_f(x^+,\beta) -
    V_f(\overline{x}^+,\beta) \nonumber \\
  &\leq -[ c(\beta) - b(\beta)|\delta x| - a(\beta)|\delta x|^2 ]
    |\delta x|^2 \label{eq:terminal:3}
\end{align}
where \(c(\beta) \defas \underline{\sigma}(Q_K(\beta))\). The polynomial
\(p_\beta(s) = c(\beta) - b(\beta)s - a(\beta)s^2\) has roots at
\ifthenelse{\boolean{LongVersion}}{%
  \[
    s_{\pm}(\beta) \defas \frac{-b(\beta) \pm \sqrt{[b(\beta)]^2 +
        4a(\beta)c(\beta)}}{2a(\beta)}
  \]
}{%
  \(s_{\pm}(\beta) \defas \frac{-b(\beta) \pm \sqrt{[b(\beta)]^2 +
      4a(\beta)c(\beta)}}{2a(\beta)}\) %
}%
and is positive in between. Moreover, \(s_{\pm}\) are continuous over
\(\calB\) because \((a,b,c)\) are as well, and \(s_{\pm}(\beta)\) are positive
and negative, respectively. Define %
\ifthenelse{\boolean{LongVersion}}{%
  \[
    c_f \defas \min_{\beta\in\calB} \underline{\sigma}(P_f(\beta))
    [s_{+}(\beta)]^2
  \]
}{%
  \(c_f \defas \min_{\beta\in\calB} \underline{\sigma}(P_f(\beta))
  [s_{+}(\beta)]^2\) %
}%
which exists and is positive due to continuity and positivity of \(s_{+}\) and
\(\underline{\sigma}(P_f(\cdot))\) and compactness of \(\calB\). Finally, we
have that \(V_f(x,\beta)\leq c_f\) implies %
\ifthenelse{\boolean{LongVersion}}{%
  \[
    \underline{\sigma}(P_f(\beta))|\delta x|^2 \leq V_f(x,\beta) \leq c_f
  \]
}{%
  \(\underline{\sigma}(P_f(\beta))|\delta x|^2 \leq V_f(x,\beta) \leq c_f\) %
}%
and therefore %
\ifthenelse{\boolean{LongVersion}}{%
  \[
    |\delta x| \leq \sqrt{\frac{c_f}{\underline{\sigma}(P_f(\beta))}} \leq
    s_{+}(\beta)
  \]
}{%
  \(|\delta x| \leq \sqrt{\frac{c_f}{\underline{\sigma}(P_f(\beta))}} \leq
  s_{+}(\beta)\) %
}%
and \cref{eq:terminal:3} implies \Cref{assum:stabilizability} with
\(P_f(\beta)\) and \(c_f>0\) as constructed.